\documentclass[journal]{vgtc}                     

\onlineid{0}

\vgtccategory{Research}

\title{MR-Compare: A Mixed-Reality Framework for Spatially Grounded Visual Comparison of 3D Gaussian Splatting and Mesh Reconstructions with the Physical Environment}

\author{
  \authororcid{Changrui Zhu}{0000-0002-8192-5747},
  \authororcid{Ernst Kruijff}{0000-0003-1625-0955}, 
  \authororcid{Pengju Zhang}{0000-0001-6480-9480}, and 
  \authororcid{Simon Julier}{0000-0003-4380-137X}
}

\authorfooter{
  \item Changrui Zhu (Changrui.zhu.19@ucl.ac.uk), 
  Pengju Zhang (pengju.zhang.21@ucl.ac.uk), 
  and Simon Julier (s.julier@ucl.ac.uk) are with University College London (UCL). 
  \item
  	Ernst Kruijff (ernst.kruijff@h-brs.de) is with Bonn-Rhein-Sieg University of Applied Sciences.

}

\abstract{We introduce \textbf{MR-Compare}, a mixed reality framework for spatially grounded visual comparison between 3D Gaussian splatting and mesh reconstructions with live video see-through (VST). Implemented on a PC-tethered Meta Quest~3, it combines a two-stage registration pipeline with a \textit{3D Slider} for cross-media comparison.
We evaluated five representative desktop and mobile reconstruction workflows through a real-world benchmark with an exploratory user study ($n=30$) in two static indoor rooms. MR-Compare achieved centimetre-level translation error across all workflows. The two desktop 3DGS workflows showed the strongest overall pattern, with 3DGS-MCMC yielding the lowest registration error and strongest VST-referenced visual consistency. Room-session measures indicated high perceived usability and low workload. We further propose an \textit{anisotropy filter}, a zero-shot module that leverages Gaussian anisotropies to improve 3DGS registration in MR-Compare. A controlled Replica threshold sweep shows that moderate pruning can improve robustness and reduce residual errors. These results establish system-level feasibility in the tested setting rather than task-level effectiveness or standalone deployment.
The project is available at \url{https://github.com/changruizhu96/MR-Compare}.

}


\teaser{
  \centering
  \includegraphics[width=0.9\linewidth]{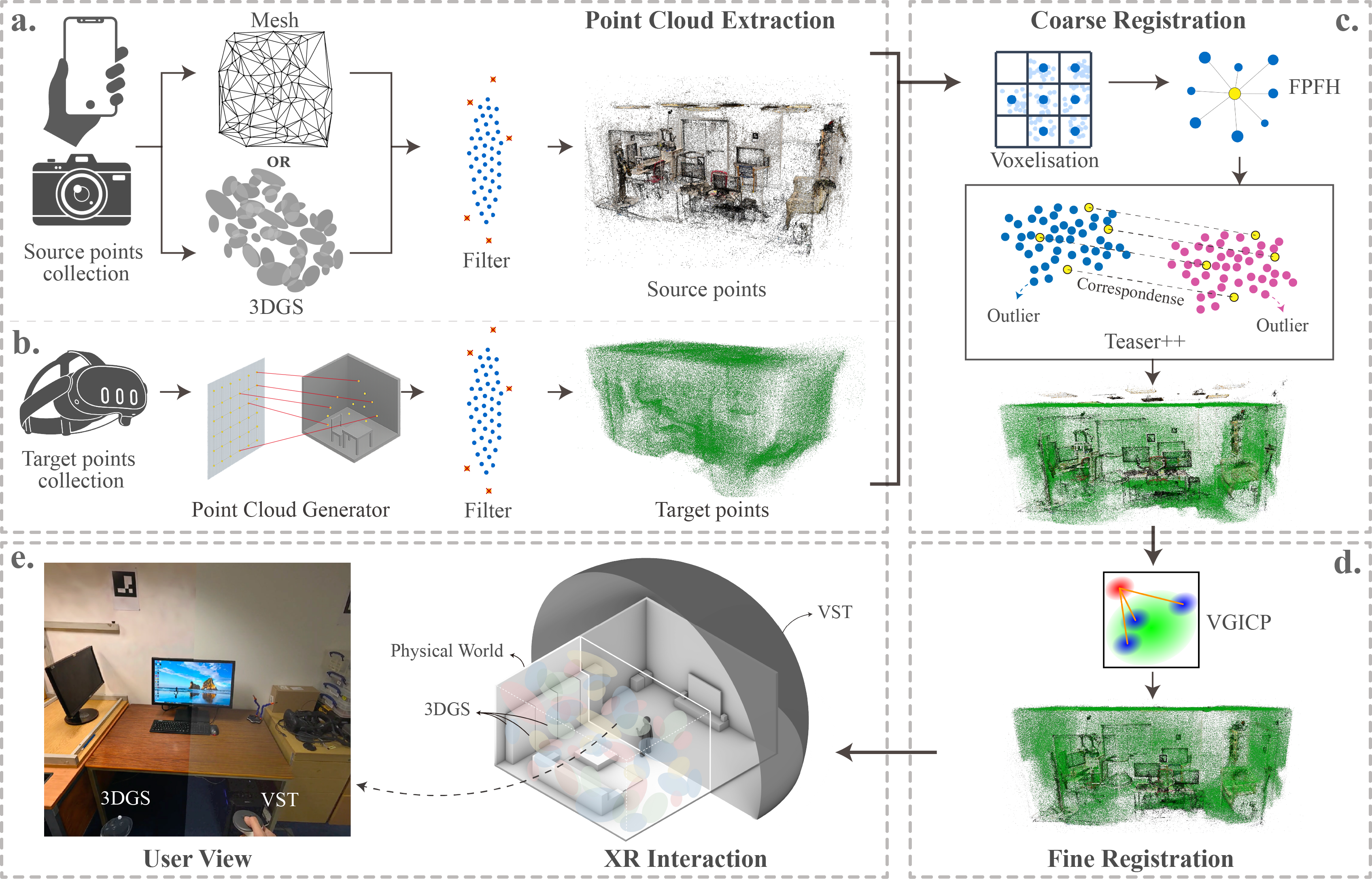}
\caption[Architecture of MR-Compare.]{
Architecture of the MR-Compare: from data collection and extraction (a: source points; b: target points), through the spatial registration (c: coarse registration; d: fine registration), to the XR interaction (e: interaction mechanism and a user view).
}
  \label{fig:teaser}}

\graphicspath{{figures/}{./}} 

\usepackage{booktabs}

\usepackage{mathptmx}                  
\usepackage{times}  
\usepackage{tabularx} 
\usepackage{makecell} 

\usepackage{amsmath}
\usepackage{amssymb}   
\usepackage{algorithm}
\usepackage{algpseudocode}
\usepackage{verbatim}
\usepackage{graphicx}
\usepackage{subcaption}  
\usepackage{pdfpages} 
\usepackage{xcolor}
\usepackage{mathtools}
\usepackage{bm}  

\usepackage{amsmath,amssymb}

\usepackage[mathcal]{eucal}

\usepackage[table]{xcolor}
\usepackage{tikz}      
\usepackage{multirow}

\newcommand{\supplementref}{%
  the Appendix%
}
\DeclareRobustCommand{\appref}[2]{%
  \iflabelexists{#1}
    {#2}
    {\supplementref}%
}
\DeclareRobustCommand{\appfullref}[2]{%
  \iflabelexists{#1}
    {#2}
    {}%
}

\usepackage{booktabs,threeparttable,siunitx,longtable,array}
\usepackage{caption}
\newcommand{\pformat}[1]{%
  \ifdim #1 pt < .001pt $<$.001\else \num[round-precision=3]{#1}\fi
}

\ExplSyntaxOn
\NewDocumentCommand{\pval}{m}
 {
  \fp_compare:nTF {#1 < 0.001}
    { $<$.001 }
    {
      \tl_set:Nx \l_tmpa_tl
        {
          \num[
            round-mode = places,
            round-precision = 3
          ]{#1}
        }
      \regex_replace_once:nnN { ^0\. } { . } \l_tmpa_tl
      \tl_use:N \l_tmpa_tl
    }
 }
\ExplSyntaxOff

\definecolor{HeatC}{HTML}{2166AC}

\definecolor{HeatR}{RGB}{178,24,43}

\definecolor{HeatCold}{RGB}{49,130,189} 
\definecolor{HeatWarm}{RGB}{239,59,44}

\def\TminSum{0}\def\TmaxSum{5}
\def\RminSum{0.6}\def\RmaxSum{3.5}

\def\TminSumTwo{1}\def\TmaxSumTwo{4.5}
\def\RminSumTwo{1.7}\def\RmaxSumTwo{2.2}

\def\PTSOneMin{0}
\def\PTSOneMax{8}

\def\PTSTwoMin{0}
\def\PTSTwoMax{8}

\def\TIMEOneMin{0.3}
\def\TIMEOneMax{3.50}
\def\TIMETwoMin{0.6}
\def\TIMETwoMax{4.5}

\def\TminDet{0}\def\TmaxDet{6}
\def\RminDet{0.1}\def\RmaxDet{5}

\newcount\heatcount

\newcommand{\calcHeat}[3]{%
  \pgfmathtruncatemacro{\heattmp}{%
    ifthenelse(#3-#2==0, 0,
      max(0, min(100, round(100*(#1-#2)/(#3-#2)))))
  }%
  \global\heatcount=\heattmp\relax
}

\newcommand{\cellTsum}[2]{%
  \pgfmathsetmacro{\reversedvalue}{\TmaxSum + \TminSum - (#1)}%
  \calcHeat{\reversedvalue}{\TminSum}{\TmaxSum}%
  \cellcolor{HeatC!\the\heatcount}\num{#1} \, $\pm$ \num{#2}%
}

\newcommand{\cellTsumTwo}[2]{%
  \pgfmathsetmacro{\reversedvalue}{\TmaxSumTwo + \TminSumTwo - (#1)}%
  \calcHeat{\reversedvalue}{\TminSumTwo}{\TmaxSumTwo}%
  \cellcolor{HeatC!\the\heatcount}\num{#1} \, $\pm$ \num{#2}%
}

\newcommand{\cellRsum}[2]{%
  \pgfmathsetmacro{\reversedvalue}{\RmaxSum + \RminSum - (#1)}%
  \calcHeat{\reversedvalue}{\RminSum}{\RmaxSum}%
  \cellcolor{HeatC!\the\heatcount}\num{#1} \, $\pm$ \num{#2}%
}

\newcommand{\cellRsumTwo}[2]{%
  \pgfmathsetmacro{\reversedvalue}{\RmaxSumTwo + \RminSumTwo - (#1)}%
  \calcHeat{\reversedvalue}{\RminSumTwo}{\RmaxSumTwo}%
  \cellcolor{HeatC!\the\heatcount}\num{#1} \, $\pm$ \num{#2}%
}

\newcommand{\cellTIMEone}[2]{%
  \calcHeat{\TIMEOneMax + \TIMEOneMin - (#1)}{\TIMEOneMin}{\TIMEOneMax}%
  \cellcolor{HeatC!\the\heatcount}\num{#1} \, $\pm$ \num{#2}%
}

\newcommand{\cellPTSone}[1]{%
  \calcHeat{#1}{\PTSOneMin}{\PTSOneMax}%
  \cellcolor{HeatC!\the\heatcount}\num{#1}
}

\newcommand{\cellTIMEtwo}[2]{%
  \calcHeatLo{#1}{\TIMETwoMin}{\TIMETwoMax}%
  \cellcolor{HeatC!\the\heatcount}\num{#1} \, $\pm$ \num{#2}%
}

\newcommand{\cellPTStwo}[1]{%
  \calcHeat{#1}{\PTSTwoMin}{\PTSTwoMax}%
  \cellcolor{HeatC!\the\heatcount}\num{#1}
}

\newcommand{\cellTdet}[1]{%
  \calcHeatLo{#1}{\TminDet}{\TmaxDet}%
  \cellcolor{HeatC!\the\heatcount}\num{#1}%
}
\newcommand{\cellRdet}[1]{%
  \calcHeatLo{#1}{\RminDet}{\RmaxDet}%
  \cellcolor{HeatC!\the\heatcount}\num{#1}%
}

\def\PSNRmin{13}\def\PSNRmax{23}
\def\SSIMmin{0.67}\def\SSIMmax{0.8}
\def\LPIPSmin{0.25}\def\LPIPSmax{0.5}
\def\DISTSmin{0.1}\def\DISTSmax{0.22}

\definecolor{HeatMono}{RGB}{244,67,54} %

\newcommand{\numThree}[1]{\num[round-mode=places, round-precision=2]{#1}}

\newcommand{\calcHeatLo}[3]{
  \pgfmathparse{int(round(100*(#3-#1)/(#3-#2)))}%
  \global\heatcount=\pgfmathresult\relax
}

\newcommand{\cellPSNR}[2]{%
  \calcHeat{#1}{\PSNRmin}{\PSNRmax}%
  \cellcolor{HeatC!\the\heatcount}\numThree{#1}~$\pm$~\numThree{#2}%
}
\newcommand{\cellSSIM}[2]{%
  \calcHeat{#1}{\SSIMmin}{\SSIMmax}%
  \cellcolor{HeatC!\the\heatcount}\numThree{#1}~$\pm$~\numThree{#2}%
}
\newcommand{\cellLPIPS}[2]{%
  \calcHeatLo{#1}{\LPIPSmin}{\LPIPSmax}%
  \cellcolor{HeatC!\the\heatcount}\numThree{#1}~$\pm$~\numThree{#2}%
}
\newcommand{\cellDISTS}[2]{%
  \calcHeatLo{#1}{\DISTSmin}{\DISTSmax}%
  \cellcolor{HeatC!\the\heatcount}\numThree{#1}~$\pm$~\numThree{#2}%
}

\usepackage{pgf} 
\definecolor{HeatT}{RGB}{244,67,54}     
\definecolor{HeatPos}{RGB}{220,20,60}   
\definecolor{HeatNeg}{RGB}{49,130,206}  

\newcommand{\clipToHeat}[1]{%
  \pgfmathsetmacro{\tclip}{max(min(#1,1),0)}%
  \pgfmathsetmacro{\pctraw}{5 + 95*\tclip}%
  \pgfmathtruncatemacro{\heatcount}{round(\pctraw)}%
}

\newcommand{\calcMono}[3]{
  \pgfmathsetmacro{\den}{(#3-#2)+1e-12}%
  \pgfmathsetmacro{\t}{(#1-#2)/\den}%
  \clipToHeat{\t}%
}

\newcommand{\calcDiverge}[2]{
  \pgfmathsetmacro{\den}{#2 + 1e-12}%
  \pgfmathsetmacro{\t}{abs(#1)/\den}%
  \clipToHeat{\t}%
  \pgfmathparse{ifthenelse(#1>=0,1,0)}%
  \ifnum\pgfmathresult=1
    \def\heatcol{HeatPos}%
  \else
    \def\heatcol{HeatNeg}%
  \fi
}

\def\dEmin{9.946}\def\dEmax{16.237}

\def\dLmaxabs{14.186}
\def\damaxabs{2.192}
\def\dbmaxabs{10.098}

\newcommand{\cellDE}[2]{
  \calcMono{#1}{\dEmin}{\dEmax}%
  \cellcolor{HeatT!\the\heatcount}\numThree{#1}~$\pm$~\numThree{#2}%
}

\newcommand{\cellDL}[2]{%
  \calcDiverge{#1}{\dLmaxabs}%
  \cellcolor{\heatcol!\the\heatcount}\numThree{#1}~$\pm$~\numThree{#2}%
}
\newcommand{\cellDa}[2]{%
  \calcDiverge{#1}{\damaxabs}%
  \cellcolor{\heatcol!\the\heatcount}\numThree{#1}~$\pm$~\numThree{#2}%
}
\newcommand{\cellDb}[2]{%
  \calcDiverge{#1}{\dbmaxabs}%
  \cellcolor{\heatcol!\the\heatcount}\numThree{#1}~$\pm$~\numThree{#2}%
}

\begin{document}


\firstsection{Introduction}

\maketitle

Visual comparison is a fundamental activity across both everyday activities and professional domains. Applications include inspecting construction progress~\cite{eppler2006comparison}, preserving cultural heritage~\cite{bellavia2022challenges}, and detecting changes in tasks that require comparing scenes across time, modalities, or physical and digital forms~\cite{zhang2020extended}.
Many of these comparison tasks involve complex 3D spatial structures, for which 2D displays offer limited support. Traditional displays flatten depth and restrict viewpoint flexibility, forcing users to infer spatial relationships from limited projections~\cite{kim2017comparison,lyi2020comparative}. This often makes accurate comparison, especially in indoor or geometrically detailed settings, challenging and cognitively demanding.
In recent years, Virtual Reality (VR) has been increasingly adopted to overcome the constraints of 2D displays~\cite{ZHU2026103379}. VR offers immersion, 3D depth perception, natural viewpoint navigation, and embodied interaction, thereby facilitating visual comparison~\cite{fernandez2017access,joos2022visual,kim2015bema}.
However, VR isolates users from the physical world, making it inadequate for tasks that require a direct comparison between a reconstruction and the live physical environment. Mixed Reality (MR) addresses this by embedding virtual content within the real-world context~\cite{speicher2019mixed}. 
In particular, video see-through (VST) MR is well-suited due to its wider field of view and a fully digital rendering pipeline. To enable spatially grounded visual comparison in MR, the reconstruction must be precisely registered to the headset’s live-world coordinate system, so that the digital model and the physical scene are spatially co-located.

In this paper, we introduce MR-Compare, a spatially grounded MR framework that enables visual comparison between the physical world and 3D reconstructions on Meta Quest~3, supporting PC-tethered rendering of both high-fidelity mesh-based models and 3D Gaussian Splatting (3DGS).
To solve the coordinate unification problem, MR-Compare integrates a robust coarse-to-fine registration pipeline that utilises TEASER++ for coarse registration and V-GICP for fine registration. We also introduce a geometry-driven \textit{3D Slider} that enables users to dynamically slide between the 3D reconstructions and the live VST stream for seamless comparison.
MR-Compare’s utility is shaped by both registration accuracy and cross-media visual discrepancies between reconstructions and the live VST stream. We therefore conduct two complementary evaluations. The real-world comparative evaluation is a benchmark with an exploratory user study across five low-friction workflows representative of practical XR usage: RealityScan~\cite{RealityScan2025} (desktop photogrammetry) and Polycam~\cite{polycam} (mobile scanning) for mesh reconstruction, alongside standard 3DGS~\cite{kerbl2023gaussian} and 3DGS-MCMC~\cite{kheradmand2024mcmc} via Nerfstudio (desktop 3DGS workflows), and Scaniverse~\cite{scaniverse} (a mobile 3DGS workflow). Complementing this real-world evaluation, a controlled Replica evaluation introduces a compatible, training-free anisotropy filter for more surface-proximal source selection and evaluates registration design choices across Replica scenes~\cite{straub2019replica}.

Therefore, our contributions are threefold: (1) a PC-tethered, self-contained Unity MR pipeline for spatially grounded visual comparison between heterogeneous 3D reconstructions and the physical world; (2) a real-world comparative evaluation across five representative workflows spanning desktop and mobile capture, mesh and 3DGS representations, and standard versus MCMC-based 3DGS, showing centimetre-level ArUco-referenced translation error in two static indoor rooms and the strongest overall pattern for the desktop 3DGS workflows; and (3) a zero-shot, model-independent \textit{anisotropy filter} for 3DGS-to-scan registration, validated across multiple Replica scenes~\cite{straub2019replica}, where a threshold sweep shows that moderate pruning can improve robustness and reduce residual alignment error.

\section{Related Work}
\subsection{Visual Comparison in XR Environments}
Visual comparison typically relies on three core paradigms: juxtaposition, superposition, and interchange \cite{gleicher2011visual, kim2017comparison}. These paradigms are frequently combined to minimise cognitive load and optimise visual clarity \cite{lyi2020comparative}. 
Recent efforts have brought these paradigms into real-world MR by overlaying synthetic reference models onto physical scenes. For example, DRCmpVis replaces selected real objects with virtual avatars based on extracted attributes to support contextual filtering and re-ranking~\cite{liu2024drcmpvis}. Similarly, holographic anatomical models have been overlaid to aid localisation of brain tumours~\cite{fick2023comparing}. In industrial settings, AR-based inspection tools overlay CAD models onto physical products to support in-situ deviation checking and annotation~\cite{marino2021augmented}. However, these systems largely rely on idealised or synthesised content rather than high-fidelity reconstructions of real environments.

\subsection{Scene Representation in 3D Reconstruction}

In 3D reconstruction, the chosen scene representation directly impacts geometric fidelity, visual quality, and practical deployability (e.g., storage, rendering efficiency, and compatibility with standard pipelines). Point clouds, as a common intermediate geometric representation, are obtained either from active depth sensing (e.g., LiDAR or structured-light sensors) or from Structure-from-Motion (SfM)~\cite{snavely2006photo} followed by Multi-View Stereo (MVS)~\cite{schonberger2016structure}. While straightforward, point clouds lack an explicit surface, motivating surface-based representations such as meshes, which have become a dominant practical choice~\cite{kazhdan2006poisson}. Meshes integrate well with standard rendering pipelines and underpin widely used consumer photogrammetry tools (e.g., RealityScan and Polycam~\cite{polycam}). However, meshes can struggle in texture-poor regions and do not naturally model view-dependent appearance effects~\cite{pharr2023pbr}.
To overcome these limitations, Radiance Fields emerged, mapping 3D positions and viewing directions directly to emitted radiance. Neural Radiance Fields (NeRF) \cite{mildenhall2020nerf} pioneered an implicit neural formulation that achieved unprecedented photorealism. Although subsequent NeRF variants improved training speed and anti-aliasing, they generally remain constrained by slow inference and high computational demands during rendering~\cite{muller2022instantngp,barron2022mipnerf360}.
\paragraph{3D Gaussian Splatting} (3DGS) addresses these bottlenecks by representing scenes as explicitly optimizable, anisotropic 3D Gaussians, thereby combining the photorealism of radiance fields with the real-time efficiency of point-based rasterisation~\cite{kerbl2023gaussian}.
Amidst the rapid proliferation of 3DGS variants, several core advancements have emerged as foundational pillars. 3DGS-MCMC \cite{kheradmand2024mcmc} resolves local minima by modelling Gaussian densification as a rigorous probabilistic sampling process. Mip-Splatting \cite{yu2024mipsplatting} introduces 3D filters to enable scale-invariant, alias-free rendering across varying camera distances. Scaffold-GS \cite{scaffoldgs} leverages neural anchors to dynamically guide the Gaussian distributions, significantly reducing spatial redundancy. On the geometric front, SuGaR \cite{guedon2023sugar} applies signed distance function (SDF)-based regularisation to flatten Gaussians along implicit surfaces. 2DGS \cite{Huang2DGS2024} fundamentally alters the underlying rasteriser by projecting Gaussians as 2D surfels, prioritising precise mesh extraction over standard pipeline compatibility. More recently, 3DGS has rapidly entered consumer-facing ecosystems. Nerfstudio~\cite{nerfstudio} integrates efficient implementations such as \textit{gsplat}~\cite{ye2025gsplat}, reducing training time while maintaining competitive quality for benchmarking. 
Commercial tools further broaden access: Postshot~\cite{jawset2025} offers integrated training and basic editing with a GUI, Luma AI~\cite{lumalabsInteractiveScenes} provides cloud reconstruction, and Scaniverse~\cite{scaniverse} enables on-device mobile capture and optimisation.

\paragraph{3DGS in Extended Reality}
Existing 3DGS applications in extended reality have primarily focused on Virtual Reality (VR). Following the VR-enabled update to the original SIBR viewer \cite{sibr2020}, subsequent research has branched into two key directions: performance optimisation and interactive dynamics. Specifically, VR-Splat~\cite{tu2025vrsplat} enhanced real-time frame rates and visual fidelity, while VR-GS \cite{jiang2024vr} integrated physical dynamics into the 3DGS framework for immersive interaction. Parallel efforts have also transitioned 3DGS into game engines, notably via Unity-based viewers \cite{pranckevicius2023unitygs, kleinbeck_multi-layer_2025}. 
Beyond VR, a recent perceptual study evaluated pre-reconstructed 3DGS as an alternative to conventional VST in static environments~\cite{schleising2025virtual}. While demonstrating its perceptual potential, the study does not address spatial alignment with the physical environment. Spatially registered 3DGS for direct in-situ comparison therefore remains underexplored.

\section{System Design}
The primary objective of our system is to facilitate spatially grounded visual comparison by registering reconstructed assets, such as polygonal meshes and 3DGS, with the physical environment. We target Meta Quest~3 and design the system as a self-contained Unity pipeline, avoiding external tracking infrastructure during use. The current prototype is evaluated in a tethered configuration because high-fidelity real-time 3DGS rendering remains challenging on standalone Quest~3 hardware.
Existing alignment strategies include marker-, image-, and geometry-based registration. Marker-based methods are simple but rely on pre-defined fiducials~\cite{garrido2014automatic}. Image-based methods, often solved as a Perspective-n-Point problem (PnP), recover pose from feature correspondences but are sensitive to lighting and camera variation, though learning-based approaches (e.g., SuperPoint~\cite{detone2018superpoint}) improve robustness. Furthermore, raw image access on MR headsets is often restricted for privacy, limiting applicability. Depth sensors, increasingly available in headsets and less privacy-sensitive, make geometry-based registration via point cloud a practical solution.


\subsection{Registration Pipeline}
Reconstruction-specific registration pipelines (e.g., \emph{Automated 3D-GS Registration}~\cite{liu2025automated}, \emph{GaussReg}~\cite{chang2024gaussreg}) typically rely on symmetric, model-specific matching or continuous RGB streams, which do not fit our MR setting with cross-modal inputs (a reconstruction as source and a headset depth-scan point cloud as target) and restricted RGB access.
Therefore, to register heterogeneous reconstructions (meshes and 3DGS) to headset depth-scan point clouds, we adopt a representation-agnostic point-cloud registration pipeline. 
Next, we describe how the source and target point clouds are acquired and then detail the registration approach used in MR-Compare.
\paragraph{Source Point Cloud Acquisition}
As shown in \cref{fig:teaser}a, we use vertex positions as the source point set $P_s$ for meshes. For 3DGS, prior work suggests that centres can provide stable geometric references for pose estimation (e.g., \emph{NoPoSplat}~\cite{ye2024noposplat}). Therefore, we extract Gaussian centres $\{\mu_k\}$ and treat them as a conventional dense point cloud for registration. Many 3D reconstruction pipelines also include distant content (e.g., skyboxes) to assist visualisation, but this introduces outliers in geometry-based alignment. We therefore apply a radius-crop filter, retaining only points within a radius $r$ of the model centre $c$ (the reconstruction coordinate origin/pivot):
\[ 
P_s \leftarrow \{\, p \in P_s \mid \|p-c\|_2 \le r \,\}. 
\]
Some pipelines may generate \emph{floating noise}. We optionally apply a voxel-hash density filter. Each point \(p \in P_s\) is mapped to a grid cell: $\phi(p) = \left\lfloor \frac{p}{s} \right\rfloor,$
where \(s\) is the grid spacing. We keep \(p\) only if the occupancy of its (hashed) cell is at least \(N_{\min}\): 
\[
P_s \leftarrow \{\, p \in P_s \mid n(\phi(p)) \ge N_{\min} \,\}.
\]
This removes isolated, weakly supported points (e.g., floating noise) and yields a cleaner source set for more stable registration.

\paragraph{Target Point Cloud Acquisition}
\label{sec:target-point-cloud}
The target point cloud $P_t$ of the physical environment is acquired using our modified version of the \emph{Depth-Scanner-Project}\footnote{\url{https://github.com/Appletea0673/Depth-Scanner-Project}}. Built on the Meta Depth API, the modified implementation comprises three modules: \textit{Point Cloud Generator}, \textit{Data Manager}, and \textit{Renderer}.
As shown in \cref{fig:teaser}b, the \textit{Point Cloud Generator} uses the Depth API's \texttt{Raycast} to asynchronously map screen-space samples to world-space points with configurable density, frequency, and resolution. The \textit{Data Manager} stores points in density-capped grid chunks, loading only nearby chunks for rendering. The original \textit{Renderer} used Gaussian billboards, but scaled poorly to dense clouds.
We replace the original renderer with a GPU-accelerated \textit{Visual Effect Graph} (VFX) pipeline and support two modes: (1) \textit{Pre-Scanning}, where a dense cloud is saved relative to a Quest spatial anchor\footnote{A world-locked frame of reference with fixed position and orientation.} and can be repeatedly loaded for consistent registration, and (2) \textit{On-the-Fly Scanning}, where clouds are generated live and directly used for registration. We also add a voxel-hash density filter to remove isolated outliers from transient depth returns on the user or other objects.

\paragraph{Coarse-to-fine Registration}
In \cref{fig:teaser}c, MR-Compare follows a two-stage coarse-to-fine pipeline to balance robustness and precision under self-contained Unity constraints. After acquiring the source and target point clouds, we first apply voxelisation $\mathcal{V}(\cdot)$ to obtain $\tilde{P}_s=\mathcal{V}(P_s)$ and $\tilde{P}_t=\mathcal{V}(P_t)$, reducing computation and mitigating density/resolution mismatch between reconstructions and on-device depth scans. 
For the coarse stage, classical correspondence-based approaches often combine handcrafted descriptors, such as FPFH~\cite{rusu2009fpfh}, with RANSAC or global optimisation, as in FGR~\cite{zhou2016fgr}. Although FPFH is practical for edge computing given its efficiency and lightweight dependencies, RANSAC/FGR are less reliable at high outlier ratios~\cite{lim2025kiss}. We also avoid learning-based descriptors (e.g., SpinNet~\cite{ao2021spinnet}, Predator~\cite{huang2021predator}) and estimators (e.g., GeoTransformer~\cite{qin2022geotransformer}) because they typically require GPU inference and introduce heavier dependencies outside a self-contained Unity pipeline. We therefore establish putative correspondences with FPFH~\cite{rusu2009fpfh} and then estimate an initial transform $T_g$ using either TurboReg~\cite{turboReg2025} (a state-of-the-art estimator based on the SC2-PCR~\cite{huang2021sc2pcr} with a strong accuracy--speed trade-off) or TEASER++~\cite{yang2020teaser} (a more principled truncated least-squares solver with certifiable optimality and optional scale estimation, robust to extreme outliers up to 99\%). Together, they trade off benchmark efficiency against certifiable robustness.

For fine refinement, it is common practice to iteratively minimise spatial distances; one of the most classical approaches is iterative closest point (ICP)~\cite{besl1992method}. Later variants, such as Generalised ICP (G-ICP)~\cite{segal2009generalized}, model surface geometry probabilistically to improve robustness. Building on this, Voxelised G-ICP (V-GICP) \cite{koideVoxelizedGICPFast2021} leverages voxel structures to significantly accelerate covariance aggregation and convergence.
MR-Compare implements this with the \textit{small gicp} backend~\cite{koideVoxelizedGICPFast2021} and supports both G-ICP and V-GICP (default) to compute the update $\Delta T_\ell$.
The final transform $T_{\mathrm{total}}$ and source points $P_s'$ are calculated as:
\[
T_{\mathrm{total}} = \Delta T_\ell T_g,
\qquad
P_s' = T_{\mathrm{total}}(P_s).
\]

Both stages are integrated as native Windows DLLs via minor codebase modifications, enabling a self-contained Unity pipeline without external runtime dependencies. $T_{\mathrm{total}}$ can be bound to a spatial anchor and persisted across sessions to avoid repeated registration.

\subsection{Visual Comparison Visualisation: 3D Slider}

Following prior 3D visual-comparison work, we adopt the concept of \textit{3D Slider}~\cite{zhu2025evaluating}, as it was well-received by participants and sensitive to spatial offsets, matching our goal of inspecting residual misalignment between registered reconstructions and the physical scene.
Specifically, a hollow hemisphere mesh is rendered with a premultiplied-alpha shader that writes only alpha to the framebuffer, acting as a volumetric mask. On the Quest~3, passthrough at pixel $(x,y)$ is enabled when $\alpha(x,y)>0$.
The alpha field is sampled from a grayscale texture $T_r(u,v)$ with optional inversion $\lambda\in\{0,1\}$:
\[
\alpha(x,y)=(1-\lambda)\,T_r(u,v)+\lambda\,(1-T_r(u,v)).
\]
Since only pixels covered by the mesh geometry contribute a nonzero $\alpha$, passthrough is spatially constrained to the mesh interior. By translating the hemisphere in 3D space, users can interactively `slide' between past reconstructions and the present environment, enabling perceptually grounded visual comparison. We also prototyped mini-window/cut-plane, switching/interchange, and opacity-based superposition modes. Their design trade-offs included restricted visible regions, visual transitions, and ghosting that could confound consistency and artefact judgements. We therefore treat the 3D Slider as a controlled comparison mechanism for this study, not a universally optimal technique.

\section{System Evaluation}

\subsection{Real-World Comparative Evaluation}
We examine MR-Compare across five representative 3D reconstruction pipelines in two indoor rooms. We first benchmarked these workflows objectively using registration error and image-based visual consistency measures. We then conducted an exploratory user study in the same settings to examine how participants perceived workflow-specific qualities in MR and whether MR-Compare provided a usable environment for such comparisons, focusing on perceived alignment, perceived visual consistency, and overall usability rather than performance on specific downstream tasks such as inspection, change detection, or decision making.
The real-world evaluation was conducted in two indoor rooms, \textit{Office} and \textit{Reception}. \textit{Office} was measured approximately $4.20 \times 3.27$m (about 12m$^2$). \textit{Reception} covered approximately $18\,\mathrm{m}^2$, with a maximum span of $6.12\,\mathrm{m}$ (wall segments of $4.34\,\mathrm{m}$ and $3.81\,\mathrm{m}$). \appfullref{fig:floorplan-overall}{Floor plans for both rooms are provided in Appendix~\cref{fig:floorplan-overall}.}

The five reconstructions per environment were selected to reflect representative, low-friction pipelines that XR users are most likely to adopt. For desktop workflows, we used Nerfstudio to produce 3DGS and 3DGS-MCMC, and RealityScan to produce a dense mesh.
For mobile workflows, we used Scaniverse for on-device 3DGS and Polycam for mobile photogrammetry meshes. Together, these pipelines span workflows (desktop vs.\ mobile), representations (3DGS vs.\ mesh), and algorithms (3DGS vs.\ 3DGS-MCMC), while preserving realistic differences in capture and reconstruction.

\subsubsection{Objective Benchmark}

To ensure identical camera motion across pipelines, two iPhone~15 Pro Max devices were rigidly mounted on a Nikon Z7II. We recorded a three-minute video for mobile pipelines and captured $\sim$ 550 high-resolution photos per scene for desktop pipelines, reflecting typical input conditions for each pipeline and framing the evaluation as an end-to-end workflow comparison.
Mobile reconstructions used iPhone LiDAR for metric scale (Scaniverse on-device optimisation; Polycam space scanning with dense post-processing). Desktop workflows ran on a PC (RTX~3090, i7-11700K): photos were downsampled to 4K; RealityScan performed SfM/MVS (with scale set via reference objects) to produce a dense mesh and a calibrated point cloud used by Nerfstudio to train 3DGS and 3DGS-MCMC (30k iterations).
Reconstruction time showed a clear desktop--mobile divide: Polycam and Scaniverse completed in 6--10 minutes, whereas RealityScan mesh and 3DGS required approximately 50--55~minutes, and 3DGS-MCMC exceeded 75~minutes.

Regarding the MR-Compare parameters, preprocessing used a \SI{10}{\metre} radius crop and a voxel-hash density filter ($s = 0.2$, $N_{\min} = 3$). TEASER++ was used as the default coarse-registration method, as it produced more robust initial alignments and reliable downstream convergence with V-GICP across all reconstructions; coarse-registration and classical-baseline results are reported in \cref{sec:coarse-estimator-ablation}. TEASER++ used voxel size $0.1$, max iterations $200$, noise bound $0.1$, normal/FPFH radii $0.4/0.8$, matcher ratio $0.8$, and disabled scaling. V-GICP used voxel size $0.1$, max iterations $300$, voxel resolution $0.3$, and max correspondence distance $0.2$; all other parameters remained at default.

Registration error was measured using a $20\,\mathrm{cm}\times20\,\mathrm{cm}$ ArUco marker (evaluation only) as the relative pose difference of the same marker in the physical scene (proxy for ground truth) versus the reconstruction. ArUco pose estimation can achieve $\sim$1--2\,mm translation error and sub-degree rotation error in controlled setups~\cite{poroykov2020modeling,leon2023accuracy}. Accordingly, we interpret this measure as a sparse pose reference for characterising registration, rather than as dense local ground truth throughout the scene.
Six markers were installed per environment (IDs 0–5 in \textit{Office}, 6–11 in \textit{Reception}), mounted at eye level on all four walls (two on each longer wall, one on each shorter wall). For each marker, poses were continuously recorded over a five-second interval, and the registration error was reported as the mean~$\pm$~SD of these samples. Registration itself was performed in a pre-scanning mode and repeated six times, with registration time recorded as mean~$\pm$~SD.

To evaluate how each reconstruction visually matched the live VST appearance, we used four image quality metrics, PSNR~\cite{huynh2008scope}, SSIM~\cite{wang2004ssim}, LPIPS~\cite{zhang2018lpips}, and DISTS~\cite{ding2020dists}, to quantify the visual agreement between the reconstructed renderings and the reference VST images. We refer to this measure as \emph{VST-referenced visual consistency}. Higher PSNR/SSIM and lower LPIPS/DISTS indicate closer agreement, though in this cross-media setting, they are treated as relative indicators rather than absolute fidelity measures. After registering each reconstruction within MR-Compare, we selected 30 \emph{evaluation} viewpoints in the physical environment to assess visual consistency against the corresponding live VST views. The viewpoints were chosen to cover all objects in the scene, with major objects observed from multiple angles. VST images were captured using Quest~3 screenshots, as passthrough frames were not directly accessible in our tethered configuration.

\subsubsection{Exploratory User Study}
We conducted an exploratory user study to characterise the perceived qualities of these workflows in situ, complementing our objective benchmark. A total of 30 participants (22--32 years; $M = 26.47$, $SD = 2.99$; 18 male, 12 female; all had normal or corrected-to-normal vision) took part in the study. All participants used a Meta Quest~3 tethered to a desktop PC (i7-12700K, RTX~5090), maintained at 72~Hz throughout. The experiment was approved by the UCL Computer Science Research Ethics Committee with ID \textbf{UCL/CSREC/R/39}.
This exploratory user study examined how five reconstruction workflows were perceived relative to live VST in MR. Rather than assessing perceptual performance, it focused on perceived system characteristics, including perceived alignment, perceived visual consistency, and overall reconstruction quality. These subjective assessments were subsequently compared with objective registration and image-based visual consistency measures. 
Using a within-subjects design across \textit{Office} and \textit{Reception}, each participant completed ten conditions (two environments $\times$ five reconstructions). Presentation order was counterbalanced using a Latin square, such that within each group of five participants, each reconstruction appeared once in each ordinal position. Each session lasted approximately one hour, including a 2-minute familiarisation phase. During the verbally administered questionnaire, participants could freely switch between reconstructions before providing Likert ratings, which the experimenter recorded.

\begin{figure*}[t]
    \centering
    {\small\bfseries Each panel: Reconstruction (left) \textbar~VST (right)\par}
    \vspace{2pt}
    \includegraphics[width=0.95\linewidth]{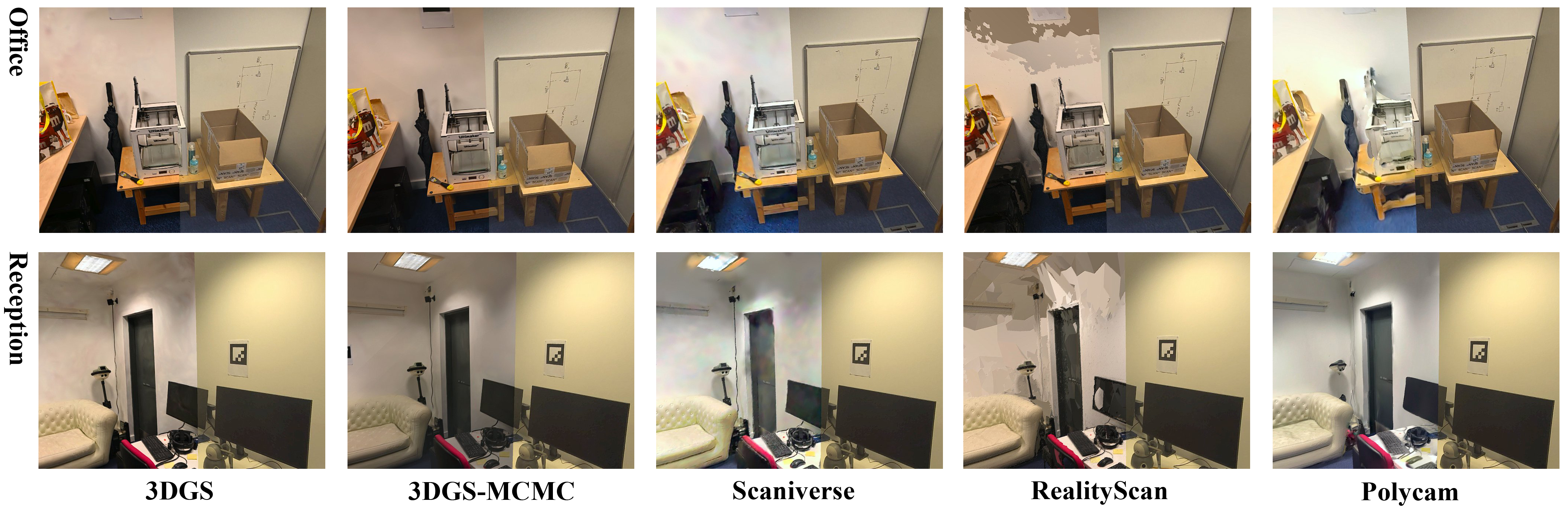}
    \caption[Reconstruction renderings and live VST views in the two physical rooms.]{Reconstruction renderings and spatially registered live VST views in \textit{Office} (top) and \textit{Reception} (bottom). Within each panel, the reconstruction is on the left and live VST is on the right.\appfullref{sec:screenshots}{ Additional screenshots are provided in Appendix~\cref{sec:screenshots}.}}
    \label{fig:example-screenshots}
\end{figure*}

\begin{table*}[t]
\centering
\resizebox{0.95\linewidth}{!}{%
\begin{tabular}{lcccc|cccc}
\toprule
\multirow{2}{*}{Method}
 & \multicolumn{4}{c|}{\textbf{Office} (Scanned points: 1.521M)} & \multicolumn{4}{c}{\textbf{Reception} (Scanned points: 1.802M)} \\
\cmidrule(lr){2-5}\cmidrule(lr){6-9}
 & Points & Time (s) & $\|\Delta p\|$ (cm) & $\left| \Delta \theta \right|$ (°)
 & Points & Time (s) & $\|\Delta p\|$ (cm) & $\left| \Delta \theta \right|$ (°) \\
\midrule
3DGS
 & 1.09M   & \cellTIMEone{2.81}{0.29} & \cellTsum{1.614}{0.688}     & \cellRsum{1.15400}{0.63212}
 & 0.73M   & \cellTIMEtwo{3.56}{0.41} & \cellTsumTwo{2.69000}{0.27000} & \cellRsumTwo{1.85089}{0.61699} \\
3DGS-MCMC
 & 0.89M   & \cellTIMEone{1.03}{0.03} & \cellTsum{0.892}{0.308}     & \cellRsum{0.86066}{0.40379}
 & 0.78M   & \cellTIMEtwo{1.94}{0.17} & \cellTsumTwo{1.64900}{0.76200} & \cellRsumTwo{1.78516}{0.81137} \\
Scaniverse
 & 0.53M   & \cellTIMEone{1.10}{0.07} & \cellTsum{3.538}{0.916}     & \cellRsum{2.88263}{1.02424}
 & 0.43M   & \cellTIMEtwo{1.77}{0.06} & \cellTsumTwo{2.56100}{0.38000} & \cellRsumTwo{2.14952}{0.92660} \\
RealityScan
 & 5.00M & \cellTIMEone{1.64}{0.03} & \cellTsum{2.423}{0.608}     & \cellRsum{1.53379}{0.88552}
 & 5.00M & \cellTIMEtwo{1.99}{0.07} & \cellTsumTwo{2.91500}{0.74100} & \cellRsumTwo{2.17338}{1.23700} \\
Polycam
 & 0.09M   & \cellTIMEone{0.48}{0.02} & \cellTsum{2.745}{1.162}     & \cellRsum{1.93061}{1.11886}
 & 0.17M   & \cellTIMEtwo{0.86}{0.01} & \cellTsumTwo{4.35000}{1.39600} & \cellRsumTwo{2.11964}{1.35205} \\
\bottomrule
\end{tabular}%
}
\caption{Summary of point count, registration time, and pose error across reconstruction workflows. Values are mean $\pm$ SD; time is in seconds. Darker colour indicates better performance.}
\label{tab:pose_points_time_summary}
\end{table*}

\begin{table*}[!t]
\centering
\renewcommand{\arraystretch}{1}
\resizebox{0.95\linewidth}{!}{%
\begin{tabular}{lcccccccc}
\toprule
& \multicolumn{4}{c}{\textbf{Office}} & \multicolumn{4}{c}{\textbf{Reception}} \\
\cmidrule(r){2-5} \cmidrule(l){6-9}
Method & PSNR ($\uparrow$) & SSIM ($\uparrow$) & LPIPS ($\downarrow$) & DISTS ($\downarrow$) 
       & PSNR ($\uparrow$) & SSIM ($\uparrow$) & LPIPS ($\downarrow$) & DISTS ($\downarrow$) \\
\midrule
3DGS       & \cellPSNR{18.799}{1.195} & \cellSSIM{0.740}{0.046} & \cellLPIPS{0.307}{0.034} & \cellDISTS{0.157}{0.029} 
           & \cellPSNR{17.911}{1.171} & \cellSSIM{0.717}{0.042} & \cellLPIPS{0.345}{0.034} & \cellDISTS{0.149}{0.025} \\
3DGS-MCMC  & \cellPSNR{20.700}{1.543} & \cellSSIM{0.790}{0.056} & \cellLPIPS{0.260}{0.031} & \cellDISTS{0.108}{0.021} 
           & \cellPSNR{19.205}{1.379} & \cellSSIM{0.779}{0.044} & \cellLPIPS{0.295}{0.033} & \cellDISTS{0.113}{0.017} \\
Scaniverse & \cellPSNR{14.284}{1.137} & \cellSSIM{0.720}{0.048} & \cellLPIPS{0.401}{0.051} & \cellDISTS{0.171}{0.033} 
           & \cellPSNR{13.607}{0.751} & \cellSSIM{0.715}{0.039} & \cellLPIPS{0.415}{0.032} & \cellDISTS{0.187}{0.030} \\
RealityScan& \cellPSNR{18.213}{1.459} & \cellSSIM{0.690}{0.057} & \cellLPIPS{0.420}{0.037} & \cellDISTS{0.189}{0.025} 
           & \cellPSNR{17.809}{1.291} & \cellSSIM{0.682}{0.037} & \cellLPIPS{0.475}{0.023} & \cellDISTS{0.208}{0.033} \\
Polycam    & \cellPSNR{13.391}{0.798} & \cellSSIM{0.710}{0.044} & \cellLPIPS{0.359}{0.035} & \cellDISTS{0.164}{0.023} 
           & \cellPSNR{14.308}{0.636} & \cellSSIM{0.703}{0.041} & \cellLPIPS{0.420}{0.030} & \cellDISTS{0.167}{0.019} \\
\bottomrule
\end{tabular}%
}
\caption{VST-referenced visual-consistency metrics across reconstruction workflows in \textit{Office} and \textit{Reception}. Values are mean $\pm$ SD; darker colour indicates closer agreement with live VST.}
\label{tab:Objective_visual_all}
\end{table*}

During the procedure, participants compared and rated all reconstructions for each question participants comparatively rated all reconstructions for each question with a think-aloud protocol. Retrospective edits were allowed (validated in a pilot study, $n{=}5$). The questionnaire was administered throughout the study to characterise the perceived qualities of each reconstruction workflow relative to the corresponding live VST view in MR.
It covered three areas. First, participants rated \textbf{perceived visual consistency} across eight dimensions\footnote{These dimensions were informed by standard computer graphics and image quality measures; brief definitions were provided before scoring.
}, using a 7-point bidirectional Likert scale (1: much worse; 4: about the same; 7: much better): \textit{visual clarity, 3D depth, lighting and colour, visual artefacts, completeness, geometry accuracy, smoothness, and text/pattern recognisability}. Second, participants rated \textbf{perceived alignment} in terms of \textit{translation}, \textit{rotation}, and \textit{scale} misalignment, and reported overall \textit{registration satisfaction}, using a 5-point Likert scale (1: not at all; 5: completely). After experiencing all five workflows in each room, participants completed the NASA-TLX~\cite{hart2006nasa}, SSQ~\cite{kennedy1993simulator}, and two-item UMUX-LITE~\cite{lewis2013umux} to assess workload, simulator sickness, perceived ease of use, and perceived usefulness (1: strongly disagree; 5: strongly agree). Room floor plans, additional screenshots, the full questionnaire, and runtime and resource-usage details are provided in~\appref{appendix_Questionnaire}{the appendices (\cref{fig:floorplan-overall}, \cref{sec:screenshots}, \cref{appendix_Questionnaire}, and \cref{app:runtime})}.

\subsection{Real-World Evaluation Results}
\label{sec:study1-results}

In our tests, the system typically required 3–5 seconds to start up and initialise before MR-Compare was ready to run. The runtime of the registration components is reported in \cref{tab:pose_points_time_summary}. Detailed runtime characteristics and resource usage are reported in \appref{app:runtime}{Appendix~\ref{app:runtime}}. \Cref{fig:example-screenshots} illustrates reconstruction quality and registration, showing representative VST screenshots of all methods in both rooms, identical to what participants saw during the experiment. Visually, the two desktop 3DGS variants show the strongest consistency, with 3DGS-MCMC slightly better overall. Standard 3DGS still shows local wall artefacts, Scaniverse exhibits strong colour distortion, RealityScan suffers from incomplete and spiky surfaces, and Polycam is mainly limited by substantial geometric distortion.

\begin{figure*}[t]
    \centering
    \includegraphics[width=0.95\linewidth]{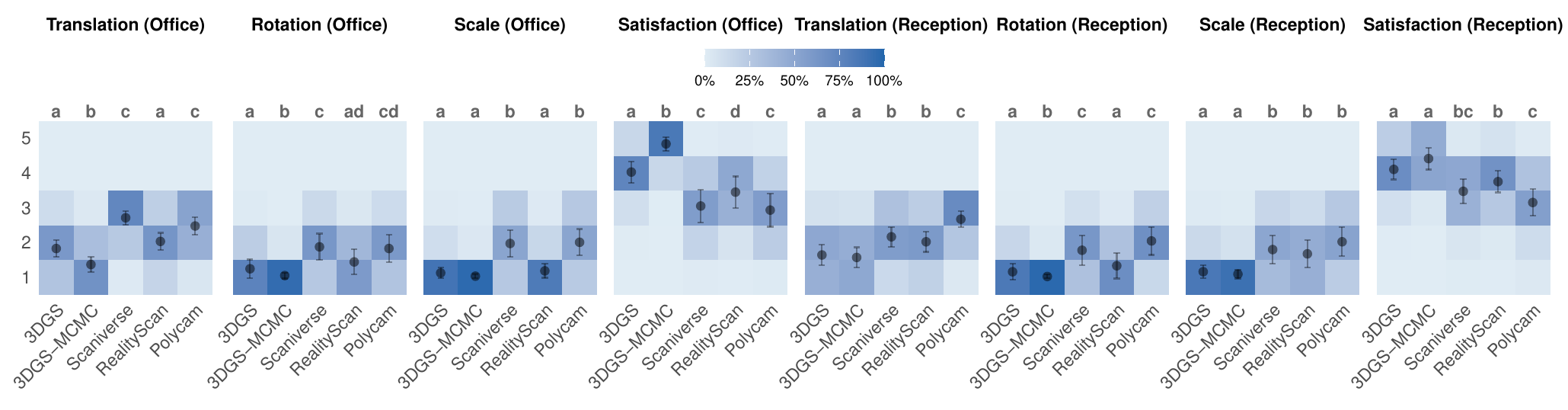}
    \caption[Predicted distributions for perceived misalignment and registration satisfaction.]{Participants' predicted response distributions for perceived translation, rotation, and scale misalignment ($\downarrow$), and registration satisfaction ($\uparrow$), in \textit{Office} and \textit{Reception}. Misalignment items range from 1 (not at all misaligned) to 5 (completely misaligned); satisfaction ranges from 1 (not at all satisfied) to 5 (completely satisfied).
     Colours show CLMM-based probabilities, black dots with error bars indicate expected ratings (95\% CI), 
     and letters mark non-significant groups (FDR–BH corrected).}
    \label{fig:Room_Subjective_align}
\end{figure*}

\begin{figure*}[!t]
    \centering
    \includegraphics[width=0.95\linewidth]{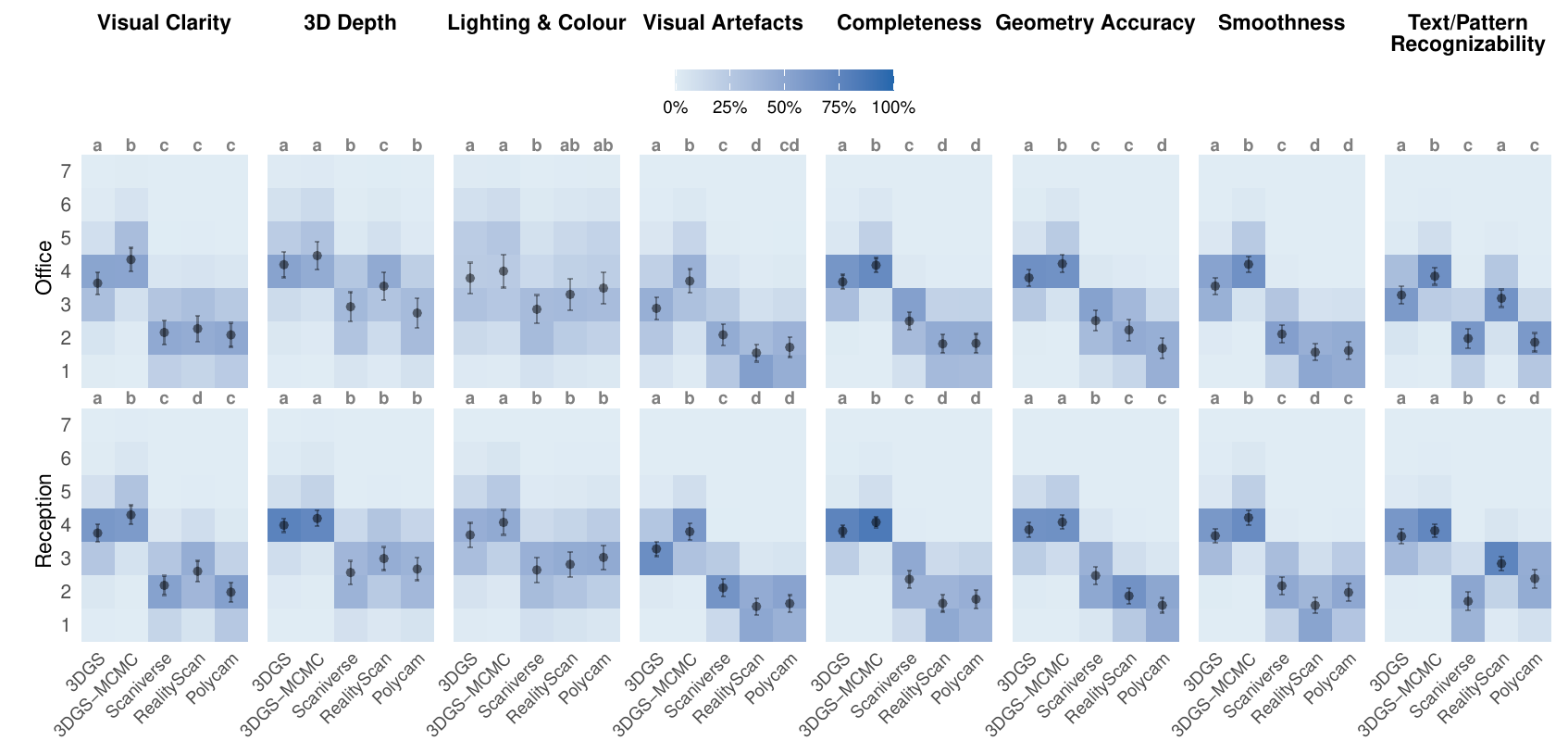}
    \caption[Predicted distributions for VST-referenced visual-consistency ratings.]{Participants' predicted response distributions for each visual consistency score in \textit{Office} (first row) and \textit{Reception} (second row). Ratings use a 7-point bidirectional scale relative to VST (1: much worse; 4: about the same; 7: much better). Colours show CLMM-based probabilities, black dots with error bars indicate expected ratings (95\% CI), and letters above the heatmap mark non-significant groups (FDR--BH corrected) if they share the same letters.}
    \label{fig:Room_visual_consistency}
\end{figure*}

\subsubsection{Objective Benchmark Results}
\Cref{tab:pose_points_time_summary} reports translation ($\|\Delta p\|$) and rotation ($|\Delta \theta|$) errors and registration time.
Across both scenes, 3DGS-MCMC achieves the lowest errors, with translation errors of $0.89 \pm 0.31$ cm and $1.65 \pm 0.76$ cm, and rotation errors of $0.86 \pm 0.40^\circ$ and $1.79 \pm 0.81^\circ$ in \textit{Office} and \textit{Reception}, respectively.
3DGS follows with slightly higher errors, while RealityScan and Polycam generally perform worse, particularly in \textit{Reception}, where Polycam reaches the largest translation error ($4.35 \pm 1.40$ cm).
Scaniverse exhibits the highest errors in \textit{Office} ($3.54 \pm 0.92$ cm, $2.88 \pm 1.024^\circ$), but shows moderately better performance ($2.56 \pm 0.38$ cm, $2.15 \pm 0.93^\circ$) in \textit{Reception} compared to RealityScan and Polycam. Regarding VST-referenced visual consistency, \cref{tab:Objective_visual_all} reports relative agreement with live VST in both scenes. Across both rooms, 3DGS-MCMC achieves the closest agreement across all four metrics, with 3DGS second. For PSNR, RealityScan ranks third in both rooms, but on SSIM, LPIPS, and DISTS, it performs worst. Polycam and Scaniverse remain similar, with their relative ranking varying across metrics and scenes.

\subsubsection{Exploratory User Study Results}
We analysed ordinal questionnaire responses using cumulative link mixed models (CLMMs)~\cite{christensen2019ordinal}, with \textit{reconstruction method} as a fixed effect and \textit{participant} as a random intercept. From each model, we report three complementary outputs. (1) We visualise the predicted category probabilities as heatmaps (darker indicates higher probability), obtained by mapping the latent linear predictor through the cumulative link. (2) We overlay each scene’s \textbf{expected rating} (weighted mean of response levels) with approximate confidence intervals. (3) For inference, we test the fixed effect using a Type II Wald $\chi^2$ test; only significant effects are followed by post-hoc pairwise comparisons on the latent scale, with Benjamini--Hochberg FDR correction (FDR-BH)~\cite{benjamini1995controlling}. Pairwise differences are summarised using \textit{compact letter displays (CLDs)}~\cite{piepho2004algorithm}, where groups sharing a letter are not significantly different. Below, we report the main patterns and key pairwise results; complete omnibus and post-hoc statistics from the CLMMs are provided in \appref{tab:align_clmm_rooms}{the Appendix (\cref{tab:align_clmm_rooms,tab:clmm_questionnaire_visual_consistency} and \cref{sec:pairwise-comparisons})}.

Regarding the \textbf{registration ratings}, our CLMMs indicated that reconstruction method significantly affected \textit{translation}, \textit{rotation}, \textit{scale}, and \textit{registration satisfaction} in both scenes (all $\chi^2(4) \geq 28.27$, $p < .001$\appfullref{tab:align_clmm_rooms}{; \cref{tab:align_clmm_rooms}}).
\cref{fig:Room_Subjective_align} shows a clear three-tier pattern; 3DGS workflows were rated best overall, whereas Scaniverse and Polycam were generally rated worst, with RealityScan in between.
Post-hoc tests show that Scaniverse and Polycam significantly underperform RealityScan on translation, scale, and satisfaction in the \textit{Office} and rotation in \textit{Reception} (all $p \leq .011$, $OR \leq 0.23$), and both 3DGS and 3DGS-MCMC on all measures (all $p \leq .006$, $OR \leq 0.208$).  
Within the top two, 3DGS-MCMC significantly outperformed 3DGS in \textit{translation} (\textit{Office}), \textit{rotation} (both scenes), and \textit{satisfaction} (\textit{Office}) with all $p \leq .035$, $OR \geq 4.97$. 3DGS-MCMC significantly outperformed RealityScan on all four dimensions across scenes except for scale in \textit{Reception} (all $p \leq .018$, $OR \geq 4.17$), while 3DGS was significantly better on satisfaction for both scenes in translation and scale (all $p \leq .029$, $OR \geq 3.66$). \appfullref{tab:pairwise_alignment_side_by_side}{Full post-hoc statistics are reported in Appendix~\cref{tab:pairwise_alignment_side_by_side}.}

Regarding the \textbf{visual consistency ratings},
\Cref{fig:Room_visual_consistency} shows a consistent ordering in both rooms: 3DGS-MCMC was rated highest, followed by 3DGS, with Scaniverse, RealityScan, and Polycam lower overall. CLMMs confirmed a significant effect of reconstruction method on all eight dimensions in both rooms (all $\chi^2(4) \ge 97.4$, $p<.001$\appfullref{tab:clmm_questionnaire_visual_consistency}{; \cref{tab:clmm_questionnaire_visual_consistency}}).
Across both rooms, the two desktop 3DGS workflows (3DGS and 3DGS-MCMC) received significantly higher ratings than Scaniverse, RealityScan, and Polycam on most criteria (all $p \leq .010$, $OR \geq 3.76$), with the main exceptions being \textit{Lighting and Colour} and \textit{Text/Pattern Recognisability}. Within the top tier, 3DGS-MCMC generally scored slightly higher than standard 3DGS, and it was significantly higher in both rooms on \textit{Visual Clarity}, \textit{Visual Artefacts}, \textit{Completeness}, and \textit{Smoothness} (all $p \leq .007$; $OR \geq 4.55$). Differences among Scaniverse, RealityScan, and Polycam were also present and varied by criterion, but were typically less pronounced than the gap between these methods and the desktop 3DGS workflows. \appfullref{tab:pairwise_vc_side_by_side}{Full details are shown in \cref{tab:pairwise_vc_side_by_side}.}
\subsubsection{Objective-Subjective Measures Consistency}
Subjective ratings showed monotonic relationships with objective measures for registration errors. In the \textit{Office}, correlations were consistent for both translation and rotation ($\rho = 1.00$, $p=.0167$),
and satisfaction was negatively associated with both translation and rotation errors ($\rho=-0.90$ for both), although neither exact test was significant (both $p=.083$). In the \textit{Reception}, correlations were moderate for translation ($\rho=0.70$, $p=.233$) and rotation ($\rho=0.60$, $p=.3500$), and satisfaction decreased with increasing errors, though not significantly.
\appfullref{tab:alignment-correlation-spearman}{}
For visual consistency, because subjective judgments were bidirectional and relative rather than continuous, we compared subjective and objective rankings using $\Delta$ rank instead of correlation\appfullref{fig:heatmap_vc_rank}{}. Overall, 3DGS and 3DGS-MCMC showed near-perfect subjective-objective agreement, whereas Polycam, RealityScan, and Scaniverse showed systematic deviations. SSIM and PSNR showed the strongest agreement with subjective ratings, followed by LPIPS, while DISTS showed the weakest. At the dimension level, PSNR aligned best with \textit{3D Depth} and \textit{Text/Pattern Recognisability}, and SSIM with \textit{Visual Artefacts} and \textit{Completeness}. Detailed registration-correlation and visual-consistency ranking results are provided in \appref{tab:alignment-correlation-spearman}{the Appendix (\cref{tab:alignment-correlation-spearman} and \cref{fig:alignment_cor_figure,fig:heatmap_vc_rank})}.

\subsubsection{\texorpdfstring{System-Level Usability and Comfort}{System-Level Usability and Comfort}}

\begin{table}[htbp]
\centering
\Large
\resizebox{\linewidth}{!}{%
\begin{tabular}{lcccccccc}
\toprule
 & \multicolumn{4}{c}{Office} & \multicolumn{4}{c}{Reception} \\
\cmidrule(lr){2-5}\cmidrule(lr){6-9}
Metric ($n=30$) & Mean & SD & Median & IQR & Mean & SD & Median & IQR \\
\midrule
\multicolumn{9}{l}{\textbf{NASA TLX (0-10)}} \\
Mental Demand   & 3.80 & 2.55 & 4.00 & 3.75 & 3.13 & 2.60 & 2.50 & 4.00 \\
Physical Demand & 2.53 & 2.13 & 2.50 & 2.75 & 3.00 & 2.00 & 3.00 & 2.75 \\
Temporal Demand & 2.23 & 2.11 & 2.00 & 3.00 & 1.90 & 1.69 & 2.00 & 2.75 \\
Performance     & 2.17 & 2.12 & 2.00 & 3.00 & 2.13 & 2.22 & 2.00 & 3.00 \\
Effort          & 3.13 & 2.16 & 3.00 & 4.00 & 2.60 & 2.18 & 2.00 & 3.00 \\
Frustration     & 0.87 & 1.55 & 0.00 & 1.00 & 1.03 & 1.71 & 0.00 & 1.00 \\
\midrule
\multicolumn{9}{l}{\textbf{SSQ (Nausea + Oculomotor + Disorientation) $\times$~3.74}} \\
Total      & 18.33 & 19.18 & 13.09 & 25.24 & 17.45 & 26.25 & 11.22 & 18.70 \\
\midrule
\multicolumn{9}{l}{\textbf{UMUX-LITE (1-5)}} \\
Usefulness      & 4.27 & 0.45 & 4.00 & 0.75 & 4.30 & 0.47 & 4.00 & 1.00 \\
Ease of Use     & 4.50 & 0.57 & 5.00 & 1.00 & 4.47 & 0.51 & 4.00 & 1.00 \\
\bottomrule
\end{tabular}%
}
\caption[Descriptive room-session NASA-TLX, SSQ, and UMUX-LITE statistics.]{Descriptive statistics for NASA-TLX, SSQ, and UMUX-LITE. Measures were collected after participants experienced all five workflows in a room and do not represent workflow-level comparisons.}
\label{tab:nasatlx_ssq_umux}
\end{table}

Participants provided overall post-experiment ratings of the usability of MR-Compare for visual comparison. These measures were intended to capture the usability of the full system with all reconstruction workflows. \Cref{tab:nasatlx_ssq_umux} shows that NASA-TLX subscales were generally low and UMUX-LITE ratings were high in both scenes. The SSQ total is the standard aggregate score, and its interpretation is discussed in \cref{sec:system-usability}.

\subsection{Registration Ablation and Refinement}

The controlled Replica evaluation narrows the focus to the two 3DGS workflows and examines two aspects of the registration pipeline: coarse registration and source-point filtering. The coarse-registration ablation evaluates ICP, G-ICP, and V-GICP independently as single-stage local baselines and compares four coarse-registration methods (TEASER++, TurboReg, RANSAC, and FGR), each followed by the same V-GICP refinement. Beyond the choice of coarse-registration method, the controlled evaluation also examines the source points supplied to registration. The real-world evaluation used all Gaussian centres as a compatible, model-independent approach for registration. However, these centres do not necessarily lie close to the underlying scene surface, leaving room for further improvement in registration precision. We therefore introduce a training-free, zero-shot \textit{anisotropy filter} to extract surface proxy points from Gaussian centres. Within each controlled comparison, registration settings not explicitly varied are held fixed.
For the experimental design, we used the eight Replica indoor scenes from NICE-SLAM~\cite{zhu2022nice} to enable a controlled, reproducible assessment. For each scene, we generated 3DGS and 3DGS-MCMC in Nerfstudio and used their Gaussian centres as the source point set $P_s$. The target point cloud $P_t$ was generated in Unity by scanning the Replica mesh with the target-acquisition pipeline in \cref{sec:target-point-cloud}, with added Quest~3-like noise (quantisation, axial noise, adaptive Gaussian noise, and burst drift\appfullref{app:scanner_impl}{}). The source and target point sets remain broadly comparable in size. Scanner implementation, baseline settings, the controlled evaluation protocol, and scene-wise results are provided in \appref{appendix:study2-evaluation}{the Appendix (\cref{app:scanner_impl,appendix:study2-evaluation} and \cref{tab:success_filter_on_off})}.


\subsubsection{\texorpdfstring{Coarse Registration Ablation}{Coarse Registration Ablation}}
\label{sec:coarse-estimator-ablation}

The ablation compares three fine-registration-only baselines (ICP, G-ICP, and V-GICP) with four coarse-to-fine variants: TEASER++, TurboReg, RANSAC, or FGR for coarse registration, followed by the same V-GICP refinement.
TEASER++ parameters match those used in the real-world evaluation, and TurboReg follows the official recommendation~\cite{turboReg2025}.
FGR retains the default optimisation parameters of the original standalone implementation, whereas RANSAC uses a practical configuration.
All variants use the same final-success criterion: $\leq 5$~cm translation error and $\leq 5^\circ$ rotation error after the tested registration pipeline.
With radius crop and density filtering, TEASER++ and TurboReg each succeeded in 15/16 scene--reconstruction pairs. RANSAC and FGR each achieved 12/16, whereas all three fine-registration-only baselines achieved 0/16. Without radius crop and density filtering, TEASER++ retained 8/16 successes and TurboReg achieved 2/16; RANSAC and FGR fell to 1/16 and 2/16, respectively, while ICP, G-ICP, and V-GICP remained at 0/16. \appfullref{tab:success_filter_on_off}{Scene-wise results are reported in Appendix~\cref{tab:success_filter_on_off}.}

\subsubsection{Anisotropy Filter}
\begin{figure}[t]
    \centering
    \includegraphics[width=1\linewidth]{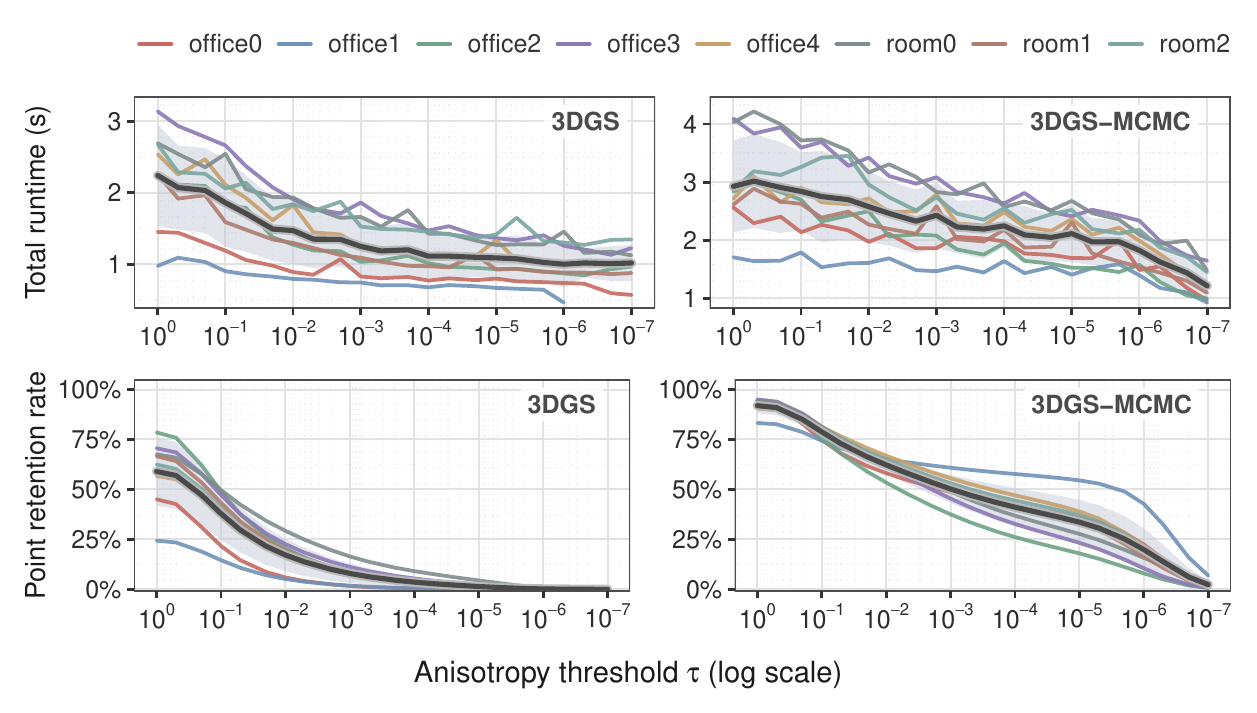}
    \caption[Runtime and point retention across anisotropy threshold $\tau$.]{Runtime and point retention across $\tau$ for 3DGS and 3DGS-MCMC. Coloured lines show individual scenes; the black line and grey band indicate the mean and SD. Scene labels \textit{office0}--\textit{office4} and \textit{room0}--\textit{room2} are Replica identifiers.}
    \label{fig:tau_time_point_ablation}
\end{figure}
Local flatness is often quantified via PCA on neighbourhood geometry~\cite{pauly2002efficient}. PCA yields eigenvalues of the local covariance, and planar regions typically satisfy $\lambda_3 \ll \lambda_1,\lambda_2$, motivating eigenvalue-ratio flatness measures (e.g., $\sqrt{\lambda_{\min}/\lambda_{\max}}$) to distinguish surface-like structure from volumetric noise.
3DGS already provides an ellipsoidal covariance for each Gaussian ($\mu\in\mathbb{R}^3$) parameterised as $\Sigma = R\,S^{2}R^{\mathsf T}$ with $S^{2}=\mathrm{diag}(\sigma_x^{2},\sigma_y^{2},\sigma_z^{2})$, where $R$ is a rotation matrix and $(\sigma_x,\sigma_y,\sigma_z)$ are the principal-axis scales. This form is already an eigendecomposition, so we can measure flatness directly from the scales without explicit eigen/SVD computation. We define:
\begin{equation}
\rho = \frac{\min(\sigma_x,\sigma_y,\sigma_z)}{\max(\sigma_x,\sigma_y,\sigma_z)} \in (0,1],
\label{eq:scale_ratio}
\end{equation}
which is monotonically equivalent to $\sqrt{\lambda_{\min}/\lambda_{\max}}$. Near-spherical Gaussians are pruned, retaining only $P_s \leftarrow \{\,\mu \in P_s \mid \rho(\mu)\le\tau\,\}$, where smaller $\tau$ enforces stricter pruning by discarding more weakly geometry-constrained points and retaining a flatter, surface-proxy subset.
We implement this in a compute shader with linear complexity $O(N)$, incurring $<\!5$\,ms on our hardware; overall preprocessing time is dominated by CPU-side point extraction/saving and scales with the retained point count (up to $\sim$700\,ms when $\tau{=}1$). We also provide an inspector-level debug mode in Unity to visualise pruning outcomes and facilitate filter tuning.

\begin{figure*}[t]
    \centering
    \includegraphics[width=0.95\linewidth]{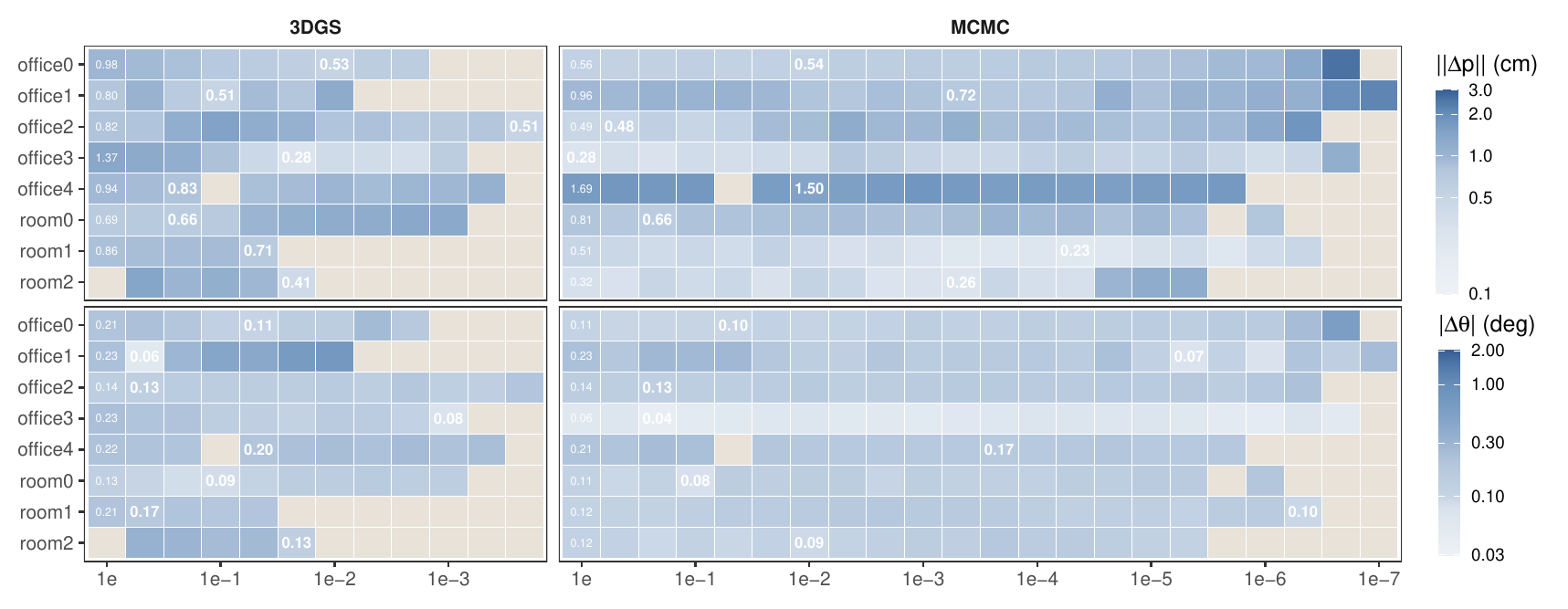}
    \caption[Registration-error heatmaps across anisotropy threshold $\tau$.]{$\tau$-sweep registration error heatmaps. Rows show translation (top, cm) and rotation (bottom, °) errors across scenes and anisotropy thresholds $\tau$. Numbers are shown for $\tau{=}1$ (anisotropy filter disabled) and per-scene minima; bold enlarged numbers mark the minimum. Non-convergent cells are masked in a neutral colour. The heatmap is truncated at the largest $\tau$ for which at least one scene remains non-failed. Scene labels \textit{office0}--\textit{office4} and \textit{room0}--\textit{room2} are Replica identifiers, not the physical rooms used in the real-world evaluation.}
    \label{fig:ablation_results}
\end{figure*}

\paragraph{Ablation}
We tuned the anisotropy filter by sweeping $\tau$ on top of radius crop and density control. We compare the baseline $\tau{=}1$ (filter disabled) against a log-spaced range $\tau\in[1,10^{-7}]$. Across the eight Replica scenes, 3DGS-MCMC took longer to reconstruct than standard 3DGS ($M{=}31.20$, $SD{=}1.95$ vs.\ $M{=}18.09$, $SD{=}1.93$ minutes). It also exhibited smaller anisotropy scores $\rho$ ($\textit{Mdn}=4.78\times10^{-4}$, $\textit{IQR}=3.27\times10^{-2}$ vs.\ $\textit{Mdn}=4.86\times10^{-2}$, $\textit{IQR}=1.61\times10^{-1}$), computed over all Gaussians and aggregated across scenes.
\Cref{fig:ablation_results} shows two main effects of anisotropy filtering. First, moderate pruning improves robustness by rescuing otherwise failed registrations --- in \textit{room2}, standard 3DGS fails at $\tau{=}1$ but converges under moderate pruning. Second, among successful runs, decreasing $\tau$ from 1 to a moderate range typically reduces both translation and rotation errors. These gains are larger for standard 3DGS ($-0.35$\,cm/$-37.6\%$; $-0.08^{\circ}$/$-38.7\%$) than for 3DGS-MCMC ($-0.12$\,cm/$-16.9\%$; $-0.04^{\circ}$/$-29.1\%$). Consistent with the real-world results, 3DGS-MCMC remains particularly accurate for rotation, while translation advantages are more mixed. 
The optimal $\tau$ range is more concentrated for standard 3DGS (typically up to $\sim10^{-2}$), whereas a broader optimum in 3DGS-MCMC remains feasible under more aggressive pruning (down to $\sim10^{-6}$). Yet these different thresholds map to a similar retention regime in \cref{fig:tau_time_point_ablation} (roughly $\lesssim$20--25\%), where failures occur under aggressive pruning, consistent with reduced geometric coverage. This also shows the efficiency trade-off: as $\tau$ decreases, retention and runtime both fall, with a steeper drop for standard 3DGS, while 3DGS-MCMC retains more points and supports a broader feasible regime.

\section{Discussion}
Several limitations should be noted before interpreting the results further. The system was evaluated in a PC-tethered Quest~3 configuration in two static indoor rooms, since high-fidelity 3DGS remains impractical on the standalone Quest~3. MR-Compare is currently intended for static scenes, rather than environments with substantial dynamic change. Outdoor illumination and dynamic objects may introduce additional reconstruction--VST differences. Because only the Quest~3 was tested, differences in sensing and image processing across HMDs may limit generalisability. In the real-world evaluation, VST-referenced visual consistency is most appropriate for relative comparison, while its absolute values may still reflect headset-specific imaging effects such as exposure and tone mapping. Without a dedicated VST-free Model-vs-Model condition, these effects cannot be fully disentangled from workflow differences. The ArUco-based registration error provides a sparse pose proxy rather than dense local ground truth, motivating the complementary full-ground-truth controlled Replica evaluation. The exploratory user study assesses perceived workflow characteristics and room-session usability rather than task accuracy or decision quality.

With these limitations in mind, the overall findings support MR-Compare as a practical framework for spatially grounded visual comparison in mixed reality. Across reconstruction methods, objective registration errors broadly align with participants' subjective judgments, and the two desktop 3DGS workflows show the strongest overall pattern in both registration and VST-referenced visual consistency. Participants also reported low workload and high perceived usability in room-session measures. The controlled Replica evaluation further shows that TEASER++ provides more robust coarse registration than TurboReg when preprocessing is disabled. The proposed anisotropy-based filtering can reduce translation and rotation errors and recover convergence in challenging cases, with particularly strong gains for standard 3DGS.

\subsection{Workflow-Level Performance Differences}
Across both objective registration error and VST-referenced visual consistency, the two desktop 3DGS workflows showed the strongest overall pattern among the tested workflows. Within the desktop setting, 3DGS and 3DGS-MCMC also achieved better results than the RealityScan mesh workflow, suggesting that the advantage may relate to downstream representation and optimisation. With the same images and SfM basis, the desktop comparison isolates downstream reconstruction differences: photogrammetry meshes provide explicit geometry, but can inherit SfM/MVS failure modes in weakly textured regions, where pose noise and default hole filling introduce local artefacts~\cite{snavely2006photo, schonberger2016structure}. By contrast, 3DGS optimises Gaussian primitives through multi-view photometric consistency, which can yield smoother and more coherent reconstructions~\cite{kerbl2023gaussian}. 
The broader desktop--mobile performance gap also reflects workflow-level differences: mobile pipelines prioritise speed and accessibility through shorter captures, LiDAR-assisted reconstruction, and on-device optimisation under tighter runtime and compute budgets. This practical speed--fidelity trade-off is consistent with their higher registration errors and lower visual-consistency ratings in our study. Accordingly, the result is specific to the tested applications, scenes, and capture protocol, rather than evidence that 3DGS workflows generally outperform meshes or mobile reconstruction.

\subsection{Strong 3DGS Subjective--Objective Consistency}
For registration\appfullref{fig:alignment_cor_figure}{ (Appendix \cref{fig:alignment_cor_figure})}, both desktop 3DGS workflows showed strong subjective--objective consistency in both rooms, whereas the other three workflows were less consistent in \textit{Reception}, especially for rotation. This room-dependent pattern may have several causes, which the real-world evaluation does not isolate.
For VST-referenced visual consistency\appfullref{fig:heatmap_vc_rank}{ (Appendix \cref{fig:heatmap_vc_rank})}, the $\Delta$Rank analysis showed perfect subjective--objective agreement for 3DGS and 3DGS-MCMC (all zeros), whereas the other workflows diverged across metrics (e.g., Polycam had a higher PSNR but worse perceptual scores; RealityScan had the opposite). This highlights that global image metrics compress complex cross-media appearance differences into single values and may not match perception when trade-offs exist. 

Objective scores also indicate a non-trivial reconstruction--VST discrepancy (e.g., PSNR peaking at 20.7\,dB and LPIPS reaching 0.26), whose absolute values are not directly comparable to conventional rendering benchmarks because we compare a rendered reconstruction against a live VST stream rather than a matched ground-truth render; the two streams differ in resolution, camera pipeline, and colour processing (cf.\ \cref{fig:Room_visual_consistency}). Despite this measurable cross-media gap, participants rated 3DGS-MCMC consistently high (mostly around 4, i.e., \textit{about the same}), suggesting once reconstructions reach a high-fidelity regime, substantial objective discrepancies can correspond to relatively small perceived differences in MR.

\subsection{3DGS-MCMC under the Tested Protocol}
In the real-world evaluation, the desktop 3DGS-MCMC workflow achieved the best registration and VST-referenced visual-consistency results under the tested protocol. Its ratings were significantly higher than standard 3DGS for visual clarity, artefacts, completeness, and smoothness in both scenes, while lighting and colour did not differ significantly.
These findings align with the original motivation of 3DGS-MCMC, which targets improved rendering quality and optimisation robustness by reformulating densification within a Markov chain Monte Carlo view (implemented via stochastic gradient Langevin dynamics and more principled sample updates)~\cite{kheradmand2024mcmc}.

The controlled Replica evaluation provides an additional alignment-relevant insight: 3DGS-MCMC shows substantially smaller anisotropy scores $\rho$ than standard 3DGS, indicating flatter Gaussians. This is consistent with a potentially more surface-proximal source set for correspondence-based registration. However, surface distance was not measured directly. Meanwhile, radius crop and density control have markedly different effects on these two. \cref{fig:tau_time_point_ablation} shows a large retention difference when $\tau{=}1$ (radius cropping and density filtering only). This might suggest that 3DGS contains more isolated outlier noise that is removed by preprocessing, whereas 3DGS-MCMC yields a cleaner source set.
Despite these advantages, 3DGS-MCMC incurs practical costs. Training is substantially slower (around 20 minutes on the real-world datasets and 13 minutes on Replica, longer than 3DGS), and some participants reported fog-like, view-dependent artefacts that affected perceived \textit{Lighting and Colour}. Accordingly, while 3DGS-MCMC outperformed 3DGS on most subjective measures, no significant difference was observed for \textit{Lighting and Colour}.

\subsection{Robustness Trade-offs and Refinement}
\label{study2_reveals_the_robustness_tradeoffs_and_refinement_potential_of_mrcompare}
The controlled Replica evaluation clarifies two practical aspects of MR-Compare's cross-media registration design. First, it highlights the importance of preprocessing: without radius crop and density control, both TEASER++ and TurboReg exhibit substantially lower success rates, whereas enabling these two filters yields successful alignment in 15/16 cases and the difference between the two estimators becomes small. The complete failure of all three fine-registration-only baselines supports the need for a coarse-registration stage, while the lower overall success of RANSAC and FGR indicates that robustness also depends on the chosen coarse-registration method.
Second, and more importantly, moderate anisotropy-based pruning can improve the robustness and accuracy of 3DGS-to-scan alignment using Gaussian centres as surface proxies. The error reduction is particularly pronounced for standard 3DGS ($-0.35$\,cm/$-37.6\%$; $-0.08^{\circ}$/$-38.7\%$), consistent with its Gaussians being less anisotropic overall.

The pruning effect also has direct computational implications. Decreasing $\tau$ removes more splats and reduces the source point count, lowering the cost of voxelisation, correspondence search, and subsequent optimisation; accordingly, total runtime decreases monotonically with $\tau$. The retention curves clarify the associated robustness trade-off: while moderate pruning removes weakly supported points, overly aggressive pruning begins to cause alignment failures once retention falls below $\sim$20--25\%. The heatmap analysis further shows that the best-performing thresholds for both standard 3DGS and 3DGS-MCMC cluster around the median anisotropy value. We therefore use the automatically extracted median as a heuristic default and provide an inspector debug mode to visualise pruning outcomes and support manual tuning. Surface-aware variants such as SuGaR~\cite{guedon2023sugar} and 2DGS~\cite{Huang2DGS2024} instead improve surface alignment through specialised optimisation, producing more explicitly surface-oriented reconstructions. Our method acts as a lightweight post hoc \emph{selection heuristic} for off-the-shelf 3DGS assets, prioritising compatibility with existing workflows.

More generally, the controlled Replica evaluation complements the real-world evaluation with a broader, repeatable stress test across eight Replica scenes. Trends are consistent with the real-world evaluation, but absolute errors are typically lower (often sub-centimetre), plausibly because of clean captures, controlled trajectories, and simulated scan noise. Real deployments introduce additional, hard-to-simulate variability from capture conditions (e.g., illumination, exposure/white balance, motion blur) and on-device depth sensing (e.g., edge artefacts, body occlusions, drift). Together, the controlled Replica ablations and the real-room user study provide complementary evidence: the controlled evaluation isolates algorithmic effects under repeatable conditions, while the real-world evaluation tests MR-Compare under real sensing and capture variability.

\subsection{System-Level Workload, Usability, and Comfort}
\label{sec:system-usability}
Overall workload (NASA TLX) was relatively low across both rooms, with average scores below 4 on a 0--10 scale. This indicates that participants did not perceive MR-Compare as a high-workload platform. UMUX-LITE results confirmed usability, with usefulness and ease of use receiving consistently high ratings (mean scores above 4.2 on a 1--5 scale), suggesting that participants found MR-Compare useful and easy to use for visual comparison. SSQ totals (\textit{Office} = 18.33; \textit{Reception} = 17.45) fall within the 15--20 band classified as \textbf{a concern}~\cite{kennedy2003configural}. However, these magnitudes are typical for HMD use, matching our task profile --- a meta-analysis for motion sickness reports pooled SSQ-total means of 17.33 for \textit{scenic} content and 16.99 for \textit{walking} locomotion, whereas exposures of $\geq 20$ minutes average 27.35 and the all-studies pooled mean is 28.00~\cite{saredakis2020factors}. Our 25--30 minute per-room, inspection-oriented, low-locomotion protocol produced discomfort scores close to scenic/walking benchmarks and below the $\geq 20$-minute exposure and overall means, indicating that the observed discomfort was within the range commonly reported for HMD-based experiences rather than unusually elevated during MR-Compare use, thereby supporting its practical usability under the tested protocol.

\section{Conclusion}
We introduced \textbf{MR-Compare}, a PC-tethered, self-contained Unity MR framework for Meta Quest~3 that registers past 3D reconstructions with the current environment for spatially grounded visual comparison. In a real-world comparative evaluation across two static indoor rooms and five reconstruction workflows, we assessed MR-Compare using both objective benchmarks and an exploratory user study ($n = 30$). The objective benchmark showed that MR-Compare completed registration within seconds with centimetre-level ArUco-referenced translation error. The two desktop 3DGS workflows showed the strongest overall pattern, with 3DGS-MCMC yielding the lowest registration error (\SI{0.89}{cm} and \ang{0.86}) and the strongest VST-referenced objective and perceived visual consistency. MR-Compare also demonstrated practical usability, with participants self-reporting low workload and high usability.
Complementing this real-world evaluation, the controlled Replica evaluation provided a broader assessment of MR-Compare across eight scenes, including a compatible, training-free anisotropy filter and additional ablations of the registration pipeline. The results showed that anisotropy-based filtering can improve fine-grained 3DGS-to-scan alignment by reducing residual errors under controlled conditions, while the broader ablations helped characterise the robustness of key design choices. Taken together, these findings establish system-level feasibility in the evaluated setting for spatially grounded visual comparison in mixed reality.

Future work will extend MR-Compare to larger outdoor and dynamic environments and develop a broader comparative benchmark that combines objective and subjective evaluations across diverse reconstruction workflows. We plan to apply MR-Compare to task-driven scenarios, such as real-world inspection and change-detection workflows. We will also explore broader visual comparison techniques, including continuous-alpha local superposition and photometric normalisation between reconstruction renderings and live VST.

\acknowledgments{
This research received no external funding. The authors would like to thank the UCL Virtual Environments and Computer Graphics (VECG) group for their support of this experiment. In particular, we are grateful to Prof. Anthony Steed and Dr. David Swapp for providing the facilities and equipment. 
The authors used ChatGPT to assist with grammar checking and content condensation. The authors are fully responsible for the final content of this manuscript.
}

\bibliographystyle{abbrv-doi-hyperref}

\bibliography{template}

\appendix
\newpage
\section{Scene Design}
The floor plans of the \textit{Office} and \textit{Reception} study spaces are shown in \cref{fig:floorplan-overall}.
\renewcommand{\thefigure}{A.\arabic{figure}}
\setcounter{figure}{0}
\renewcommand{\theHfigure}{A.\arabic{figure}}
\renewcommand{\thetable}{A.\arabic{table}}
\setcounter{table}{0}
\renewcommand{\theHtable}{A.\arabic{table}}

\begin{figure}[htbp]
    \centering
    \begin{subfigure}{0.32\textwidth}
        \centering
        \includegraphics[width=\linewidth]{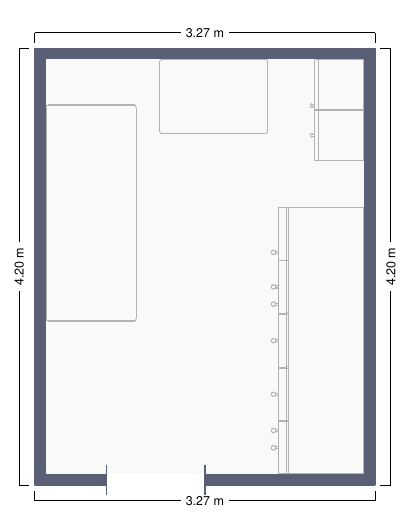} 
        \caption{\textit{Office} floor plan.}
        \label{fig:sub1}
    \end{subfigure}
    \hfill
    \begin{subfigure}{0.32\textwidth}
        \centering
        \includegraphics[width=\linewidth]{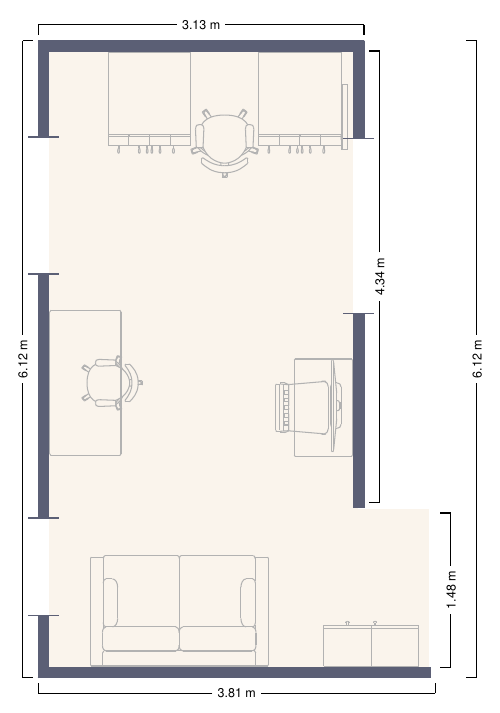} 
        \caption{\textit{Reception} floor plan.}
        \label{fig:sub2}
    \end{subfigure}
    
    \caption{Floor plans of the physical study spaces.}
    \label{fig:floorplan-overall}
\end{figure}

\section{Real-World Evaluation Details}
This section provides full objective and subjective results for the real-world evaluation.
\renewcommand{\thefigure}{B.\arabic{figure}}
\setcounter{figure}{0}
\renewcommand{\theHfigure}{B.\arabic{figure}}
\renewcommand{\thetable}{B.\arabic{table}}
\setcounter{table}{0}
\renewcommand{\theHtable}{B.\arabic{table}}

\subsection{Runtime Characteristics and Resource Usage}
\label{app:runtime}

MR-Compare is deployed in a PC-tethered configuration using a Meta Quest~3 headset connected through Quest Link. In this setup,
rendering is executed on the PC, while the headset mainly performs
tracking and video decoding of the streamed frames. Consequently,
most of the computational workload is concentrated on the PC side.
During typical operation, GPU utilisation on the PC dominates the 
runtime cost due to stereo rendering and video encoding for the 
Quest Link streaming pipeline. Across our experiments, GPU utilisation 
on the PC side generally ranges between approximately 
\textbf{60\%--90\%}, depending on scene complexity and rendering load. 
CPU utilisation on the PC side is usually more moderate, typically 
around \textbf{15\%--40\%}.

By contrast, the headset itself operates under relatively low
computational load because it primarily acts as a display and 
tracking device in the streaming pipeline. In practice, the headset 
typically runs at approximately \textbf{15\%--30\% CPU} and 
\textbf{10\%--25\% GPU} utilisation during MR-Compare operation.
Before the alignment process begins, the system undergoes a short 
initialisation stage that includes Unity runtime startup, Quest Link 
stream establishment, and scene loading. In our setup, this 
initialisation typically takes around \textbf{3--5 seconds} before 
MR-Compare starts executing the alignment pipeline.
These values should be interpreted as indicative runtime ranges 
rather than fixed benchmark measurements, as the exact utilisation 
may vary across runs depending on system scheduling, Unity runtime 
behaviour, and scene complexity.

\subsection{Objective Features}
\subsubsection{Registration Error}
We present the registration error ($\|\Delta p\|$ (cm) and $\left| \Delta \theta \right|$ (°)) of each ArUco marker in \textit{Office} and \textit{Reception}. The trend  is generally consistent, with 3DGS-MCMC and 3DGS generally showing lower errors than all other workflows. Between the two, 3DGS-MCMC is generally more accurate, whereas 3DGS may be more accurate in a few cases (e.g., M6 in \textit{Reception}).
\begin{table}[htbp]
\small
\centering
\setlength{\tabcolsep}{4pt} 
\resizebox{\linewidth}{!}{
\begin{tabular}{lcccccc}
\toprule
Method & M0 & M1 & M2 & M3 & M4 & M5 \\
\midrule
3DGS-MCMC  & \cellTdet{0.927}  & \cellTdet{1.452}  & \cellTdet{0.6214} & \cellTdet{0.7637} & \cellTdet{0.9564} & \cellTdet{0.6335} \\
3DGS       & \cellTdet{2.2615} & \cellTdet{2.2823} & \cellTdet{0.6124} & \cellTdet{0.9613} & \cellTdet{1.8677} & \cellTdet{1.6975} \\
RealityScan       & \cellTdet{2.6564} & \cellTdet{3.0494} & \cellTdet{1.3594} & \cellTdet{2.5225} & \cellTdet{2.1197} & \cellTdet{2.8334} \\
Scaniverse & \cellTdet{5.0997} & \cellTdet{2.4275} & \cellTdet{3.1812} & \cellTdet{3.6317} & \cellTdet{3.0146} & \cellTdet{3.8745} \\
Polycam    & \cellTdet{4.2277} & \cellTdet{3.2396} & \cellTdet{3.3139} & \cellTdet{1.7941} & \cellTdet{0.9970} & \cellTdet{2.8962} \\
\bottomrule
\end{tabular}%
}
\caption{Registration error in translation $\|\Delta p\|$ (cm) of each ArUco marker (IDs 0-5) in \textit{Office}. Two decimals; darker cells indicate smaller errors.}
\label{tab:pose-error-translation-office}
\end{table}

\begin{table}[htbp]
\small
\centering
\setlength{\tabcolsep}{4pt}
\resizebox{\linewidth}{!}{
\begin{tabular}{lcccccc}
\toprule
Method & M0 & M1 & M2 & M3 & M4 & M5 \\
\midrule
3DGS-MCMC  & \cellRdet{0.353847} & \cellRdet{0.872859} & \cellRdet{0.772332} & \cellRdet{1.305510} & \cellRdet{1.343829} & \cellRdet{0.515555} \\
3DGS       & \cellRdet{0.962007} & \cellRdet{0.709699} & \cellRdet{0.530916} & \cellRdet{1.110537} & \cellRdet{2.317667} & \cellRdet{1.293199} \\
RealityScan       & \cellRdet{0.735795} & \cellRdet{0.689456} & \cellRdet{2.696363} & \cellRdet{0.842023} & \cellRdet{1.954118} & \cellRdet{2.284976} \\
Scaniverse & \cellRdet{1.412564} & \cellRdet{2.130547} & \cellRdet{2.913632} & \cellRdet{4.359835} & \cellRdet{3.038749} & \cellRdet{3.440478} \\
Polycam    & \cellRdet{0.611505} & \cellRdet{1.931203} & \cellRdet{3.622324} & \cellRdet{2.621678} & \cellRdet{1.952231} & \cellRdet{0.844694} \\
\bottomrule
\end{tabular}%
}
\caption{Registration error in rotation $\left| \Delta \theta \right|$ (°) of each ArUco marker (IDs 0-5) in \textit{Office}. Two decimals; darker cells indicate smaller errors.}
\label{tab:pose-error-rotation-office}
\end{table}

\begin{table}[htbp]
\small
\centering
\setlength{\tabcolsep}{4pt}
\resizebox{\linewidth}{!}{%
\begin{tabular}{lcccccc}
\toprule
Method & M6 & M7 & M8 & M9 & M10 & M11 \\
\midrule
3DGS-MCMC   & \cellTdet{2.809} & \cellTdet{0.829} & \cellTdet{1.047} & \cellTdet{2.167} & \cellTdet{1.191} & \cellTdet{1.854} \\
3DGS        & \cellTdet{2.281} & \cellTdet{2.730} & \cellTdet{2.916} & \cellTdet{2.569} & \cellTdet{2.602} & \cellTdet{3.041} \\
RealityScan & \cellTdet{2.732} & \cellTdet{3.584} & \cellTdet{2.329} & \cellTdet{2.738} & \cellTdet{2.092} & \cellTdet{4.016} \\
Polycam     & \cellTdet{5.592} & \cellTdet{4.394} & \cellTdet{1.766} & \cellTdet{5.360} & \cellTdet{4.974} & \cellTdet{4.016} \\
Scaniverse  & \cellTdet{2.773} & \cellTdet{2.730} & \cellTdet{2.048} & \cellTdet{2.171} & \cellTdet{2.602} & \cellTdet{3.041} \\
\bottomrule
\end{tabular}%
}
\caption{Registration error in translation $\|\Delta p\|$ (cm) of each ArUco marker (IDs 6-11) in \textit{Reception}. Two decimals; darker cells indicate smaller errors.}
\label{tab:pose-error-translation-reception}
\end{table}

\begin{table}[!htbp]
\small
\centering
\setlength{\tabcolsep}{4pt}
\resizebox{\linewidth}{!}{%
\begin{tabular}{lcccccc}
\toprule
Method & M6 & M7 & M8 & M9 & M10 & M11 \\
\midrule
3DGS-MCMC   & \cellRdet{2.87811} & \cellRdet{2.77874} & \cellRdet{1.26962} & \cellRdet{1.23890} & \cellRdet{1.17132} & \cellRdet{1.37425} \\
3DGS        & \cellRdet{1.87878} & \cellRdet{2.84951} & \cellRdet{1.64741} & \cellRdet{0.93050} & \cellRdet{1.82170} & \cellRdet{1.97742} \\
RealityScan & \cellRdet{3.06384} & \cellRdet{3.11435} & \cellRdet{0.81321} & \cellRdet{3.57975} & \cellRdet{1.67148} & \cellRdet{0.79764} \\
Polycam     & \cellRdet{4.11213} & \cellRdet{2.15483} & \cellRdet{1.36067} & \cellRdet{0.95385} & \cellRdet{3.33873} & \cellRdet{0.79764} \\
Scaniverse  & \cellRdet{3.24083} & \cellRdet{2.84951} & \cellRdet{2.40970} & \cellRdet{0.59799} & \cellRdet{1.82170} & \cellRdet{1.97742} \\
\bottomrule
\end{tabular}%
}
\caption{Registration error in rotation $\left| \Delta \theta \right|$ (°) of each ArUco marker (IDs 6-11) in \textit{Reception}. Two decimals; darker cells indicate smaller errors.}
\label{tab:pose-error-rotation-reception}
\end{table}
\newpage
\subsection{Subjective Measures}
In this section, we present the full details of the statistical analysis of participants' subjective ratings of registration and visual consistency between each 3D reconstruction and VST, complementing the main manuscript.
\Cref{tab:align_clmm_rooms} reports the omnibus tests from the cumulative link mixed models (CLMMs) in both rooms. \Cref{tab:clmm_questionnaire_visual_consistency} reports the corresponding omnibus tests from cumulative link mixed models (CLMMs) for visual-consistency ratings in both rooms.

\begin{table}[ht]
\centering
\setlength{\tabcolsep}{8pt} 
\renewcommand{\arraystretch}{1.1} 
\resizebox{\linewidth}{!}{%
\begin{tabular}{lrrrrrrrr}
\toprule
& \multicolumn{4}{c}{\textbf{Office}} & \multicolumn{4}{c}{\textbf{Reception}} \\
\cmidrule(lr){2-5} \cmidrule(lr){6-9}
\textbf{Questionnaire} & \textbf{Df} & \textbf{$\chi^2$} & \textbf{p} & \textbf{Sig.} & 
\textbf{Df} & \textbf{$\chi^2$} & \textbf{p} & \textbf{Sig.} \\
\midrule
Translation & 4 & 46.58 & $<$.001 & *** & 4 & 35.01 & $<$.001 & *** \\
Rotation    & 4 & 28.27 & $<$.001 & *** & 4 & 34.69 & $<$.001 & *** \\
Scale       & 4 & 36.13 & $<$.001 & *** & 4 & 29.26 & $<$.001 & *** \\
Satisfaction     & 4 & 58.10 & $<$.001 & *** & 4 & 42.23 & $<$.001 & *** \\
\bottomrule
\end{tabular}}
\caption{Omnibus Type~II Wald $\chi^2$ tests from the CLMMs for perceived translation, rotation, and scale misalignment and registration satisfaction in \textit{Office} and \textit{Reception}.}
\label{tab:align_clmm_rooms}
\end{table}

\begin{table}[htbp]
\centering
\setlength{\tabcolsep}{6pt}
\renewcommand{\arraystretch}{1.0}
\resizebox{\linewidth}{!}{%
\begin{tabular}{lrrrrrrrr}
\toprule
 & \multicolumn{4}{c}{\textbf{Office}} & \multicolumn{4}{c}{\textbf{Reception}} \\
\cmidrule(lr){2-5} \cmidrule(lr){6-9}
\textbf{Questionnaire} & \textbf{Df} & $\boldsymbol{\chi^2}$ & \textbf{p} & \textbf{Sig.} &
\textbf{Df} & $\boldsymbol{\chi^2}$ & \textbf{p} & \textbf{Sig.} \\
\midrule
Visual Clarity                  & 4 & 121.0 & $<$.001 & *** & 4 & 122.0 & $<$.001 & *** \\
3D Depth                        & 4 & 139.0 & $<$.001 & *** & 4 & 123.0 & $<$.001 & *** \\
Lighting and Colour             & 4 & 149.0 & $<$.001 & *** & 4 & 138.0 & $<$.001 & *** \\
Visual Artefacts                & 4 & 110.0 & $<$.001 & *** & 4 &  84.6 & $<$.001 & *** \\
Completeness                    & 4 & 100.0 & $<$.001 & *** & 4 &  86.9 & $<$.001 & *** \\
Geometry Accuracy               & 4 & 107.0 & $<$.001 & *** & 4 & 103.0 & $<$.001 & *** \\
Smoothness                      & 4 &  97.4 & $<$.001 & *** & 4 & 115.0 & $<$.001 & *** \\
Text/Pattern Recognisability    & 4 & 107.0 & $<$.001 & *** & 4 &  94.1 & $<$.001 & *** \\
\bottomrule
\end{tabular}}
\caption{Omnibus (Type~II Wald $\chi^2$ tests) from the CLMMs for visual-consistency ratings of each 3D reconstruction relative to VST in \textit{Office} and \textit{Reception}.}
\label{tab:clmm_questionnaire_visual_consistency}

\end{table}

\begin{table}[htbp]
\centering
\begin{threeparttable}
\resizebox{\linewidth}{!}{
\begin{tabular}{l r r r c}
\toprule
Pair & $n$ & Spearman $\rho$ & $p$ & Sig. \\
\midrule
\multicolumn{5}{l}{\textit{Office}} \\
Subj.\ vs Obj.\ (translation)       & 5 & 1.00  & .017 & * \\
Subj.\ vs Obj.\ (rotation)          & 5 & 1.00  & .017 & * \\
Satisfaction vs Obj.\ (translation) & 5 & -0.90 & .083 &  \\
Satisfaction vs Obj.\ (rotation)    & 5 & -0.90 & .083 &  \\
\midrule
\multicolumn{5}{l}{\textit{Reception}} \\
Subj.\ vs Obj.\ (translation)       & 5 & 0.70  & .233 &     \\
Subj.\ vs Obj.\ (rotation)          & 5 & 0.60  & .350 &     \\
Satisfaction vs Obj.\ (translation) & 5 & -0.70 & .233 &     \\
Satisfaction vs Obj.\ (rotation)    & 5 & -0.60 & .350 &     \\
\bottomrule
\end{tabular}
}

\end{threeparttable}
\caption{Spearman correlations between subjective (Subj.) and objective (Obj.) measures across \textit{Office} and \textit{Reception}. The CLMM-estimated expected scores were derived from 30 participants; $n=5$ denotes the five workflow-level pairs within each room. Two-sided exact $p$-values are reported. Sig. column: * $p<0.05$, ** $p<0.01$, *** $p<0.001$.}
\label{tab:alignment-correlation-spearman}
\end{table}

\subsection{Pairwise Comparisons}
\label{sec:pairwise-comparisons}
This section reports pairwise comparisons from the fitted CLMMs.
\Cref{tab:pairwise_alignment_side_by_side} reports post-hoc pairwise comparisons for subjective registration measures, and \cref{tab:pairwise_vc_side_by_side} reports the corresponding comparisons for visual-consistency ratings.

\begin{table*}[t]
\centering
\centering
\resizebox{1\linewidth}{!}{
\begin{tabular}[t]{llrrrrrrrrrrrr}
\toprule
\multicolumn{2}{c}{ } & \multicolumn{6}{c}{Office} & \multicolumn{6}{c}{Reception} \\
\cmidrule(l{3pt}r{3pt}){3-8} \cmidrule(l{3pt}r{3pt}){9-14}
\textbf{Questionnaire} & \textbf{Contrast} & \textbf{Estimate} & \textit{OR} & \textit{95\% CI} & \textit{SE} & \textit{p (adj)} & \textbf{Sig.} & \textbf{Estimate} & \textit{OR} & \textit{95\% CI} & \textit{SE} & \textit{p (adj)} & \textbf{Sig.}\\
\midrule
 & 3DGS - 3DGS-MCMC & \num{1.56690320359862} & \num{4.79178600740617} & {}[\num{0.0207911752927636}, \num{3.11301523190448}] & \num{0.55079922647959} & \pval{0.00634885136103804} & ** & \num{0.226122936666197} & \num{ 1.2537297851646} & {}[\num{-1.31979754292107}, \num{1.77204341625346}] & \num{0.550730987643012} & \pval{0.681375460773654} & \\

 & 3DGS - Scaniverse & \num{-3.07065065068989} & \num{0.0463909607249796} & {}[\num{-4.84356336507839}, \num{-1.29773793630138}] & \num{0.631596503890494} & \pval{3.87884070197036e-06} & *** & \num{-1.61818997145237} & \num{0.198257225752363} & {}[\num{-3.15344723844687}, \num{-0.0829327044578705}] & \num{0.546932240113495} & \pval{0.00547610253876611} & **\\

 & 3DGS - RealityScan & \num{-0.682174598168258} & \num{0.505516500978934} & {}[\num{-2.14247901618307}, \num{0.778129819846555}] & \num{0.520230442000139} & \pval{0.189758752710713} &  & \num{-1.17311782121376} & \num{0.309400779574321} & {}[\num{-2.70526326597561}, \num{0.359027623548084}] & \num{0.545823659850676} & \pval{0.03951708787757} & *\\

 & 3DGS - Polycam & \num{-2.1491163834423} & \num{0.116587130591559} & {}[\num{-3.75759715534035}, \num{-0.540635611544239}] & \num{0.573017962960627} & \pval{0.00029411155808558} & *** & \num{-3.32333767539252} & \num{0.0360323664859442} & {}[\num{-5.13482518098543}, \num{-1.51185016979961}] & \num{0.645338693827583} & \pval{1.30411842808723e-06} & ***\\

 & 3DGS-MCMC - Scaniverse & \num{-4.63755385428851} & \num{0.00968135068078538} & {}[\num{-6.68073380030497}, \num{-2.59437390827205}] & \num{0.727878648649807} & \pval{1.87397974879802e-09} & *** & \num{-1.84431290811857} & \num{0.158133936114738} & {}[\num{-3.4244378152176}, \num{-0.26418800101954}] & \num{0.562916244513627} & \pval{0.00262891633091043} & **\\

 & 3DGS-MCMC - RealityScan & \num{-2.24907780176688} & \num{0.105496468372671} & {}[\num{-3.88303121225628}, \num{-0.615124391277474}] & \num{  0.582092538008} & \pval{0.000223291100891256} & *** & \num{-1.39924075787996} & \num{0.246784261836533} & {}[\num{-2.97384641111092}, \num{0.175364895350999}] & \num{0.560950021687842} & \pval{0.0180236887260475} & *\\

 & 3DGS-MCMC - Polycam & \num{-3.71601958704092} & \num{0.0243306212780292} & {}[\num{-5.57671304190794}, \num{-1.8553261321739}] & \num{0.662868211936353} & \pval{1.03545666989433e-07} & *** & \num{-3.54946061205872} & \num{0.0287401375578021} & {}[\num{-5.41111818617752}, \num{-1.68780303793991}] & \num{0.663211677434582} & \pval{8.70224131676953e-07} & ***\\

 & Scaniverse - RealityScan & \num{2.38847605252163} & \num{10.8968750178682} & {}[\num{0.672383107986517}, \num{4.10456899705675}] & \num{0.611354577877999} & \pval{0.000223291100891256} & *** & \num{0.445072150238608} & \num{1.56060278963443} & {}[\num{-1.02724705121914}, \num{1.91739135169635}] & \num{0.524510683862004} & \pval{0.440148016559123} & \\

 & Scaniverse - Polycam & \num{0.921534267247592} & \num{2.51314326691195} & {}[\num{-0.728746157452588}, \num{2.57181469194777}] & \num{0.58790900320158} & \pval{0.130003409646617} &  & \num{-1.70514770394015} & \num{0.181745539660436} & {}[\num{-3.33336229928712}, \num{-0.0769331085931737}] & \num{0.580048096930322} & \pval{0.00547610253876611} & **\\

\multirow{-10}{*}{\raggedright\arraybackslash Translation} & RealityScan - Polycam & \num{-1.46694178527404} & \num{0.230629722997742} & {}[\num{-3.03980553394498}, \num{0.105921963396907}] & \num{0.560329471775098} & \pval{0.0110558952692346} & * & \num{-2.15021985417876} & \num{0.116458551059626} & {}[\num{-3.83743914752886}, \num{-0.463000560828651}] & \num{0.601068399097169} & \pval{0.00115707396118034} & **\\
\cmidrule{1-14}
 & 3DGS - 3DGS-MCMC & \num{1.69807122699185} & \num{5.46339956520154} & {}[\num{-0.418636923996049}, \num{3.81477937797976}] & \num{0.754072920268713} & \pval{0.0347581951291424} & * & \num{1.86216985102339} & \num{6.43769045647067} & {}[\num{-0.291867930646009}, \num{4.01620763269278}] & \num{0.767371524333429} & \pval{0.0206938323848855} & *\\

 & 3DGS - Scaniverse & \num{-2.26205833634174} & \num{0.104135917261506} & {}[\num{-4.13476166664155}, \num{-0.389355006041943}] & \num{0.667146705329708} & \pval{0.00215484997621881} & ** & \num{-2.52049184797105} & \num{0.0804200425857756} & {}[\num{-4.57922005675027}, \num{-0.461763639191829}] & \num{0.733417685243561} & \pval{0.000981633279703645} & ***\\

 & 3DGS - RealityScan & \num{-0.825061250630845} & \num{0.438208151117235} & {}[\num{-2.60559118149222}, \num{0.955468680230528}] & \num{0.634310121574318} & \pval{0.214837293218547} &  & \num{-0.920345293456222} & \num{0.398381458821942} & {}[\num{-2.7895795316093}, \num{0.948888944696856}] & \num{0.665910848395657} & \pval{0.166945650387089} & \\

 & 3DGS - Polycam & \num{-2.10563606441923} & \num{0.121768197213852} & {}[\num{-4.00592411089838}, \num{-0.205348017940068}] & \num{0.676973703668823} & \pval{0.00373711173514178} & ** & \num{-3.47671329793865} & \num{0.0309088323936958} & {}[\num{-5.69865183904952}, \num{-1.25477475682778}] & \num{0.791561030069778} & \pval{3.73990168812098e-05} & ***\\

 & 3DGS-MCMC - Scaniverse & \num{-3.9601295633336} & \num{0.0190606445709713} & {}[\num{-6.2757655960986}, \num{-1.6444935305686}] & \num{0.824940568538748} & \pval{1.58267710527756e-05} & *** & \num{-4.38266169899444} & \num{0.0124920642161264} & {}[\num{-6.88054002006193}, \num{-1.88478337792694}] & \num{0.889864008490816} & \pval{4.21604358152366e-06} & ***\\

 & 3DGS-MCMC - RealityScan & \num{-2.5231324776227} & \num{0.080207963171566} & {}[\num{-4.64867583227845}, \num{-0.397589122966952}] & \num{0.757220443382786} & \pval{0.00215484997621881} & ** & \num{-2.78251514447961} & \num{0.0618826676298982} & {}[\num{-5.00515275259141}, \num{-0.559877536367808}] & \num{0.791810071249408} & \pval{0.000882428834887675} & ***\\

 & 3DGS-MCMC - Polycam & \num{-3.80370729141108} & \num{0.0222879904280551} & {}[\num{-6.13019955422677}, \num{-1.47721502859539}] & \num{0.828808078140208} & \pval{2.2229242280286e-05} & *** & \num{-5.33888314896203} & \num{0.00480122997567064} & {}[\num{-8.02530065246659}, \num{-2.65246564545748}] & \num{0.957030704012366} & \pval{2.42474213363973e-07} & ***\\

 & Scaniverse - RealityScan & \num{ 1.4369970857109} & \num{4.20804044023359} & {}[\num{-0.29555281377947}, \num{3.16954698520127}] & \num{0.617217334194238} & \pval{0.0331703348979244} & * & \num{1.60014655451483} & \num{4.95375836685277} & {}[\num{-0.274158538273646}, \num{ 3.4744516473033}] & \num{0.667717329918103} & \pval{0.0206938323848855} & *\\

 & Scaniverse - Polycam & \num{0.156422271922519} & \num{ 1.1693198698012} & {}[\num{-1.51816098794837}, \num{ 1.8310055317934}] & \num{0.596566838189096} & \pval{0.793164091281043} &  & \num{-0.956221449967596} & \num{0.384342402712963} & {}[\num{-2.75518242457083}, \num{0.842739524635637}] & \num{0.640876142955933} & \pval{0.150761264356158} & \\

\multirow{-10}{*}{\raggedright\arraybackslash Rotation} & RealityScan - Polycam & \num{-1.28057481378838} & \num{0.277877526703686} & {}[\num{-3.05045616756626}, \num{0.489306539989504}] & \num{0.630516587914845} & \pval{0.0528196233615887} &  & \num{-2.55636800448242} & \num{0.0775860214104759} & {}[\num{-4.56562881265138}, \num{-0.547107196313471}] & \num{0.71579502563464} & \pval{0.000882428834887675} & ***\\
\cmidrule{1-14}
 & 3DGS - 3DGS-MCMC & \num{1.29306411765694} & \num{ 3.6439349123979} & {}[\num{-1.06653594278185}, \num{3.65266417809573}] & \num{0.840602662871061} & \pval{ 0.1549818047768} &  & \num{0.598470726726388} & \num{1.81933441239545} & {}[\num{-1.3351207012614}, \num{2.53206215471417}] & \num{0.68883796475617} & \pval{0.427722154609969} & \\

 & 3DGS - Scaniverse & \num{-3.0866946576812} & \num{0.045652602775164} & {}[\num{-5.13130179914161}, \num{-1.04208751622079}] & \num{0.728387084087971} & \pval{5.64523015446466e-05} & *** & \num{-2.35799540660949} & \num{0.0946096871873394} & {}[\num{-4.20892710924909}, \num{-0.507063703969886}] & \num{0.659390607805792} & \pval{0.000697724262561147} & ***\\

 & 3DGS - RealityScan & \num{-0.425922604315308} & \num{0.653166892463321} & {}[\num{-2.33389713444305}, \num{1.48205192581244}] & \num{0.679711997641369} & \pval{0.589896594292125} &  & \num{-2.00960902667256} & \num{0.134041070908878} & {}[\num{-3.85700535101366}, \num{-0.16221270233146}] & \num{0.65813113656666} & \pval{0.0037697051010881} & **\\

 & 3DGS - Polycam & \num{-3.1806649014992} & \num{0.0415580139470851} & {}[\num{-5.24290336825595}, \num{-1.11842643474246}] & \num{0.73466820742007} & \pval{4.98410668170632e-05} & *** & \num{-2.98228508645794} & \num{0.0506769003575642} & {}[\num{-4.99797142756936}, \num{-0.966598745346506}] & \num{0.718084108514562} & \pval{0.000109327131749109} & ***\\

 & 3DGS-MCMC - Scaniverse & \num{-4.37975877533814} & \num{0.0125283804109229} & {}[\num{-6.90470248400008}, \num{-1.8548150666762}] & \num{0.899505997090907} & \pval{5.6061993344331e-06} & *** & \num{-2.95646613333587} & \num{0.0520023622610263} & {}[\num{-4.95435341808325}, \num{-0.958578848588503}] & \num{0.711743231334961} & \pval{0.000109327131749109} & ***\\

 & 3DGS-MCMC - RealityScan & \num{-1.71898672197224} & \num{0.179247683662248} & {}[\num{-4.03743956780773}, \num{0.599466123863237}] & \num{0.825944052394998} & \pval{0.0534458047736368} &  & \num{-2.60807975339895} & \num{0.0736758838812879} & {}[\num{-4.5937996548462}, \num{-0.622359851951701}] & \num{0.707408626088191} & \pval{0.000567711480811731} & ***\\

 & 3DGS-MCMC - Polycam & \num{-4.47372901915614} & \num{0.0114047080823784} & {}[\num{-7.01792718078697}, \num{-1.92953085752531}] & \num{0.906365356314166} & \pval{5.6061993344331e-06} & *** & \num{-3.58075581318432} & \num{0.027854637395024} & {}[\num{-5.75354376172399}, \num{-1.40796786464466}] & \num{0.774051232672434} & \pval{3.72807356801703e-05} & ***\\

 & Scaniverse - RealityScan & \num{2.66077205336589} & \num{14.3073308586615} & {}[\num{0.733042069637879}, \num{4.58850203709391}] & \num{0.686749837307945} & \pval{0.000178115869054984} & *** & \num{0.348386379936926} & \num{1.41677955919524} & {}[\num{-1.24101282509836}, \num{1.93778558497221}] & \num{0.566220194056682} & \pval{0.538367002483706} & \\

 & Scaniverse - Polycam & \num{-0.0939702438180015} & \num{0.91030984918331} & {}[\num{-1.65006562186115}, \num{1.46212513422514}] & \num{0.554355774266751} & \pval{0.865393525975881} &  & \num{-0.624289679848449} & \num{0.535641770564333} & {}[\num{-2.25580469909995}, \num{1.00722533940305}] & \num{0.581223866150395} & \pval{0.353475101427564} & \\

\multirow{-10}{*}{\raggedright\arraybackslash Scale} & RealityScan - Polycam & \num{-2.7547422971839} & \num{0.0636254140046126} & {}[\num{-4.69902802344027}, \num{-0.810456570927522}] & \num{0.692647786493689} & \pval{0.000139510840443547} & *** & \num{-0.972676059785374} & \num{0.378069945382745} & {}[\num{-2.65489843655526}, \num{0.709546316984511}] & \num{0.599288257854634} & \pval{0.149397528491248} & \\
\cmidrule{1-14}
 & 3DGS - 3DGS-MCMC & \num{-3.48530606874051} & \num{0.0306443777076174} & {}[\num{-5.58616958566595}, \num{-1.38444255181507}] & \num{0.748428302009699} & \pval{5.35182937550581e-06} & *** & \num{-1.12179520262427} & \num{0.325694581724518} & {}[\num{-2.79593188567106}, \num{0.552341480422522}] & \num{0.59640774611506} & \pval{0.0749785184249486} & \\

 & 3DGS - Scaniverse & \num{3.27600177041427} & \num{26.4697287944085} & {}[\num{ 1.4767661310714}, \num{5.07523740975714}] & \num{0.640973991703864} & \pval{6.41002263998601e-07} & *** & \num{2.15749266858375} & \num{8.64942347629922} & {}[\num{0.536147308219519}, \num{3.77883802894798}] & \num{0.577600946112184} & \pval{0.000375035054844899} & ***\\

 & 3DGS - RealityScan & \num{2.12635737533165} & \num{8.38427036952026} & {}[\num{ 0.4646721357634}, \num{3.78804261489991}] & \num{0.591971945014639} & \pval{0.000468803284980315} & *** & \num{ 1.2980868104057} & \num{3.66228331848889} & {}[\num{-0.270329413024169}, \num{2.86650303383557}] & \num{0.558745050065879} & \pval{0.0288105999959024} & *\\

 & 3DGS - Polycam & \num{3.60801937566395} & \num{36.8929094097248} & {}[\num{1.73184244111648}, \num{5.48419631021142}] & \num{0.66838416969043} & \pval{1.68355058546953e-07} & *** & \num{3.05230569896443} & \num{21.1640862206724} & {}[\num{1.33197828123477}, \num{4.77263311669409}] & \num{0.612863100234338} & \pval{2.11505818898909e-06} & ***\\

 & 3DGS-MCMC - Scaniverse & \num{6.76130783915478} & \num{863.771131101441} & {}[\num{4.07594185762792}, \num{9.44667382068164}] & \num{0.956656101473004} & \pval{7.87912643762474e-12} & *** & \num{3.27928787120802} & \num{26.5568540640174} & {}[\num{1.45345368463413}, \num{5.10512205778191}] & \num{0.65044966938576} & \pval{2.11505818898909e-06} & ***\\

 & 3DGS-MCMC - RealityScan & \num{5.61166344407216} & \num{273.598976279298} & {}[\num{3.13846687800503}, \num{8.08486001013929}] & \num{0.881071184094218} & \pval{6.33647468568462e-10} & *** & \num{2.41988201302997} & \num{11.2445325282892} & {}[\num{0.679847620626817}, \num{4.15991640543312}] & \num{0.619883669383073} & \pval{0.000236766455554849} & ***\\

 & 3DGS-MCMC - Polycam & \num{7.09332544440446} & \num{1203.90466929123} & {}[\num{4.33435754776219}, \num{9.85229334104672}] & \num{0.982876632178922} & \pval{5.31848336558439e-12} & *** & \num{ 4.1741009015887} & \num{ 64.981388725016} & {}[\num{2.22201239085609}, \num{6.12618941232132}] & \num{0.695427512396611} & \pval{1.94651558568345e-08} & ***\\

 & Scaniverse - RealityScan & \num{-1.14964439508262} & \num{0.316749386993771} & {}[\num{-2.6739816333522}, \num{0.374692843186966}] & \num{0.54304200236571} & \pval{0.0380622474518499} & * & \num{-0.859405858178049} & \num{0.423413575312173} & {}[\num{-2.31259692571279}, \num{0.59378520935669}] & \num{0.517696325538736} & \pval{0.096902811732472} & \\

 & Scaniverse - Polycam & \num{0.332017605249678} & \num{ 1.3937773860954} & {}[\num{-1.16539053556859}, \num{1.82942574606795}] & \num{0.533448566848473} & \pval{0.533679845059053} &  & \num{0.894813030380683} & \num{2.44687825479529} & {}[\num{-0.553126259949602}, \num{2.34275232071097}] & \num{0.515825390723601} & \pval{0.0919891907537277} & \\

\multirow{-10}{*}{\raggedright\arraybackslash Satisfaction} & RealityScan - Polycam & \num{ 1.4816620003323} & \num{4.40025282865917} & {}[\num{-0.0972559368263652}, \num{3.06057993749096}] & \num{0.562486263957647} & \pval{0.010544144846652} & * & \num{1.75421888855873} & \num{5.77893198863845} & {}[\num{0.232867519094295}, \num{3.27557025802317}] & \num{0.541978292752089} & \pval{0.00201537032745166} & **\\
\bottomrule
\end{tabular}}
\caption{Post-hoc pairwise comparisons for registration questionnaire items in \textit{Office} and \textit{Reception} (latent scale; $p$ values are FDR--BH adjusted). Odds ratios (ORs) are directional and may appear as reciprocals of those in the main text when the contrast direction is reversed.}
\label{tab:pairwise_alignment_side_by_side}
\end{table*}

\begin{table*}
\centering
\centering
\resizebox{\ifdim\width>\linewidth\linewidth\else\width\fi}{!}{
\begin{tabular}[t]{llrrrrrrrrrrrr}
\toprule
\multicolumn{2}{c}{ } & \multicolumn{6}{c}{Office} & \multicolumn{6}{c}{Reception} \\
\cmidrule(l{3pt}r{3pt}){3-8} \cmidrule(l{3pt}r{3pt}){9-14}
\textbf{Questionnaire} & \textbf{Contrast} & \textbf{Estimate} & \textit{OR} & \textit{95\% CI} & \textit{SE} & \textit{p (adj)} & \textbf{Sig.} & \textbf{Estimate} & \textit{OR} & \textit{95\% CI} & \textit{SE} & \textit{p (adj)} & \textbf{Sig.}\\
\midrule
 & 3DGS - 3DGS-MCMC & \num{-1.74981092070416} & \num{0.173806803611798} & {}[\num{-3.22188693755688}, \num{-0.277734903851439}] & \num{0.524424049847206} & \pval{0.0012114236008485} & ** & \num{-1.63765697126252} & \num{0.194435075974807} & {}[\num{-3.21470999819616}, \num{-0.0606039443288708}] & \num{0.561821893530029} & \pval{0.00444747242098963} & **\\

 & 3DGS - Scaniverse & \num{3.39241633213035} & \num{29.7377217385335} & {}[\num{1.81679161679757}, \num{4.96804104746314}] & \num{0.561313060463263} & \pval{3.0121033984084e-09} & *** & \num{4.00291342432833} & \num{54.7574495526019} & {}[\num{2.28019828849885}, \num{ 5.7256285601578}] & \num{0.613713719890839} & \pval{1.38341702880976e-10} & ***\\

 & 3DGS - RealityScan & \num{3.13206784136965} & \num{22.9213282415549} & {}[\num{1.53915955719986}, \num{4.72497612553945}] & \num{0.567470296272794} & \pval{5.67089393622368e-08} & *** & \num{2.94747094402385} & \num{19.0576947523234} & {}[\num{1.33509917635504}, \num{4.55984271169266}] & \num{0.574404122191995} & \pval{4.7944194455719e-07} & ***\\

 & 3DGS - Polycam & \num{3.55915640300395} & \num{35.1335460861539} & {}[\num{1.93919678475388}, \num{5.17911602125401}] & \num{ 0.5771072783374} & \pval{1.73735725088409e-09} & *** & \num{4.54071324058567} & \num{93.7576480371128} & {}[\num{2.72223732421836}, \num{6.35918915695297}] & \num{0.647828300776104} & \pval{7.99344465487029e-12} & ***\\

 & 3DGS-MCMC - Scaniverse & \num{5.14222725283452} & \num{171.096419245782} & {}[\num{3.22226737474895}, \num{7.06218713092009}] & \num{0.683981753172274} & \pval{2.7793954272592e-13} & *** & \num{5.64057039559085} & \num{281.623309364699} & {}[\num{3.58631477264402}, \num{7.69482601853767}] & \num{0.731824335750286} & \pval{6.41295854565651e-14} & ***\\

 & 3DGS-MCMC - RealityScan & \num{4.88187876207381} & \num{ 131.87819904191} & {}[\num{2.97114670982385}, \num{6.79261081432378}] & \num{0.680694359219387} & \pval{2.46528320261485e-12} & *** & \num{4.58512791528636} & \num{98.0157240496706} & {}[\num{2.65252362270788}, \num{6.51773220786485}] & \num{0.688486299799218} & \pval{6.85861258735877e-11} & ***\\

 & 3DGS-MCMC - Polycam & \num{5.30896732370811} & \num{202.141373962699} & {}[\num{3.35022283807469}, \num{7.26771180934152}] & \num{0.69779868974968} & \pval{2.7793954272592e-13} & *** & \num{6.17837021184818} & \num{ 482.20542290045} & {}[\num{4.03612280046673}, \num{8.32061762322964}] & \num{0.763171229195942} & \pval{5.69575068919936e-15} & ***\\

 & Scaniverse - RealityScan & \num{-0.260348490760701} & \num{0.770782928264944} & {}[\num{-1.67009202318842}, \num{1.14939504166702}] & \num{0.502218230619825} & \pval{0.671312460892746} &  & \num{-1.05544248030448} & \num{0.34803839309601} & {}[\num{-2.44775474518546}, \num{ 0.3368697845765}] & \num{0.496008377449078} & \pval{0.0370529710083299} & *\\

 & Scaniverse - Polycam & \num{0.166740070873591} & \num{1.18144713287261} & {}[\num{-1.19155207414263}, \num{1.52503221588981}] & \num{0.483888779798198} & \pval{0.730407540715807} &  & \num{0.537799816257338} & \num{1.71223548217026} & {}[\num{-0.835332947282882}, \num{1.91093257979756}] & \num{0.489175719589006} & \pval{0.271593603913759} & \\

\multirow{-10}{*}{\raggedright\arraybackslash Visual Clarity} & RealityScan - Polycam & \num{0.427088561634292} & \num{1.53278840195914} & {}[\num{-1.02085420693999}, \num{1.87503133020857}] & \num{0.515826629840862} & \pval{0.509609929309861} &  & \num{1.59324229656182} & \num{4.91967413979502} & {}[\num{0.136842163127401}, \num{3.04964242999624}] & \num{0.518839548657699} & \pval{0.00305000200957523} & **\\
\cmidrule{1-14}
 & 3DGS - 3DGS-MCMC & \num{-0.547789337312375} & \num{0.578226662623732} & {}[\num{-1.87781304704566}, \num{0.782234372420912}] & \num{0.473818207936281} & \pval{0.275148064771503} &  & \num{-0.796000056777653} & \num{0.451129853788947} & {}[\num{-2.35557254291113}, \num{0.763572429355823}] & \num{0.555594486864207} & \pval{0.168826304649529} & \\

 & 3DGS - Scaniverse & \num{2.50941321119398} & \num{12.2977117833013} & {}[\num{ 1.0730659663833}, \num{3.94576045600467}] & \num{0.511695748376463} & \pval{2.34627443347462e-06} & *** & \num{ 3.8647182714837} & \num{47.6898349534728} & {}[\num{2.08870714720927}, \num{5.64072939575812}] & \num{0.632700306032406} & \pval{3.35676449088173e-09} & ***\\

 & 3DGS - RealityScan & \num{1.32565977655821} & \num{3.76466837573413} & {}[\num{-0.0482420891304225}, \num{2.69956164224683}] & \num{0.489449710645716} & \pval{0.00965632237969509} & ** & \num{2.91954604516097} & \num{18.5328724627415} & {}[\num{1.23869222658776}, \num{4.60039986373418}] & \num{0.598800711815072} & \pval{1.80746119310662e-06} & ***\\

 & 3DGS - Polycam & \num{2.86860923281089} & \num{17.6125062627611} & {}[\num{1.36716326110161}, \num{4.37005520452016}] & \num{0.534887035789083} & \pval{2.72808269745674e-07} & *** & \num{3.62990327389711} & \num{37.7091689798149} & {}[\num{1.91441251636943}, \num{5.34539403142478}] & \num{0.611140049996563} & \pval{7.14468184529387e-09} & ***\\

 & 3DGS-MCMC - Scaniverse & \num{3.05720254850636} & \num{21.2679777295288} & {}[\num{1.50682044567196}, \num{4.60758465134075}] & \num{0.552320431737859} & \pval{1.55439122914418e-07} & *** & \num{4.66071832826135} & \num{105.711990800289} & {}[\num{2.70962432711381}, \num{6.61181232940889}] & \num{0.695073220404726} & \pval{2.00906158831478e-10} & ***\\

 & 3DGS-MCMC - RealityScan & \num{1.87344911387058} & \num{6.51071391044432} & {}[\num{0.403934201048964}, \num{ 3.3429640266922}] & \num{0.523511661809703} & \pval{0.000690827485024262} & *** & \num{3.71554610193862} & \num{41.0810153819076} & {}[\num{1.86350448541161}, \num{5.56758771846564}] & \num{0.659786012342755} & \pval{3.57419908911605e-08} & ***\\

 & 3DGS-MCMC - Polycam & \num{3.41639857012326} & \num{30.4595194259003} & {}[\num{1.79870570041666}, \num{5.03409143982987}] & \num{0.576299753836261} & \pval{3.06343330203298e-08} & *** & \num{4.42590333067476} & \num{ 83.588280986291} & {}[\num{2.53488288731273}, \num{6.31692377403678}] & \num{0.673672139141297} & \pval{2.51884030012531e-10} & ***\\

 & Scaniverse - RealityScan & \num{-1.18375343463578} & \num{0.306127549748406} & {}[\num{-2.58455110861725}, \num{0.217044239345697}] & \num{0.499031286968794} & \pval{0.0221089705174649} & * & \num{-0.945172226322727} & \num{0.388612635812695} & {}[\num{-2.37643064763415}, \num{0.486086194988697}] & \num{0.509882865483264} & \pval{0.0911163279666719} & \\

 & Scaniverse - Polycam & \num{0.359196021616906} & \num{1.43217751180968} & {}[\num{-1.01804798807185}, \num{1.73644003130566}] & \num{0.490640342563941} & \pval{0.464109739514796} &  & \num{-0.23481499758659} & \num{0.79071712067372} & {}[\num{-1.62138200705229}, \num{1.15175201187911}] & \num{0.49396164203746} & \pval{0.634522621674368} & \\

\multirow{-10}{*}{\raggedright\arraybackslash 3D Depth} & RealityScan - Polycam & \num{1.54294945625268} & \num{4.67836858520814} & {}[\num{0.0905457567408385}, \num{2.99535315576452}] & \num{0.517415827301852} & \pval{0.00477240087525681} & ** & \num{0.710357228736137} & \num{2.03471798857007} & {}[\num{-0.66584650667139}, \num{2.08656096414366}] & \num{0.490269746993287} & \pval{0.168826304649529} & \\
\cmidrule{1-14}
 & 3DGS - 3DGS-MCMC & \num{-0.294083117396595} & \num{0.745214548575877} & {}[\num{-1.58405508750481}, \num{0.995888852711617}] & \num{0.459549858165517} & \pval{0.558519306929823} &  & \num{-0.783130928102523} & \num{0.456973019509804} & {}[\num{-2.1392610859742}, \num{0.572999229769157}] & \num{0.48311857632971} & \pval{0.13127442563362} & \\

 & 3DGS - Scaniverse & \num{1.38610238043381} & \num{3.99923215096411} & {}[\num{0.0556968144327037}, \num{2.71650794643491}] & \num{0.473954243445409} & \pval{0.0172477916058788} & * & \num{2.18777331934373} & \num{8.91533938293496} & {}[\num{0.753700939137893}, \num{3.62184569954956}] & \num{0.510885332545166} & \pval{5.64229507354434e-05} & ***\\

 & 3DGS - RealityScan & \num{0.698587549629498} & \num{2.01091038950178} & {}[\num{-0.599952597566308}, \num{ 1.9971276968253}] & \num{0.462602253610211} & \pval{0.212073692659195} &  & \num{1.83212142208318} & \num{6.24712540086134} & {}[\num{0.414767832287915}, \num{3.24947501187845}] & \num{0.504929297887118} & \pval{0.000570209066088471} & ***\\

 & 3DGS - Polycam & \num{0.422886807884007} & \num{1.52636151409832} & {}[\num{-0.867335007609571}, \num{1.71310862337759}] & \num{0.459638865069597} & \pval{0.446938873774819} &  & \num{1.39611279707302} & \num{4.03946717974277} & {}[\num{0.0348149177399324}, \num{2.75741067640612}] & \num{0.484959566459466} & \pval{0.0066527030112001} & **\\

 & 3DGS-MCMC - Scaniverse & \num{ 1.6801854978304} & \num{5.36655136243212} & {}[\num{0.290362556917523}, \num{3.07000843874328}] & \num{0.495121560911217} & \pval{0.00690104025537087} & ** & \num{2.97090424744625} & \num{19.5095530858659} & {}[\num{1.43208224024842}, \num{4.50972625464408}] & \num{0.548202171470761} & \pval{5.98131776861642e-07} & ***\\

 & 3DGS-MCMC - RealityScan & \num{0.992670667026093} & \num{2.69843146962802} & {}[\num{-0.357257301981523}, \num{2.34259863603371}] & \num{0.480909059317835} & \pval{0.104622130161798} &  & \num{2.61525235018571} & \num{13.6706657376898} & {}[\num{1.10308548952545}, \num{4.12741921084596}] & \num{0.53870633040245} & \pval{6.0287542505754e-06} & ***\\

 & 3DGS-MCMC - Polycam & \num{0.716969925280602} & \num{2.04821754622911} & {}[\num{-0.62251381024491}, \num{2.05645366080611}] & \num{0.477188322645591} & \pval{0.212073692659195} &  & \num{2.17924372517555} & \num{8.83961854919994} & {}[\num{0.735766161352324}, \num{3.62272128899877}] & \num{0.514235909842616} & \pval{5.64229507354434e-05} & ***\\

 & Scaniverse - RealityScan & \num{-0.687514830804309} & \num{0.502824120629507} & {}[\num{-2.02304935500109}, \num{0.648019693392475}] & \num{0.475781424241566} & \pval{0.212073692659195} &  & \num{-0.355651897260544} & \num{0.700716499118261} & {}[\num{-1.73472831809413}, \num{1.02342452357305}] & \num{0.491293135261165} & \pval{0.469121135619495} & \\

 & Scaniverse - Polycam & \num{-0.963215572549799} & \num{0.381663643539761} & {}[\num{-2.29184061081085}, \num{0.365409465711248}] & \num{0.473319934104305} & \pval{0.104622130161798} &  & \num{-0.791660522270702} & \num{0.453091801247051} & {}[\num{-2.15370949238133}, \num{0.57038844783992}] & \num{0.485227141002388} & \pval{0.13127442563362} & \\

\multirow{-10}{*}{\raggedright\arraybackslash Lighting and Colour} & RealityScan - Polycam & \num{-0.275700741745491} & \num{0.759040045775727} & {}[\num{-1.59851339228164}, \num{1.04711190879066}] & \num{0.471249282945617} & \pval{0.558519306929823} &  & \num{-0.436008625010158} & \num{0.646612148875036} & {}[\num{-1.7915888177245}, \num{0.919571567704182}] & \num{0.482922652374843} & \pval{0.407337125316008} & \\
\cmidrule{1-14}
 & 3DGS - 3DGS-MCMC & \num{-1.58910351173945} & \num{0.204108510622261} & {}[\num{-3.00483683286424}, \num{-0.173370190614666}] & \num{0.50435208050956} & \pval{0.00232619882614583} & ** & \num{-1.76420355518522} & \num{0.171323181621786} & {}[\num{-3.32545220457948}, \num{-0.20295490579095}] & \num{0.55619161657447} & \pval{0.00189283736785506} & **\\

 & 3DGS - Scaniverse & \num{1.62170012147305} & \num{5.06168849081563} & {}[\num{0.260918689548575}, \num{2.98248155339752}] & \num{0.484775583133565} & \pval{0.00137009366498734} & ** & \num{4.73718766933013} & \num{114.112826982868} & {}[\num{2.26497392641178}, \num{7.20940141224848}] & \num{ 0.8807210553712} & \pval{1.25000634269513e-07} & ***\\

 & 3DGS - RealityScan & \num{2.93139417354967} & \num{18.7537582728545} & {}[\num{1.40566332817315}, \num{ 4.4571250189262}] & \num{0.543538472027977} & \pval{1.7306564388606e-07} & *** & \num{6.61628240814059} & \num{747.162283211994} & {}[\num{3.93181559483571}, \num{9.30074922144548}] & \num{0.956335774645406} & \pval{1.52292535267747e-11} & ***\\

 & 3DGS - Polycam & \num{2.48469261329986} & \num{11.9974318369924} & {}[\num{0.995868906214803}, \num{3.97351632038492}] & \num{0.530390380007251} & \pval{5.60880969925278e-06} & *** & \num{6.32315178834883} & \num{557.326803304414} & {}[\num{ 3.6739021452931}, \num{8.97240143140456}] & \num{0.943789730259937} & \pval{5.21954094929641e-11} & ***\\

 & 3DGS-MCMC - Scaniverse & \num{ 3.2108036332125} & \num{24.7990075248904} & {}[\num{1.61515624322659}, \num{4.80645102319842}] & \num{0.568446097079684} & \pval{5.39867029956365e-08} & *** & \num{6.50139122451534} & \num{666.067638381734} & {}[\num{3.68901487144037}, \num{9.31376757759031}] & \num{1.00190328480954} & \pval{1.72769224245066e-10} & ***\\

 & 3DGS-MCMC - RealityScan & \num{4.52049768528912} & \num{91.8813145795849} & {}[\num{2.73673482581126}, \num{6.30426054476699}] & \num{0.635461845737009} & \pval{1.12956142233956e-11} & *** & \num{8.38048596332581} & \num{4361.12775947293} & {}[\num{5.31253724886878}, \num{11.4484346777828}] & \num{1.09295041230201} & \pval{1.75020357181551e-13} & ***\\

 & 3DGS-MCMC - Polycam & \num{4.07379612503931} & \num{ 58.779674597674} & {}[\num{2.34121959183284}, \num{5.80637265824578}] & \num{0.617226822400758} & \pval{2.05356374174958e-10} & *** & \num{8.08735534353405} & \num{ 3253.0729235158} & {}[\num{5.05846926786765}, \num{11.1162414192004}] & \num{1.07903442766686} & \pval{3.31509831254748e-13} & ***\\

 & Scaniverse - RealityScan & \num{1.30969405207662} & \num{3.70503998949816} & {}[\num{-0.101162212747059}, \num{ 2.7205503169003}] & \num{0.502614639244653} & \pval{0.0114587718229312} & * & \num{1.87909473881047} & \num{6.54757491306535} & {}[\num{0.258403760313484}, \num{3.49978571730745}] & \num{0.577367823919417} & \pval{0.00162223903050213} & **\\

 & Scaniverse - Polycam & \num{0.86299249182681} & \num{ 2.3702430243903} & {}[\num{-0.531803779108422}, \num{2.25778876276204}] & \num{0.49689329949108} & \pval{0.0915851877845606} &  & \num{ 1.5859641190187} & \num{4.88399786457035} & {}[\num{-0.00891128360415605}, \num{3.18083952164157}] & \num{0.568171078171199} & \pval{0.00583208255061452} & **\\

\multirow{-10}{*}{\raggedright\arraybackslash Visual Artefacts} & RealityScan - Polycam & \num{-0.446701560249811} & \num{0.639734802082755} & {}[\num{-1.8743935796201}, \num{0.980990459120482}] & \num{0.508612342135327} & \pval{0.379794440260438} &  & \num{-0.293130619791762} & \num{0.745924701804417} & {}[\num{-1.79199223972656}, \num{1.20573100014304}] & \num{0.533966365790872} & \pval{0.583027189561115} & \\
\cmidrule{1-14}
 & 3DGS - 3DGS-MCMC & \num{-1.96862925401747} & \num{0.139648147216178} & {}[\num{-3.67476773048213}, \num{-0.262490777552804}] & \num{0.607808319125179} & \pval{0.001333248939367} & ** & \num{-1.63178249547437} & \num{0.195580641625423} & {}[\num{-3.68799836552706}, \num{0.424433374578323}] & \num{ 0.7325226697454} & \pval{0.02878426034468} & *\\

 & 3DGS - Scaniverse & \num{4.24813842527456} & \num{69.9750272813717} & {}[\num{2.19538389530776}, \num{6.30089295524137}] & \num{0.731289574467059} & \pval{1.04679056429565e-08} & *** & \num{6.35585838092151} & \num{575.856432744908} & {}[\num{3.62822214622639}, \num{9.08349461561664}] & \num{0.971714792125382} & \pval{1.22325435076675e-10} & ***\\

 & 3DGS - RealityScan & \num{6.03438425103232} & \num{417.541629639881} & {}[\num{3.82172111798314}, \num{ 8.2470473840815}] & \num{0.788256685046824} & \pval{6.42599713388696e-14} & *** & \num{8.25093949199861} & \num{3831.22353501974} & {}[\num{5.35326563784271}, \num{11.1486133461545}] & \num{1.03229034393326} & \pval{1.31856989742331e-14} & ***\\

 & 3DGS - Polycam & \num{5.99419613886478} & \num{401.094130381153} & {}[\num{3.76472216224823}, \num{8.22367011548133}] & \num{0.794245513452436} & \pval{1.11322481454711e-13} & *** & \num{7.90838737754359} & \num{2720.00059364971} & {}[\num{5.05273607905978}, \num{10.7640386760274}] & \num{1.01731989500386} & \pval{3.80989587764684e-14} & ***\\

 & 3DGS-MCMC - Scaniverse & \num{6.21676767929203} & \num{501.080957222075} & {}[\num{3.54637518690166}, \num{ 8.8871601716824}] & \num{0.95132182680009} & \pval{1.27326759422993e-10} & *** & \num{7.98764087639588} & \num{2944.34269137837} & {}[\num{4.35782237641599}, \num{11.6174593763758}] & \num{1.29311536644661} & \pval{1.08849271144137e-09} & ***\\

 & 3DGS-MCMC - RealityScan & \num{8.00301350504978} & \num{2989.95466795215} & {}[\num{5.16571522469886}, \num{10.8403117854007}] & \num{1.01078167008478} & \pval{2.42045366986208e-14} & *** & \num{9.88272198747298} & \num{19588.9710923298} & {}[\num{6.10782101866251}, \num{13.6576229562834}] & \num{1.34480069722771} & \pval{6.66459539342987e-13} & ***\\

 & 3DGS-MCMC - Polycam & \num{7.96282539288224} & \num{2872.17652633982} & {}[\num{5.10828149519659}, \num{10.8173692905679}] & \num{1.01692538574977} & \pval{2.43363983758019e-14} & *** & \num{9.54016987301796} & \num{13907.3099006345} & {}[\num{5.80039483687091}, \num{ 13.279944909165}] & \num{1.33228715604429} & \pval{2.00598056293834e-12} & ***\\

 & Scaniverse - RealityScan & \num{1.78624582575775} & \num{5.96700917258572} & {}[\num{0.308756881825814}, \num{3.26373476968969}] & \num{0.526352393973402} & \pval{0.00098532213158073} & *** & \num{1.89508111107709} & \num{6.65308802188356} & {}[\num{0.392906217464772}, \num{3.39725600468941}] & \num{0.535146712716117} & \pval{0.000568937598410721} & ***\\

 & Scaniverse - Polycam & \num{1.74605771359021} & \num{ 5.7319610433068} & {}[\num{0.261362302888461}, \num{3.23075312429197}] & \num{0.528919682921288} & \pval{0.00120349973257509} & ** & \num{1.55252899662208} & \num{4.72340055434374} & {}[\num{0.0636689261622301}, \num{3.04138906708192}] & \num{0.530403334384644} & \pval{0.00427711428727154} & **\\

\multirow{-10}{*}{\raggedright\arraybackslash Completeness} & RealityScan - Polycam & \num{-0.0401881121675398} & \num{0.960608719966647} & {}[\num{-1.43885628035372}, \num{1.35848005601864}] & \num{0.498272654914092} & \pval{0.93571643479856} &  & \num{-0.342552114455017} & \num{0.709956119445193} & {}[\num{-1.72151480818456}, \num{1.03641057927453}] & \num{0.491252620214525} & \pval{0.485612950773255} & \\
\cmidrule{1-14}
 & 3DGS - 3DGS-MCMC & \num{-1.49708960388707} & \num{0.223780503218118} & {}[\num{-3.04112305572584}, \num{0.046943847951691}] & \num{0.550058737893192} & \pval{0.0072455777890716} & ** & \num{-0.777170770954449} & \num{0.459704783302454} & {}[\num{-2.21000133140615}, \num{0.655659789497248}] & \num{0.51044293681479} & \pval{0.127873349089201} & \\

 & 3DGS - Scaniverse & \num{ 4.2895155605388} & \num{ 72.931129223508} & {}[\num{2.31900279095705}, \num{6.26002833012054}] & \num{0.701991116674162} & \pval{1.65534173881649e-09} & *** & \num{4.13884218830182} & \num{62.7301496860761} & {}[\num{2.05053841602308}, \num{6.22714596058056}] & \num{0.743953918841125} & \pval{4.41188972931062e-08} & ***\\

 & 3DGS - RealityScan & \num{ 5.0398874370457} & \num{154.452628403212} & {}[\num{2.95971981296189}, \num{7.12005506112951}] & \num{0.74105543279985} & \pval{2.07850343664934e-11} & *** & \num{6.14938688723637} & \num{ 468.43009824918} & {}[\num{3.72300350185239}, \num{8.57577027262036}] & \num{0.864394084869023} & \pval{2.81611453968103e-12} & ***\\

 & 3DGS - Polycam & \num{6.47634804168719} & \num{649.594317869534} & {}[\num{4.18565821747615}, \num{8.76703786589823}] & \num{0.816053533108219} & \pval{1.04251355167397e-14} & *** & \num{7.11028515739487} & \num{1224.49667059227} & {}[\num{4.55804112995492}, \num{9.66252918483482}] & \num{0.909231679441397} & \pval{2.63901185703486e-14} & ***\\

 & 3DGS-MCMC - Scaniverse & \num{5.78660516442587} & \num{325.904751194622} & {}[\num{3.51584635065924}, \num{ 8.0573639781925}] & \num{0.808953151677409} & \pval{2.11942033610282e-12} & *** & \num{4.91601295925627} & \num{136.457465670537} & {}[\num{2.73214072946175}, \num{7.09988518905079}] & \num{0.777999985045794} & \pval{5.27262840091916e-10} & ***\\

 & 3DGS-MCMC - RealityScan & \num{6.53697704093278} & \num{690.196984018165} & {}[\num{4.13787989752716}, \num{8.93607418433839}] & \num{0.854673417349419} & \pval{6.77745041578733e-14} & *** & \num{6.92655765819082} & \num{1018.98025703375} & {}[\num{4.36366115376091}, \num{9.48945416262073}] & \num{0.913026602434519} & \pval{1.09664005583063e-13} & ***\\

 & 3DGS-MCMC - Polycam & \num{7.97343764557426} & \num{ 2902.8190951755} & {}[\num{ 5.3492556544213}, \num{10.5976196367272}] & \num{0.934859430886462} & \pval{1.47588766365145e-16} & *** & \num{7.88745592834932} & \num{ 2663.6587546374} & {}[\num{5.18000173888418}, \num{10.5949101178145}] & \num{0.964524980069117} & \pval{2.89661590845324e-15} & ***\\

 & Scaniverse - RealityScan & \num{0.750371876506907} & \num{2.11778742558435} & {}[\num{-0.685638588463562}, \num{2.18638234147738}] & \num{0.51157577125185} & \pval{0.142434440332253} &  & \num{2.01054469893455} & \num{7.46738371569923} & {}[\num{0.393398422323696}, \num{3.62769097554541}] & \num{0.576105031171392} & \pval{0.000604018324569665} & ***\\

 & Scaniverse - Polycam & \num{2.18683248114839} & \num{8.90695543570645} & {}[\num{0.630205224653448}, \num{3.74345973764334}] & \num{0.554545254870012} & \pval{0.00011473014195443} & *** & \num{2.97144296909305} & \num{19.5200661359823} & {}[\num{1.27940188519612}, \num{4.66348405298998}] & \num{0.602786152050911} & \pval{1.17763275671818e-06} & ***\\

\multirow{-10}{*}{\raggedright\arraybackslash Geometry Accuracy} & RealityScan - Polycam & \num{1.43646060464149} & \num{4.20578351165194} & {}[\num{-0.0457653961632078}, \num{2.91868660544618}] & \num{0.528039960730231} & \pval{0.0072455777890716} & ** & \num{0.960898270158498} & \num{2.61404353641876} & {}[\num{-0.585078558683143}, \num{2.50687509900014}] & \num{0.550751061948853} & \pval{0.0900402096494052} & \\
\cmidrule{1-14}
 & 3DGS - 3DGS-MCMC & \num{-2.32701641350259} & \num{0.0975864708508858} & {}[\num{-4.06894013305026}, \num{-0.585092693954919}] & \num{0.620556738287988} & \pval{0.000252736678178853} & *** & \num{-2.12632728093805} & \num{0.119274552319054} & {}[\num{-3.97978570260726}, \num{-0.272868859268848}] & \num{0.660290746257313} & \pval{0.00182944647848793} & **\\

 & 3DGS - Scaniverse & \num{4.70470542394618} & \num{110.465739592515} & {}[\num{2.65275543752915}, \num{6.75665541036321}] & \num{0.731002957484091} & \pval{2.04492923288051e-10} & *** & \num{5.11166268693378} & \num{165.946041935218} & {}[\num{2.95896807652524}, \num{7.26435729734231}] & \num{0.766893022337477} & \pval{4.39854702192198e-11} & ***\\

 & 3DGS - RealityScan & \num{6.28188454942015} & \num{534.795563420477} & {}[\num{4.00357357059981}, \num{8.56019552824048}] & \num{0.811643594927102} & \pval{3.32150922826611e-14} & *** & \num{6.67910673474698} & \num{ 795.60810542884} & {}[\num{4.42579365349852}, \num{8.93241981599544}] & \num{0.802738145390375} & \pval{8.76630149169906e-16} & ***\\

 & 3DGS - Polycam & \num{6.13712752179012} & \num{462.722499740396} & {}[\num{3.87241636075398}, \num{8.40183868282627}] & \num{0.806798687844914} & \pval{7.02607481356312e-14} & *** & \num{5.62656844735694} & \num{ 277.70751273199} & {}[\num{3.44105891668632}, \num{7.81207797802755}] & \num{0.778583270111533} & \pval{1.64987262784209e-12} & ***\\

 & 3DGS-MCMC - Scaniverse & \num{7.03172183744877} & \num{1131.97801528563} & {}[\num{4.27867008322255}, \num{9.78477359167498}] & \num{0.980769018625154} & \pval{1.50434104302673e-12} & *** & \num{7.23798996787183} & \num{1391.29461153893} & {}[\num{ 4.2156032656891}, \num{10.2603766700546}] & \num{1.07671903924618} & \pval{3.57844023109083e-11} & ***\\

 & 3DGS-MCMC - RealityScan & \num{8.60890096292274} & \num{5480.22239924688} & {}[\num{ 5.6372400804428}, \num{11.5805618454027}] & \num{1.05864807042676} & \pval{4.22347590685184e-15} & *** & \num{8.80543401568503} & \num{6670.39271965258} & {}[\num{5.70382661432827}, \num{11.9070414170418}] & \num{1.10494125020333} & \pval{7.98897208622173e-15} & ***\\

 & 3DGS-MCMC - Polycam & \num{8.46414393529271} & \num{4741.66650054848} & {}[\num{5.50514838124533}, \num{11.4231394893401}] & \num{1.05413607325189} & \pval{4.89511226128806e-15} & *** & \num{7.75289572829499} & \num{2328.30480041656} & {}[\num{4.70315345997895}, \num{ 10.802637996611}] & \num{1.08646440335324} & \pval{2.40398771111156e-12} & ***\\

 & Scaniverse - RealityScan & \num{1.57717912547397} & \num{4.84127988816467} & {}[\num{0.086771254060082}, \num{3.06758699688786}] & \num{0.530954735287439} & \pval{0.00371694202254838} & ** & \num{ 1.5674440478132} & \num{ 4.7943783181007} & {}[\num{0.151017662036083}, \num{2.98387043359031}] & \num{0.50459898336486} & \pval{0.00236792459069342} & **\\

 & Scaniverse - Polycam & \num{1.43242209784394} & \num{4.18883267741912} & {}[\num{-0.054096918037801}, \num{2.91894111372569}] & \num{0.529569338511669} & \pval{0.00759209340271448} & ** & \num{0.514905760423157} & \num{1.67348078624497} & {}[\num{-0.84376460504245}, \num{1.87357612588876}] & \num{0.484023520054497} & \pval{0.287417869905396} & \\

\multirow{-10}{*}{\raggedright\arraybackslash Smoothness} & RealityScan - Polycam & \num{-0.144757027630027} & \num{0.86523249516299} & {}[\num{-1.59819646359892}, \num{1.30868240833887}] & \num{0.517784806281989} & \pval{0.779807530864655} &  & \num{-1.05253828739004} & \num{0.349050632889547} & {}[\num{-2.44037211203307}, \num{0.335295537252993}] & \num{0.49441294233588} & \pval{0.0369616127890651} & *\\
\cmidrule{1-14}
 & 3DGS - 3DGS-MCMC & \num{-1.83361021084658} & \num{0.159835485100898} & {}[\num{-3.40047429272747}, \num{-0.266746128965696}] & \num{0.558192102835072} & \pval{0.00127519184919647} & ** & \num{-0.693501001909393} & \num{0.499823120618972} & {}[\num{-2.27515164398698}, \num{0.888149640168195}] & \num{0.563459784458093} & \pval{0.218401194982166} & \\

 & 3DGS - Scaniverse & \num{4.19873176852396} & \num{66.6018109437301} & {}[\num{2.35302334691622}, \num{6.04444019013171}] & \num{0.657529824693468} & \pval{4.26777295868758e-10} & *** & \num{ 7.0591483823282} & \num{1163.45392583043} & {}[\num{4.59777495397589}, \num{9.52052181068052}] & \num{0.876859215628377} & \pval{4.12308245485507e-15} & ***\\

 & 3DGS - RealityScan & \num{0.324549256361388} & \num{1.38340694354694} & {}[\num{-1.13300615743812}, \num{1.78210467016089}] & \num{0.51925111490892} & \pval{0.531949179445007} &  & \num{3.49389664637312} & \num{32.9139521825731} & {}[\num{1.53077602287496}, \num{5.45701726987128}] & \num{0.699357679853076} & \pval{8.36751235563545e-07} & ***\\

 & 3DGS - Polycam & \num{ 4.5590174273888} & \num{95.4896082460779} & {}[\num{2.66943393325367}, \num{6.44860092152392}] & \num{0.673160228938042} & \pval{4.21747002244366e-11} & *** & \num{5.07377133010117} & \num{  159.7757596624} & {}[\num{2.92228098043327}, \num{7.22526167976907}] & \num{0.766464006928323} & \pval{8.99745639293008e-11} & ***\\

 & 3DGS-MCMC - Scaniverse & \num{6.03234197937055} & \num{416.689766366254} & {}[\num{3.83081098297912}, \num{8.23387297576198}] & \num{0.78429088428472} & \pval{7.2732695918012e-14} & *** & \num{ 7.7526493842376} & \num{2327.73130700642} & {}[\num{5.10876061701207}, \num{10.3965381514631}] & \num{0.941879929284022} & \pval{1.8559824883332e-15} & ***\\

 & 3DGS-MCMC - RealityScan & \num{2.15815946720797} & \num{8.65519282325604} & {}[\num{0.551403956253564}, \num{3.76491497816237}] & \num{0.572403342301941} & \pval{0.000232886348593637} & *** & \num{4.18739764828251} & \num{65.8511998040688} & {}[\num{2.06428863430315}, \num{6.31050666226187}] & \num{0.756353214529382} & \pval{6.1785951924954e-08} & ***\\

 & 3DGS-MCMC - Polycam & \num{6.39262763823538} & \num{597.424334063232} & {}[\num{4.13219188936125}, \num{ 8.6530633871095}] & \num{0.80527558106571} & \pval{2.04748618297779e-14} & *** & \num{5.76727233201056} & \num{319.664603479201} & {}[\num{3.43809900195359}, \num{8.09644566206753}] & \num{0.829763202824317} & \pval{1.21327959268175e-11} & ***\\

 & Scaniverse - RealityScan & \num{-3.87418251216258} & \num{0.0207713112292959} & {}[\num{-5.70075573454331}, \num{-2.04760928978184}] & \num{0.650712949370197} & \pval{4.36777195139471e-09} & *** & \num{-3.56525173595509} & \num{0.0282898630120487} & {}[\num{-5.38385569472701}, \num{-1.74664777718317}] & \num{0.647873915619094} & \pval{6.22295433006405e-08} & ***\\

 & Scaniverse - Polycam & \num{0.360285658864832} & \num{1.43373891629995} & {}[\num{-1.10434036524547}, \num{1.82491168297513}] & \num{0.521770005273023} & \pval{0.531949179445007} &  & \num{-1.98537705222704} & \num{0.137328824214812} & {}[\num{-3.55218308689578}, \num{-0.418571017558297}] & \num{0.558171423635271} & \pval{0.000469016422256977} & ***\\

\multirow{-10}{*}{\raggedright\arraybackslash Text/Pattern Recognisability} & RealityScan - Polycam & \num{4.23446817102741} & \num{69.0249594969144} & {}[\num{2.36200347981175}, \num{6.10693286224307}] & \num{0.667061690647366} & \pval{4.3643182706442e-10} & *** & \num{1.57987468372805} & \num{4.85434744439462} & {}[\num{-0.00786708726500529}, \num{ 3.1676164547211}] & \num{0.565629736591965} & \pval{0.00580016442052095} & **\\
\bottomrule
\end{tabular}}
\caption{Post-hoc pairwise comparisons for visual-consistency questionnaire items in \textit{Office} and \textit{Reception} (latent scale; $p$ values are FDR--BH adjusted). Please note that \textbf{Odds ratios (ORs) are directional}: their value depends on which one is used as the reference. Consequently, the ORs reported in the main text may appear as reciprocal values of those in the appendix, depending on the contrast considered.}
\label{tab:pairwise_vc_side_by_side}
\end{table*}

\subsection{Consistency between subjective and objective measures}

\subsubsection{Registration measures}
\begin{figure*}[!t]
    \centering
    \includegraphics[width=1\linewidth]{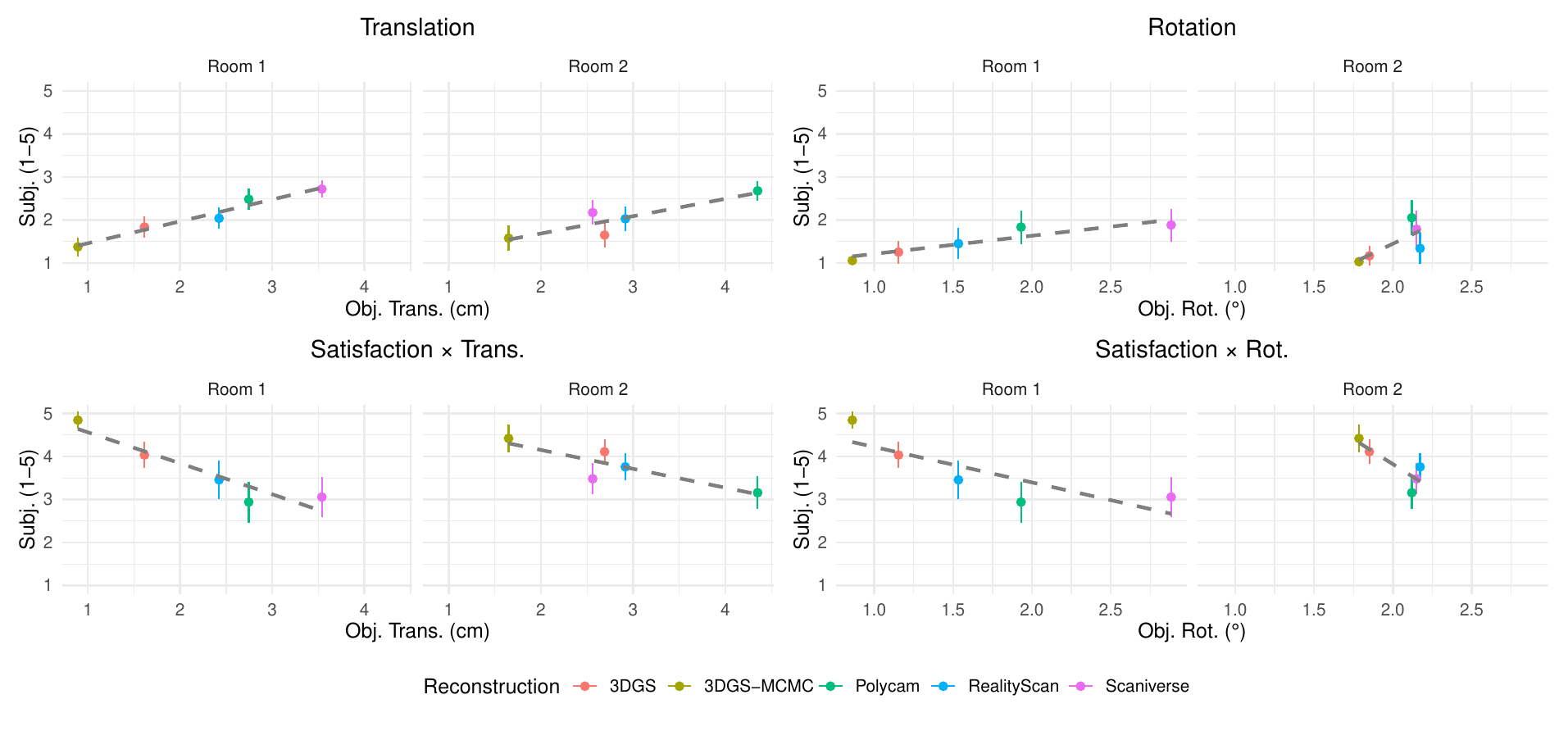}
    \caption{Correlations between subjective expected scores (1–5) and objective registration errors. Each panel shows translation error (cm), rotation error (°), satisfaction versus translation error, and satisfaction versus rotation error. Within each panel, results are faceted by \textit{Office} and \textit{Reception}. Error bars represent 95\% confidence intervals for expected scores; dashed lines indicate linear fits.}
    \label{fig:alignment_cor_figure}
\end{figure*}

\Cref{tab:alignment-correlation-spearman} shows the correlations between subjective expected scores (1–5) and objective registration error measures. \Cref{fig:alignment_cor_figure} shows correlations in a dot plot between subjective expected scores (1–5) and objective registration error measures. Each panel shows translation error (cm), rotation error (°), satisfaction versus translation error, and satisfaction versus rotation error.

For registration errors, subjective ratings showed  monotonic relationships with objective measures. In \textit{Office}, correlations were
perfect for both translation and rotation ($\rho = 1.00$, $p=.0167$),
and satisfaction was negatively associated with both translation and rotation errors ($\rho=-0.90$ for both), although neither exact test was significant (both $p=.083$). In \textit{Reception}, correlations were moderate for translation ($\rho=0.70$, $p=.233$) and rotation ($\rho=0.60$, $p=.350$), and satisfaction again tended to decrease with increasing errors, though not significantly.


\subsubsection{Visual Consistency Measures}
\begin{figure*}[t]
    \centering
    \includegraphics[width=1\linewidth]{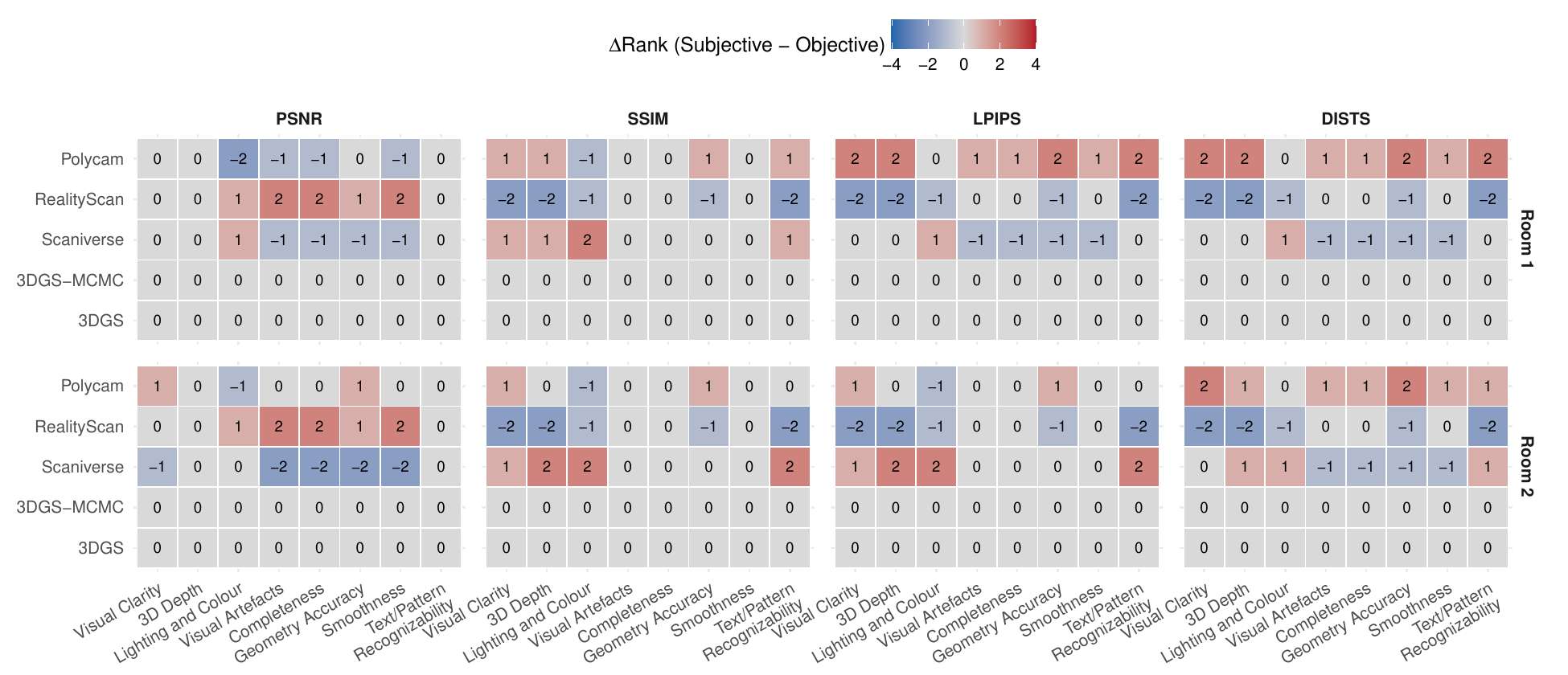}
    \caption{Rank-delta heatmaps comparing visual consistency between subjective and objective rankings of reconstruction methods in \textit{Office} and \textit{Reception}. $\Delta\mathrm{Rank}=\mathrm{Rank}_{\mathrm{subjective}}-\mathrm{Rank}_{\mathrm{objective}}$, where lower ranks indicate better performance. Grey cells denote perfect agreement ($\Delta\mathrm{Rank}=0$); red cells indicate that a method was ranked worse subjectively than objectively ($\Delta\mathrm{Rank}>0$); and blue cells indicate that a method was ranked better subjectively than objectively ($\Delta\mathrm{Rank}<0$).}
    \label{fig:heatmap_vc_rank}
\end{figure*}

Because visual-consistency judgements are bidirectional and relative rather than continuous, we compare subjective and objective rankings using $\Delta\mathrm{Rank}$ (subjective minus objective) in \cref{fig:heatmap_vc_rank}.
Overall, the $\Delta\mathrm{Rank}$ heatmap shows that 3DGS and 3DGS-MCMC achieve near-perfect consistency between subjective and objective rankings, while Polycam, RealityScan, and Scaniverse display systematic deviations.
Across metrics, PSNR and SSIM
achieved the highest agreement with subjective measures (53 and 54 of 80 exact matches),
followed by LPIPS (51), while DISTS was lowest (44). 
At the dimension level, PSNR aligned perfectly with 3D depth and
text/pattern recognisability, and SSIM aligned perfectly with visual artefacts and completeness, whereas LPIPS and DISTS showed no dimension with
full consistency, reflecting their less stable correspondence with
perception in our experiment.

\subsubsection{Screenshots}
\label{sec:screenshots}
\begin{figure*}[t]
    \centering
    \includegraphics[width=1\linewidth]{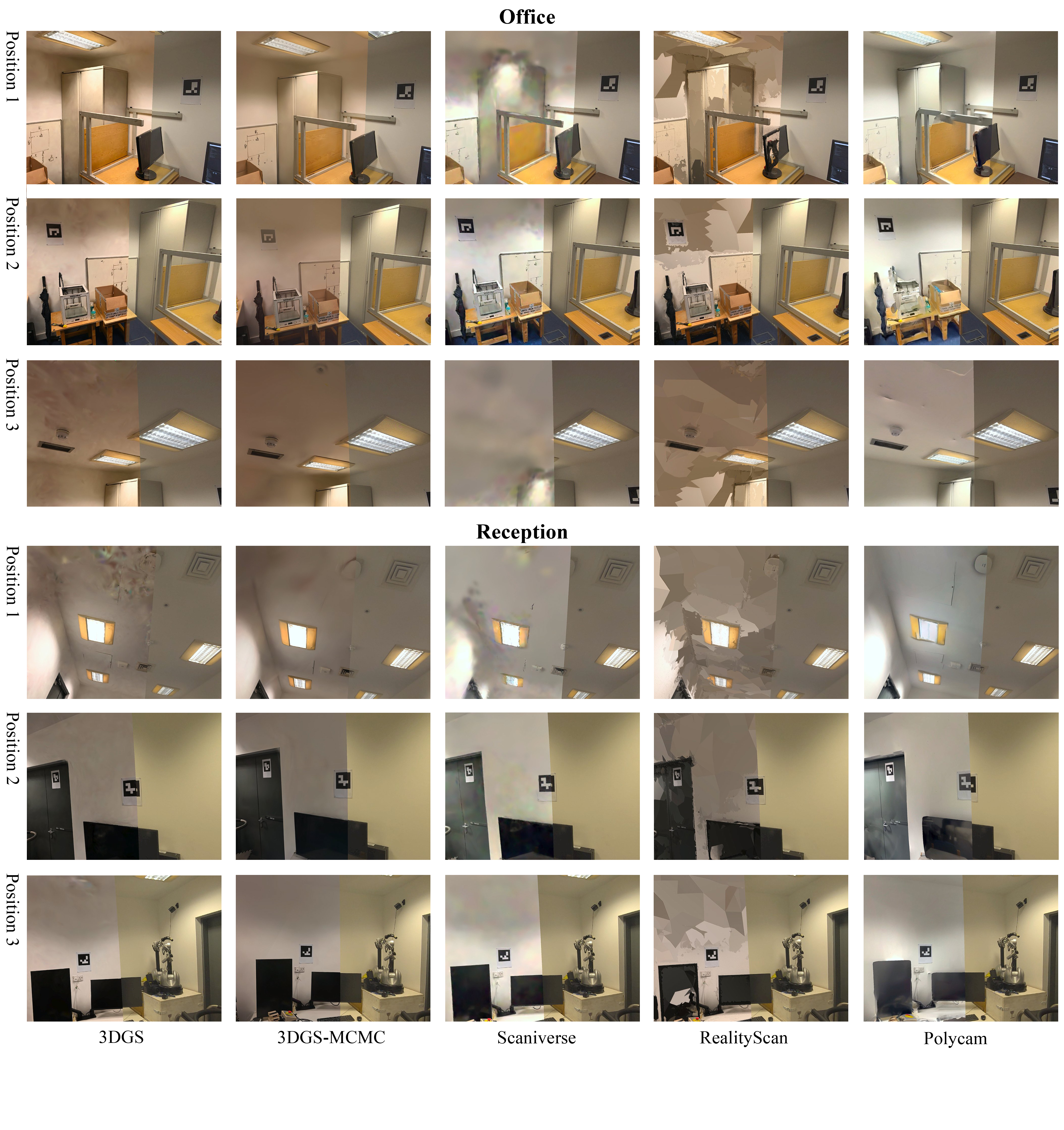}
    \caption{Reconstruction renderings (left half of each panel) and spatially registered live VST views (right half) in \textit{Office} (top three rows) and \textit{Reception} (bottom three rows).}
    \label{fig:appendix_screenshot}
\end{figure*}

Additional MR-Compare screenshots are shown in \cref{fig:appendix_screenshot}.

\section{Controlled Registration Evaluation}
\renewcommand{\thefigure}{C.\arabic{figure}}
\setcounter{figure}{0}
\renewcommand{\theHfigure}{C.\arabic{figure}}
\renewcommand{\thetable}{C.\arabic{table}}
\setcounter{table}{0}
\renewcommand{\theHtable}{C.\arabic{table}}

\begin{table*}[t]
\caption[Scene-wise registration success on Replica.]{Scene-wise registration success on eight Replica scenes. GS and MCMC denote 3DGS and 3DGS-MCMC, respectively. ICP, G-ICP, and V-GICP use fine registration only; TEASER++, TurboReg, RANSAC, and FGR perform coarse registration followed by the same V-GICP refinement. A success requires a final error of $\leq 5$~cm and $\leq 5^\circ$.}
\label{tab:success_filter_on_off}
\centering
\scriptsize
\setlength{\tabcolsep}{2pt}
\renewcommand{\arraystretch}{1.05}
\begin{tabular}{@{}l*{14}{c}@{}}
\toprule
& \multicolumn{6}{c}{Fine registration only}
& \multicolumn{8}{c}{Coarse registration + V-GICP} \\
\cmidrule(lr){2-7}\cmidrule(lr){8-15}
& \multicolumn{2}{c}{ICP}
& \multicolumn{2}{c}{G-ICP}
& \multicolumn{2}{c}{V-GICP}
& \multicolumn{2}{c}{TEASER++}
& \multicolumn{2}{c}{TurboReg}
& \multicolumn{2}{c}{RANSAC}
& \multicolumn{2}{c}{FGR} \\
\cmidrule(lr){2-3}\cmidrule(lr){4-5}\cmidrule(lr){6-7}\cmidrule(lr){8-9}\cmidrule(lr){10-11}\cmidrule(lr){12-13}\cmidrule(lr){14-15}
Scene & GS & MCMC & GS & MCMC & GS & MCMC & GS & MCMC & GS & MCMC & GS & MCMC & GS & MCMC \\
\midrule
\multicolumn{15}{l}{\textbf{With radius crop and density filtering}} \\
office0 & $\times$ & $\times$ & $\times$ & $\times$ & $\times$ & $\times$ & $\checkmark$ & $\checkmark$ & $\checkmark$ & $\checkmark$ & $\checkmark$ & $\times$ & $\checkmark$ & $\checkmark$ \\
office1 & $\times$ & $\times$ & $\times$ & $\times$ & $\times$ & $\times$ & $\checkmark$ & $\checkmark$ & $\checkmark$ & $\checkmark$ & $\times$ & $\checkmark$ & $\times$ & $\checkmark$ \\
office2 & $\times$ & $\times$ & $\times$ & $\times$ & $\times$ & $\times$ & $\checkmark$ & $\checkmark$ & $\checkmark$ & $\checkmark$ & $\checkmark$ & $\checkmark$ & $\checkmark$ & $\checkmark$ \\
office3 & $\times$ & $\times$ & $\times$ & $\times$ & $\times$ & $\times$ & $\checkmark$ & $\checkmark$ & $\checkmark$ & $\checkmark$ & $\checkmark$ & $\checkmark$ & $\checkmark$ & $\checkmark$ \\
office4 & $\times$ & $\times$ & $\times$ & $\times$ & $\times$ & $\times$ & $\checkmark$ & $\checkmark$ & $\checkmark$ & $\checkmark$ & $\checkmark$ & $\checkmark$ & $\checkmark$ & $\checkmark$ \\
room0   & $\times$ & $\times$ & $\times$ & $\times$ & $\times$ & $\times$ & $\checkmark$ & $\checkmark$ & $\checkmark$ & $\times$     & $\checkmark$ & $\checkmark$ & $\checkmark$ & $\checkmark$ \\
room1   & $\times$ & $\times$ & $\times$ & $\times$ & $\times$ & $\times$ & $\checkmark$ & $\checkmark$ & $\checkmark$ & $\checkmark$ & $\checkmark$ & $\checkmark$ & $\times$     & $\checkmark$ \\
room2   & $\times$ & $\times$ & $\times$ & $\times$ & $\times$ & $\times$ & $\times$     & $\checkmark$ & $\checkmark$ & $\checkmark$ & $\times$     & $\times$     & $\times$     & $\times$ \\
\textbf{Total} & \textbf{0/8} & \textbf{0/8} & \textbf{0/8} & \textbf{0/8} & \textbf{0/8} & \textbf{0/8} & \textbf{7/8} & \textbf{8/8} & \textbf{8/8} & \textbf{7/8} & \textbf{6/8} & \textbf{6/8} & \textbf{5/8} & \textbf{7/8} \\
\midrule
\multicolumn{15}{l}{\textbf{Without radius crop and density filtering}} \\
office0 & $\times$ & $\times$ & $\times$ & $\times$ & $\times$ & $\times$ & $\checkmark$ & $\checkmark$ & $\times$ & $\times$ & $\times$ & $\times$ & $\times$ & $\times$ \\
office1 & $\times$ & $\times$ & $\times$ & $\times$ & $\times$ & $\times$ & $\times$     & $\times$     & $\times$ & $\times$ & $\times$ & $\times$ & $\times$ & $\times$ \\
office2 & $\times$ & $\times$ & $\times$ & $\times$ & $\times$ & $\times$ & $\times$     & $\checkmark$ & $\times$ & $\times$ & $\times$ & $\times$ & $\times$ & $\times$ \\
office3 & $\times$ & $\times$ & $\times$ & $\times$ & $\times$ & $\times$ & $\checkmark$ & $\checkmark$ & $\times$ & $\checkmark$ & $\times$ & $\checkmark$ & $\checkmark$ & $\checkmark$ \\
office4 & $\times$ & $\times$ & $\times$ & $\times$ & $\times$ & $\times$ & $\checkmark$ & $\times$     & $\times$ & $\times$ & $\times$ & $\times$ & $\times$ & $\times$ \\
room0   & $\times$ & $\times$ & $\times$ & $\times$ & $\times$ & $\times$ & $\checkmark$ & $\times$     & $\times$ & $\times$ & $\times$ & $\times$ & $\times$ & $\times$ \\
room1   & $\times$ & $\times$ & $\times$ & $\times$ & $\times$ & $\times$ & $\times$     & $\times$     & $\times$ & $\checkmark$ & $\times$ & $\times$ & $\times$ & $\times$ \\
room2   & $\times$ & $\times$ & $\times$ & $\times$ & $\times$ & $\times$ & $\times$     & $\checkmark$ & $\times$ & $\times$ & $\times$ & $\times$ & $\times$ & $\times$ \\
\textbf{Total} & \textbf{0/8} & \textbf{0/8} & \textbf{0/8} & \textbf{0/8} & \textbf{0/8} & \textbf{0/8} & \textbf{4/8} & \textbf{4/8} & \textbf{0/8} & \textbf{2/8} & \textbf{0/8} & \textbf{1/8} & \textbf{1/8} & \textbf{1/8} \\
\bottomrule
\end{tabular}
\end{table*}

\subsection{Simulated Quest~3-Like Depth Scanner}
\label{app:scanner_impl}

We implement a simulated Quest~3-like depth scanner in Unity using a raycast-based acquisition pipeline for the controlled evaluation. The scanner follows a predefined waypoint trajectory and periodically captures low-resolution depth observations from the current camera pose. The aim is not to reproduce the proprietary sensing pipeline exactly, but to approximate relevant scan artefacts: low spatial resolution, depth dropout near discontinuities, spatially varying measurement quality, range- and angle-dependent noise, coarse quantisation, boundary smearing, and short-term pose drift. The pseudocode is presented in \cref{alg:scanner}, and parameters are summarised in Table~\ref{tab:scanner_params}.

At each scan step, the camera viewport is discretised into an $R \times R$ grid. For a pixel $(x,y)$ at scan step $t$, let $\mathbf{o}_t$ be the camera centre and $\mathbf{r}_{x,y,t}$ the corresponding unit viewing ray. If the ray intersects the scene mesh at depth $d_{x,y,t}$, the clean 3D point is
\begin{equation}
\mathbf{p}^{\mathrm{clean}}_{x,y,t} = \mathbf{o}_t + d_{x,y,t}\mathbf{r}_{x,y,t}.
\end{equation}
A sample is rejected if no valid intersection is found, if the depth exceeds the maximum sensing range, or if the incidence angle
\begin{equation}
\theta_{x,y,t}=\arccos\!\left(\mathbf{n}_{x,y,t}^{\top}(-\mathbf{r}_{x,y,t})\right)
\end{equation}
is larger than a prescribed threshold, where $\mathbf{n}_{x,y,t}$ is the surface normal at the hit point. We further apply a small independent dropout probability to emulate occasional missing measurements.
To model unstable sensing near depth discontinuities, we use a row-wise edge-consistency test. Let $\tilde d_{x-1,y,t}$ denote the previous valid depth on the same scanline. The current sample is discarded whenever
\begin{equation}
|d_{x,y,t}-\tilde d_{x-1,y,t}|>\tau_{\mathrm{edge}}.
\end{equation}
This simple rule suppresses abrupt depth transitions and yields fragmented observations around occlusion boundaries.
Measurement quality is made spatially non-uniform through a Perlin-noise mask defined over image coordinates:
\begin{equation}
q_{x,y}=\mathrm{Perlin}\!\left(\frac{x}{R}s+\xi_x,\;\frac{y}{R}s+\xi_y\right),
\end{equation}
where $s$ is a spatial scale and $\xi_x,\xi_y$ are random seeds sampled once at initialisation. Pixels satisfying
\begin{equation}
q_{x,y}>\tau_{\mathrm{bad}}
\end{equation}
are treated as low-quality regions. This produces coherent patches of degraded depth rather than independent pixel-wise corruption.
For each valid sample, we perturb the depth using a noise model that grows with both range and incidence angle:
\begin{equation}
\sigma(d,\theta)
=
\beta d^2
+
\sigma_0
\left(1+\alpha_d d^2\right)
\left(1+\alpha_\theta\left(\frac{\theta}{\theta_{\max}}\right)^2\right).
\end{equation}
In low-quality regions, the noise magnitude is further amplified:
\begin{equation}
\sigma'(d,\theta,q)=
\begin{cases}
\gamma\,\sigma(d,\theta), & q>\tau_{\mathrm{bad}},\\
\sigma(d,\theta), & \text{otherwise}.
\end{cases}
\end{equation}
The noisy depth is then sampled using this effective noise scale:
\begin{equation}
\hat d_{x,y,t}\sim\mathcal{N}\!\left(d_{x,y,t},\sigma'(d_{x,y,t},\theta_{x,y,t},q_{x,y})^2\right).
\end{equation}

To reproduce boundary smearing, we cast a small set of neighbouring rays around each pixel and evaluate the largest local depth discrepancy,
\begin{equation}
\delta_{x,y,t}=\max_{k\in\mathcal{N}}\left|d^{(k)}_{x,y,t}-d_{x,y,t}\right|,
\end{equation}
where $\mathcal{N}$ denotes four offset rays. If $\delta_{x,y,t}$ exceeds a threshold and the sample lies in a low-quality region, the depth is interpolated towards the neighbouring depth with the largest discrepancy:
\begin{equation}
d^{\mathrm{drag}}_{x,y,t}
=
(1-\lambda_{x,y,t})\,\hat d_{x,y,t}
+
\lambda_{x,y,t}\,d^\star_{x,y,t},
\end{equation}
where $d^\star_{x,y,t}$ is the most inconsistent neighbouring depth and $\lambda_{x,y,t}$ is a random interpolation weight controlled by the dragging strength; otherwise, $d^{\mathrm{drag}}_{x,y,t}=\hat d_{x,y,t}$. This yields sparse flying points and mild stretching near geometric boundaries.

We additionally quantise the corrupted depth to mimic the coarse, staircase-like appearance of degraded consumer depth maps. The quantisation step is chosen according to the local quality mask:
\begin{equation}
\Delta(q)=
\begin{cases}
\Delta_{\mathrm{bad}}, & q>\tau_{\mathrm{bad}},\\
\Delta_{\mathrm{base}}, & \text{otherwise},
\end{cases}
\end{equation}
and the final scalar depth value is
\begin{equation}
\bar d_{x,y,t}
=
\Delta(q_{x,y})
\cdot
\mathrm{round}\!\left(
\frac{d^{\mathrm{drag}}_{x,y,t}}{\Delta(q_{x,y})}
\right).
\end{equation}

We retain a burst-drift model to simulate short periods of tracking instability across several consecutive scans. At each scan step, a burst is triggered with probability $p_{\mathrm{burst}}$ and lasts for
\begin{equation}
L\sim\mathcal{U}\{L_{\min},\ldots,L_{\max}\}
\end{equation}
scan steps. Once triggered, a translation perturbation $\Delta\mathbf{t}_t$ and a rotation perturbation $\Delta\mathbf{R}_t$ are sampled and then decay geometrically over subsequent scans:
\begin{equation}
\Delta\mathbf{t}_{t+1}=\rho\,\Delta\mathbf{t}_t,
\qquad
\Delta\mathbf{R}_{t+1}=\mathrm{Decay}\!\left(\Delta\mathbf{R}_t,\rho\right).
\end{equation}
This produces short-lived frame-level pose offsets rather than persistent trajectory drift.

Finally, the depth sample is back-projected into 3D and re-expressed under the burst-perturbed pose. Let $\mathbf{T}_t$ be the true camera pose at scan step $t$ and $\tilde{\mathbf{T}}_t$ the perturbed pose. The exported point is
\begin{equation}
\mathbf{p}_{x,y,t}
=
\tilde{\mathbf{T}}_t\mathbf{T}_t^{-1}
\left(
\mathbf{o}_t+\bar d_{x,y,t}\mathbf{r}_{x,y,t}
\right).
\end{equation}
All valid points are accumulated over the full trajectory to form the final point cloud. The point cloud is exported in PLY format, with an optional conversion from Unity's left-handed coordinates to a right-handed convention for downstream processing.

\begin{algorithm}[htbp]
\caption{Simulated Quest~3-Like Depth Scanning}
\label{alg:scanner}
\begin{algorithmic}[1]
\Require Waypoint trajectory $\{\mathbf{T}_t\}$, scene mesh $\mathcal{M}$, scan parameters in Table~\ref{tab:scanner_params}
\State Initialise global point cloud $\mathcal{P}\leftarrow\emptyset$
\State Sample Perlin seeds $\xi_x,\xi_y$
\State Initialise burst state as inactive
\For{each scan step $t$}
    \State Move scanner to pose $\mathbf{T}_t$
    \State Update burst state with probability $p_{\mathrm{burst}}$
    \State Compute burst-perturbed pose $\tilde{\mathbf{T}}_t$
    \For{$y=1$ to $R$}
        \State $\tilde d_{\mathrm{prev}}\leftarrow\varnothing$
        \For{$x=1$ to $R$}
            \State Sample dropout event with probability $p_{\mathrm{drop}}$
            \If{dropout occurs}
                \State \textbf{continue}
            \EndIf
            \State Cast centre ray $\mathbf{r}_{x,y,t}$ into mesh $\mathcal{M}$
            \If{no hit or depth $>d_{\max}$}
                \State $\tilde d_{\mathrm{prev}}\leftarrow\varnothing$
                \State \textbf{continue}
            \EndIf
            \State Compute hit depth $d_{x,y,t}$ and incidence angle $\theta_{x,y,t}$
            \If{$\theta_{x,y,t}>\theta_{\max}$}
                \State $\tilde d_{\mathrm{prev}}\leftarrow\varnothing$
                \State \textbf{continue}
            \EndIf
            \If{$\tilde d_{\mathrm{prev}} \neq \varnothing$ and $|d_{x,y,t}-\tilde d_{\mathrm{prev}}|>\tau_{\mathrm{edge}}$}
                \State $\tilde d_{\mathrm{prev}}\leftarrow\varnothing$
                \State \textbf{continue}
            \EndIf
            \State Compute $q_{x,y}$ and $\sigma'(d_{x,y,t},\theta_{x,y,t},q_{x,y})$ as defined above
            \State Sample $\hat d_{x,y,t}\sim\mathcal{N}\!\left(d_{x,y,t},[\sigma'(d_{x,y,t},\theta_{x,y,t},q_{x,y})]^2\right)$
            \State Cast four neighbouring rays and compute local depth discrepancy $\delta_{x,y,t}$
            \If{$q_{x,y}>\tau_{\mathrm{bad}}$ and $\delta_{x,y,t}$ exceeds threshold}
                \State Compute $d^{\mathrm{drag}}_{x,y,t}$ as defined above
            \Else
                \State $d^{\mathrm{drag}}_{x,y,t}\leftarrow\hat d_{x,y,t}$
            \EndIf
            \State Compute $\Delta(q_{x,y})$ and quantise $d^{\mathrm{drag}}_{x,y,t}$ to obtain $\bar d_{x,y,t}$
            \State Compute $\mathbf{p}_{x,y,t}=\tilde{\mathbf{T}}_t\mathbf{T}_t^{-1}\!\left(\mathbf{o}_t+\bar d_{x,y,t}\mathbf{r}_{x,y,t}\right)$
            \State Add $\mathbf{p}_{x,y,t}$ to $\mathcal{P}$
            \State $\tilde d_{\mathrm{prev}}\leftarrow d_{x,y,t}$
        \EndFor
    \EndFor
\EndFor
\State Export $\mathcal{P}$ in PLY format
\end{algorithmic}
\end{algorithm}

\begin{table}[t]
\centering
\caption{Parameters of the simulated depth scanner used in our experiments.}
\label{tab:scanner_params}
\begin{tabular}{lll}
\toprule
Parameter & Symbol & Value \\
\midrule
Depth resolution & $R$ & $160$ \\
Maximum range & $d_{\max}$ & $5\,\mathrm{m}$ \\
Scan interval & $\Delta t_{\mathrm{scan}}$ & $0.15\,\mathrm{s}$ \\
Random dropout probability & $p_{\mathrm{drop}}$ & $0.005$ \\
Maximum incidence angle & $\theta_{\max}$ & $90^\circ$ \\
Edge-consistency threshold & $\tau_{\mathrm{edge}}$ & $0.05\,\mathrm{m}$ \\
Perlin spatial scale & $s$ & $30$ \\
Bad-region threshold & $\tau_{\mathrm{bad}}$ & $0.65$ \\
Base quantisation step & $\Delta_{\mathrm{base}}$ & $0.002\,\mathrm{m}$ \\
Bad-region quantisation step & $\Delta_{\mathrm{bad}}$ & $0.01\,\mathrm{m}$ \\
Base axial noise coefficient & $\beta$ & $0.002$ \\
Gaussian base scale & $\sigma_0$ & $0.0015\,\mathrm{m}$ \\
Distance factor & $\alpha_d$ & $0.08$ \\
Angle factor & $\alpha_\theta$ & $0.8$ \\
Bad-region noise multiplier & $\gamma$ & $5$ \\
Dragging intensity & --- & $0.8$ \\
Neighbour-ray offset & --- & $0.002$ \\
Dragging activation threshold & --- & $0.02\,\mathrm{m}$ \\
Burst trigger probability & $p_{\mathrm{burst}}$ & $0.015$ \\
Burst duration & $L_{\min},L_{\max}$ & $2,8$ scans \\
Burst translation strength & --- & $0.01\,\mathrm{m}$ \\
Burst rotation strength & --- & $0.25^\circ$ \\
Burst decay factor & $\rho$ & $0.9$ \\
\bottomrule
\end{tabular}
\end{table}

\subsection{Evaluation Protocol for the Controlled Ablations}
\label{appendix:study2-evaluation}

Unlike the real-world evaluation, which used ArUco markers as a sparse proxy reference in physical rooms, the controlled Replica evaluation uses the provided ground-truth camera poses. This avoids reliance on marker-based comparison and yields a simpler, reproducible protocol.
For each scene, the input image sequence is first reconstructed using Structure-from-Motion (SfM), producing a set of estimated camera poses in the COLMAP coordinate frame. The resulting 3DGS or 3DGS-MCMC reconstruction is defined in this same frame, since the Gaussian representation is optimised from these SfM estimates. Because the reconstruction is obtained from monocular images, the COLMAP frame is defined only up to an unknown global scale. To resolve scale ambiguity, we align the SfM-estimated trajectory to the Replica ground-truth trajectory using a Sim(3) Umeyama alignment implemented in \texttt{evo}~\cite{grupp2017evo} to estimate scale. The recovered scale is then applied consistently to the reconstruction and its associated camera poses, so that the source reconstruction is expressed in a metric frame compatible with the target scene.

MR-Compare then estimates a rigid alignment transform that registers the source point set to the target scan point cloud generated from the Replica mesh. To evaluate alignment accuracy, we apply this estimated alignment transform to all SfM-estimated camera poses associated with the reconstruction images. The transformed poses are then compared with the corresponding ground-truth camera poses provided by Replica. Translation error is computed as the Euclidean distance between camera centres, and rotation error as the angular difference between camera orientations. These per-image pose errors are then aggregated to obtain the alignment error for each scene and method.

For the registration ablation, ICP, G-ICP, and V-GICP are evaluated independently as fine-registration-only baselines. TEASER++, TurboReg, RANSAC, and FGR perform coarse registration, after which the same V-GICP fine-registration configuration is applied. All four coarse-registration methods use the same preprocessing and FPFH feature-extraction configuration. For the newly added RANSAC and FGR baselines, this configuration uses a voxel size of $0.1$, a normal-estimation radius of $0.4$, and an FPFH radius of $0.8$. FGR uses the original standalone implementation~\cite{zhou2016fgr} with its default optimisation parameters. RANSAC uses a practical configuration with a maximum of $200$ iterations, a maximum correspondence distance of $0.25$, and a minimum sample distance of $0.05$.
For each Replica scene, the noisy target point cloud was generated once, stored locally, and reused across all registration methods. To verify the binary classification, each scene--reconstruction--method configuration was repeated until the same success/failure outcome was observed in three consecutive runs; this stable outcome is reported in~\cref{tab:success_filter_on_off}.
This protocol evaluates end-to-end alignment of the reconstructed source representation with the target scene rather than the registration stage alone. Reported errors therefore include residual inaccuracies inherited from SfM and trajectory estimation after scale alignment. This matches the goal of assessing the full MR-Compare pipeline under controlled conditions.
\subsection{\texorpdfstring{Visualisation of Filtering Effects}{Visualisation of Filtering Effects}}

\begin{figure}[htbp]
  \centering

  \begin{subfigure}{\linewidth}
    \centering
    \includegraphics[width=0.96\linewidth]{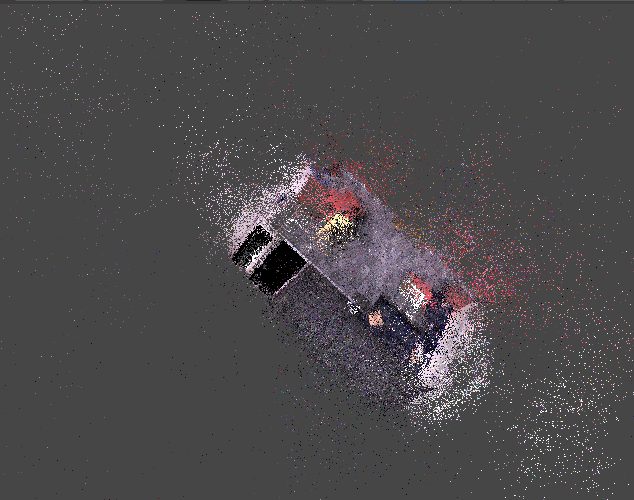}
    \caption{Raw Gaussian centres (no filtering).}
  \end{subfigure}

  \vspace{0.6em}

  \begin{subfigure}{\linewidth}
    \centering
    \includegraphics[width=0.96\linewidth]{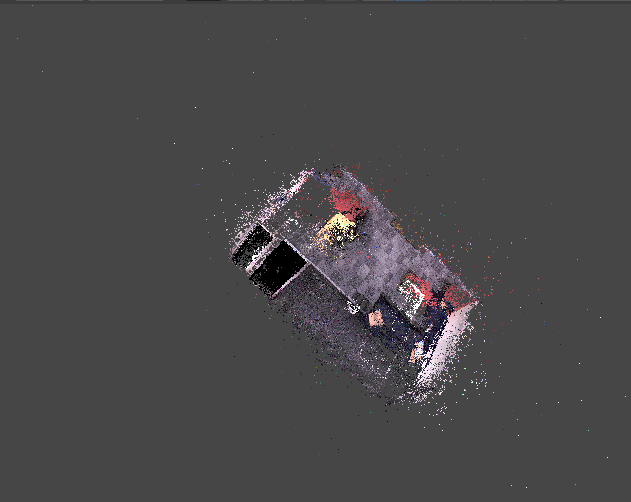}
    \caption{After radius cropping (\SI{10}{\metre}) and density filtering ($s = 0.2$, $N_{\min} = 3$).}
  \end{subfigure}

  \vspace{0.6em}

  \begin{subfigure}{\linewidth}
    \centering
    \includegraphics[width=0.96\linewidth]{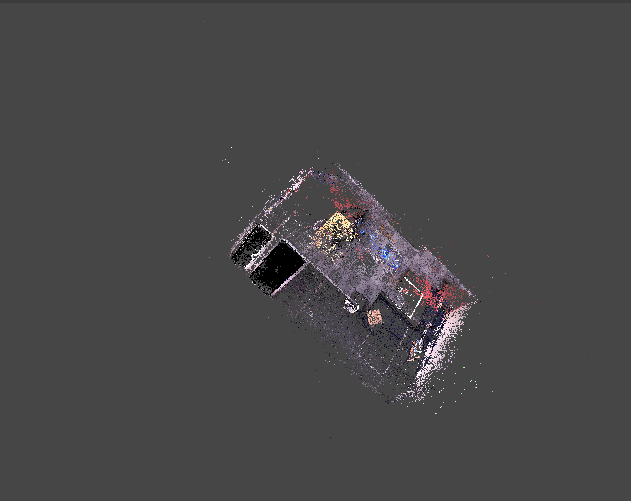}
    \caption{After additionally applying anisotropy-based pruning ($\tau$ = 0.05).}
  \end{subfigure}

  \caption{Unity debug visualisation of \textbf{3DGS} source point extraction on Replica \textit{office2}. Successive filtering removes far-field/background artefacts and isolated noise; anisotropy-based pruning further suppresses near-spherical splats to yield a cleaner surface-proxy subset for registration.}
  \label{fig:debug_filters_office2}
\end{figure}

\begin{figure}[htbp]
  \centering

  \begin{subfigure}{\linewidth}
    \centering
    \includegraphics[width=0.96\linewidth]{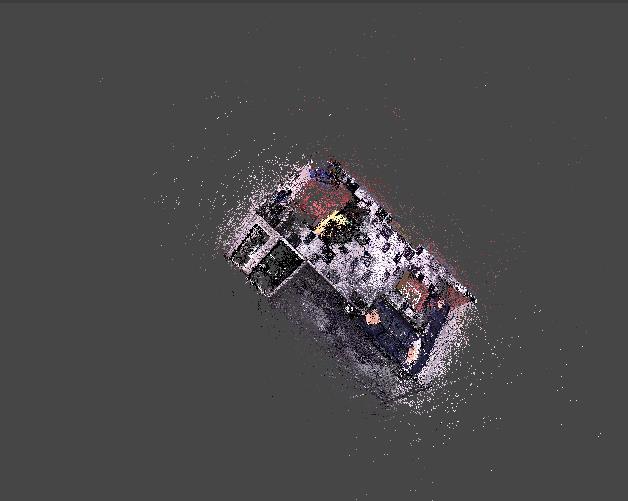}
    \caption{Raw Gaussian centres (no filtering).}
  \end{subfigure}

  \vspace{0.6em}

  \begin{subfigure}{\linewidth}
    \centering
    \includegraphics[width=0.96\linewidth]{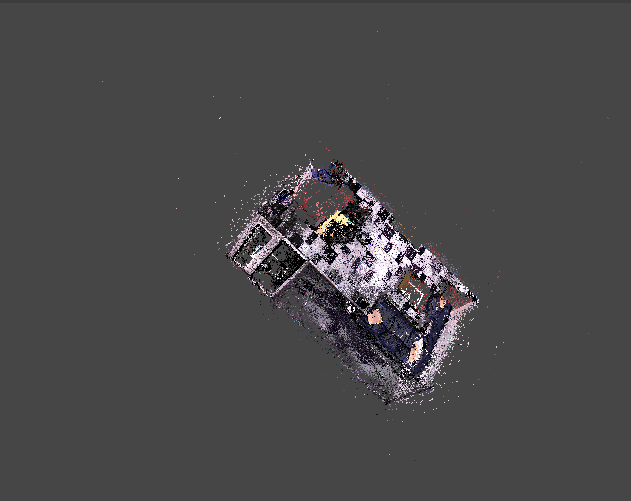}
    \caption{After radius cropping (\SI{10}{\metre}) and density filtering ($s = 0.2$, $N_{\min} = 3$).}
  \end{subfigure}

  \vspace{0.6em}

  \begin{subfigure}{\linewidth}
    \centering
    \includegraphics[width=0.96\linewidth]{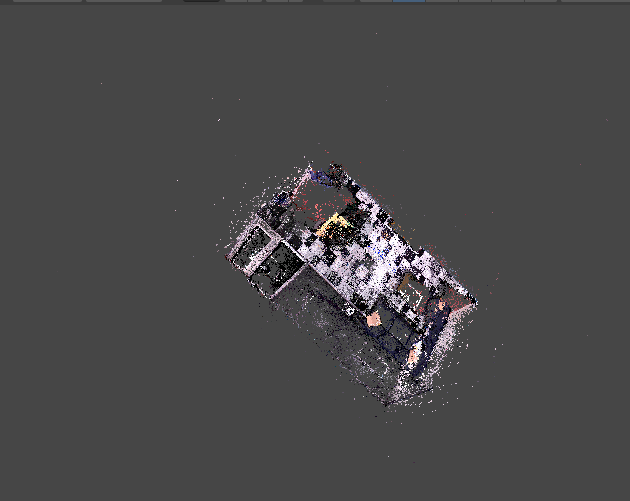}
    \caption{After additionally applying anisotropy-based pruning ($\tau$ = 0.05).}
  \end{subfigure}

  \caption{Unity debug visualisation of \textbf{3DGS-MCMC} source point extraction on Replica \textit{office2}. Successive filtering removes far-field/background artefacts and isolated noise; anisotropy-based pruning further suppresses near-spherical splats to yield a cleaner surface-proxy subset for registration.}
  \label{fig:debug_filters_office2_mcmc}
\end{figure}

To support practical tuning, MR-Compare provides a Unity Inspector debug mode that visualises the baked source points used for registration. \Cref{fig:debug_filters_office2,fig:debug_filters_office2_mcmc} show examples from Replica \textit{office2} for standard 3DGS and 3DGS-MCMC, respectively, under successive filtering: (a) raw centres, (b) after radius cropping and density filtering, and (c) after additional anisotropy-based pruning. Overall, the raw 3DGS centres exhibit substantially more far-field/background points and floating noise, whereas 3DGS-MCMC is visually cleaner before filtering. After radius cropping and density filtering, both reconstructions become relatively clean and point density is reduced, indicating that these pre-filters remove most out-of-scope content and isolated outliers. Adding anisotropy-based pruning has a stronger visible impact on standard 3DGS, further suppressing residual floating points and sharpening surface-proxy structure; for 3DGS-MCMC, the change is more subtle, mainly making some surfaces appear thinner and clearer, with the few remaining artefacts largely unchanged. However, a small sanity check with the original 3DGS~\cite{kerbl2023gaussian} achieved a similar retention level under the same preprocessing but did not exhibit such prominent far-field splats, suggesting this behaviour is likely pipeline-dependent (Nerfstudio) rather than inherent to 3DGS.

\subsection{Ablation}
\Cref{tab:success_filter_on_off} provides a scene-wise binary summary of registration success across the eight Replica scenes. For the original TEASER++ and TurboReg comparisons, radius crop and density filtering increased the overall success rate from 10/32 (31.25\%) to 30/32 (93.75\%). With radius crop and density filtering, RANSAC and FGR each achieved 12/16 successes, while ICP, G-ICP, and V-GICP each achieved 0/16; without these filters, the corresponding totals were 1/16, 2/16, 0/16, 0/16, and 0/16. This shows that radius crop and density filtering substantially improved registration feasibility across scenes.
Overall, enabling these preprocessing steps substantially improved registration stability across scenes. With filtering enabled, the coarse-to-fine variants succeeded in most scene--reconstruction pairs, whereas all fine-registration-only baselines failed; disabling filtering caused many more coarse-to-fine failures, especially for TurboReg. The pattern also suggests that 3DGS-MCMC was generally more robust than standard 3DGS under more challenging preprocessing conditions, but these experiments do not isolate the cause.

\section{Questionnaire}
\label{appendix_Questionnaire}
The full questionnaire is reproduced in \hyperref[app:mrcompare-questionnaire]{MR-Compare Visual Consistency}.
\clearpage

\includepdf[
  pages=-,
  addtotoc={1,section,1,{MR-Compare Visual Consistency},app:mrcompare-questionnaire}
]{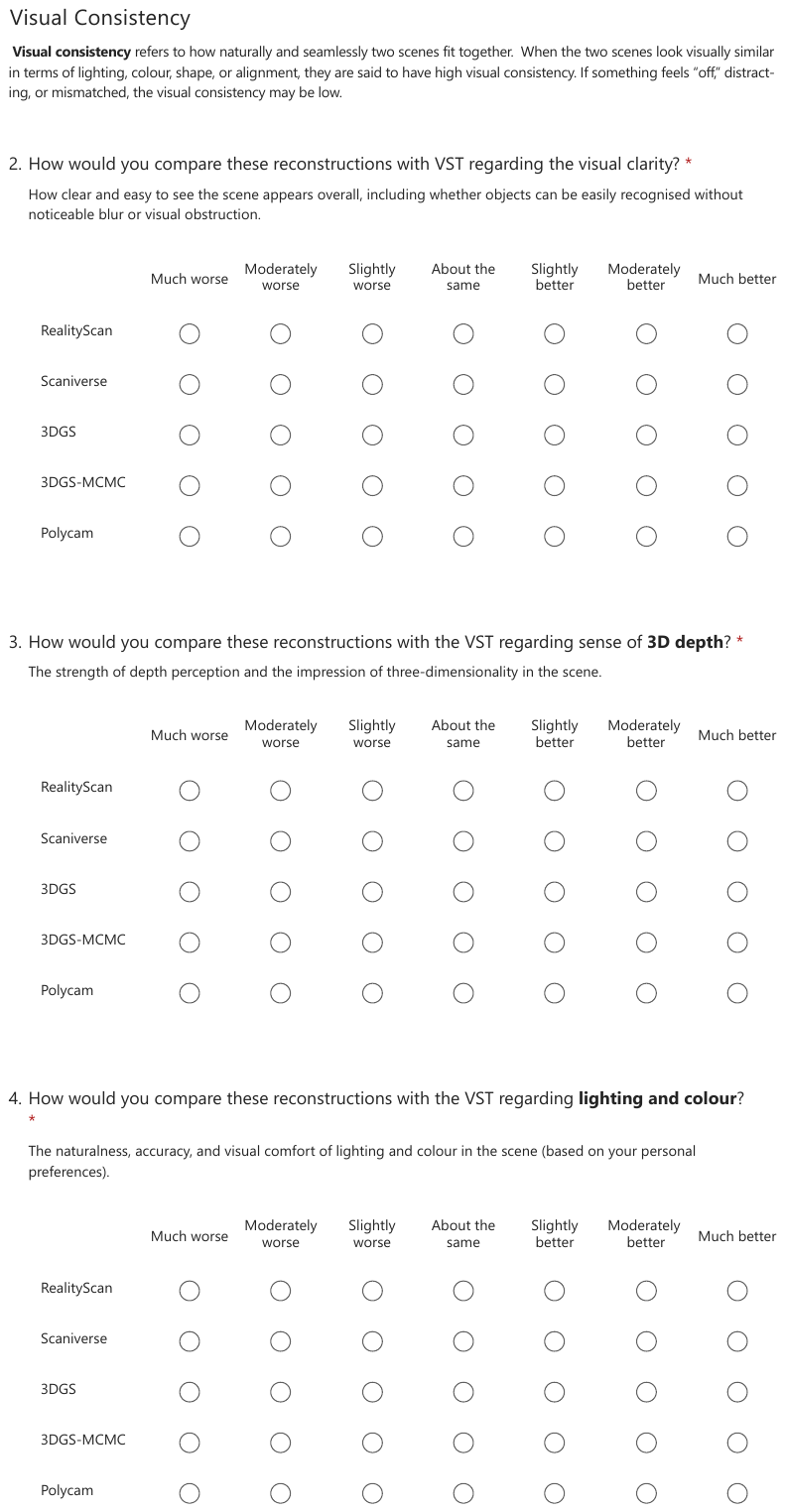}

\end{document}